\documentclass[11pt,a4paper]{article}
\pdfoutput=1
\usepackage[pdftex]{graphics}
\usepackage{jheppub}
\usepackage{amsmath,amssymb,amsfonts}
\usepackage[force]{feynmp-auto}
\usepackage{cancel}
\usepackage[justification=centering]{caption}
\usepackage{subcaption}
\usepackage[export]{adjustbox}
\usepackage{shuffle}

\newcommand{\be}{\begin{eqnarray}}
\newcommand{\ee}{\end{eqnarray}}
\newcommand{\nn}{\nonumber}
\newcommand{\bn}{\begin{enumerate}}
\newcommand{\en}{\end{enumerate}}
\newcommand{\bl}{\begin{align}}
\newcommand{\el}{\end{align}}

\def\identity{{\rlap{1} \hskip 1.6pt \hbox{1}}}
\def\iden{\identity}

\def\CN{\mathcal{N}}
\def\CO{\mathcal{O}}
\def\CP{\mathcal{P}}

\def\g{\gamma}
\def\e{\epsilon}

\def\th{\theta}

\def\k{\kappa}

\def\m{\mu}
\def\n{\nu}

\def\s{\sigma}

\def\t{\tau}

\def\w{\omega}

\def\D{\Delta}

\def\O{\Omega}

\def\identity{{\rlap{1} \hskip 1.6pt \hbox{1}}}

\def\jmath{{j}}

\def\be{{\bar{\epsilon}}}

\usepackage{graphicx}
\usepackage[export]{adjustbox}

\usepackage{bm}

\def\rmx{{
    \scalebox{1.2}[1]{$\mathrm{x}$}\kern-0.625em\scalebox{1.2}[1]{$\mathrm{x}$}
}}
\def\rmy{{
    \scalebox{1.2}[1]{$\mathrm{y}$}\kern-0.645em\scalebox{1.2}[1]{$\mathrm{y}$}\kern-0.02em
}}
\def\rmz{{
    \kern0.055em\scalebox{1.2}[1]{$\mathrm{z}$}\kern-0.528em\scalebox{1.2}[1]{$\mathrm{z}$}\kern0.04em
}}

\newcommand{\dbar}{
    d\kern-.20em\makebox[0pt][l]{$\bar{}$}\kern.20em
}
\newcommand{\deltabar}{
    \delta\kern-.20em\makebox[0pt][l]{$\bar{}$}\kern.20em
}

\usepackage{mathtools}

\newcommand{\lambdabar}{
    \lambda\kern-.20em\makebox[0pt][l]{$\bar{}$}\kern.20em
}

\usepackage[usenames,dvipsnames]{xcolor}
\definecolor{labelcolor}{RGB}{194, 175, 116}

\newif\ifToggleMacros
\ToggleMacrostrue  
\ifToggleMacros
   \newcommand{\sml}[1]{\textbf{\color{violet}{[Sangmin: #1]}}}
   \newcommand{\jwk}[1]{\textbf{\color{blue}{[JW: #1]}}}
   \newcommand{\jh}[1]{\textbf{\color[RGB]{224,26,187}{[Joonhwi: #1]}}}
   \newcommand{\ssk}[1]{\textbf{\color[RGB]{10,185,30}{[Sungsoo: #1]}}}
   \newcommand{\rred}[1]{\textcolor{red}{\; #1\; }}
\else
   \newcommand{\sml}[1]{}
   \newcommand{\jwk}[1]{}
   \newcommand{\jh}[1]{}
   \newcommand{\ssk}[1]{}
   \newcommand{\rred}[1]{}
\fi

\def\I{{\kern0.04em\scalebox{0.9}[0.65]{\textsc{I}}\kern-0.04em}}

\newcommand{\np}[1]{\vcentcolon\mathrel{#1}\vcentcolon}

\title{Classical eikonal from Magnus expansion} 

\author[a]{Joon-Hwi Kim}
\author[b]{Jung-Wook Kim}
\author[c]{Sungsoo Kim}
\author[c,d,e]{Sangmin Lee} 

\affiliation[a]{
Walter Burke Institute for Theoretical Physics,
\\ 
California Institute of Technology, Pasadena, CA 91125, U.S.A.}
\affiliation[b]{Max Planck Institute for Gravitational Physics (Albert Einstein Institute),\\
Am M\"uhlenberg 1, D-14476 Potsdam, Germany}  
\affiliation[c]{Department of Physics and Astronomy, Seoul National University, \\
1 Gwanak-ro, Gwanak-gu, Seoul 08826, Korea}
\affiliation[d]{Center for Theoretical Physics, Seoul National University, \\
1 Gwanak-ro, Gwanak-gu, Seoul 08826, Korea}
\affiliation[e]{College of Liberal Studies, Seoul National University, \\
1 Gwanak-ro, Gwanak-gu, Seoul 08826, Korea}
\emailAdd{joonhwi@caltech.edu}
\emailAdd{jung-wook.kim@aei.mpg.de}
\emailAdd{sooo4017@snu.ac.kr}
\emailAdd{sangmin@snu.ac.kr}

\abstract{
In a classical scattering problem, the classical eikonal is defined as the generator of the canonical transformation that maps in-states to out-states. It can be regarded as the classical limit of the log of the quantum $S$-matrix. 
In a classical analog of the Born approximation in quantum mechanics, the classical eikonal admits an expansion in oriented tree graphs, where oriented edges denote retarded/advanced worldline propagators. 
The Magnus expansion, which takes the log of a time-ordered exponential integral, offers an efficient method to compute the coefficients of the tree graphs to all orders. 
We exploit 
a Hopf algebra structure behind the Magnus expansion to develop a fast algorithm which can compute the tree coefficients up to the 12th order (over half a million trees) in less than an hour. 
In a relativistic setting, our methods can be applied to the post-Minkowskian (PM) expansion for gravitational binaries in the worldline formalism. We demonstrate the methods by computing the 3PM eikonal and find agreement with previous results based on amplitude methods. 
Importantly, the Magnus expansion yields a finite eikonal, while the na\"ive eikonal based on the time-symmetric propagator is infrared-divergent from 3PM on.
}

\begin{document}
\maketitle
\flushbottom

\newpage
\section{Introduction}
\label{sec:intro}

The \emph{eikonal} was originally introduced in geometric optics to study propagation of rays, which leads to equations of motion similar to that of Hamiltonian mechanics; see e.g. \textsection 53 of ref.~\cite{Landau:1975pou} for a short introduction. 
The eikonal played a crucial role in the formulation of quantum mechanics; Schr\"odinger's 
wave equation~\cite{Schrodinger:1926iou} was motivated from the similarity between the eikonal description of wave propagation and Hamiltonian mechanics of particles. After the establishment of quantum mechanics, the \emph{eikonal approximation} gained a narrower meaning of applying WKB approximation to quantum-mechanical scattering problems at small deflection angles, differing from the Born series approach in that the eikonal approximation includes all-order information of the scattering potential~\cite{Sakurai:2011zz}. 

The generalization of the eikonal approximation to quantum field theory (QFT) was initially viewed as a resummation of ladder Feynman diagrams into an exponential form for $2 \to 2$ scattering processes, where the exponent is called the \emph{eikonal phase}
~\cite{Torgerson:1966zz,Cheng:1969eh,Levy:1969cr,Abarbanel:1969ek}.
A more thorough discussion of the eikonal exponentiation can be found, {\it e.g.} in refs.~\cite{Tiktopoulos:1971hi,Meng:1972xt,Weinberg:1971cdi,Czyz:1975bf,Kabat:1992pz}.
Ignoring possible subtleties, we view the eikonal (phase) as the log of the unitary $S$-matrix:
\begin{align}
    \hat{S} = \exp( i\hat{\chi} /\hbar) \,, 
    \label{S-matrix-log}
\end{align}
an idea dating back to early days of QFT~\cite{Feynman:1951gn,Lehmann:1957zz}, 
and recently revived in refs.~\cite{Damgaard:2021ipf,Damgaard:2023ttc}. 
We may call  $\hat{\chi}/\hbar$ (dimensionless) ``phase matrix" 
and $\hat{\chi}$ (dimensionful) ``eikonal matrix". 
%

The recent invigoration of interest in the eikonal 
is partly due to their effectiveness in extracting \emph{classical} physics from QFT amplitudes. The eikonal has a well-defined classical limit in contrast to the usual amplitudes with ``superclassical'' contributions (superficially carrying negative powers of $\hbar$)~\cite{DiVecchia:2021bdo,Cristofoli:2021vyo}; see ref.~\cite{DiVecchia:2023frv} for a comprehensive review.

In ref.~\cite{Kim:2024grz}, inspired by ref.~\cite{Damgaard:2021ipf} as well as the KMOC formula~\cite{Kosower:2018adc} for classical observables, three of the authors proposed that the classical limit of the $\chi$-matrix 
should yield a generator of the canonical transformation\footnote{ 
A similar viewpoint of the radial action as the scattering generator was proposed in ref.~\cite{Gonzo:2024zxo}. However, this proposal is limited by the fact that the radial action is only well-defined for systems described by separable Hamilton-Jacobi equations.} between the two (in and out) asymptotic copies of free particle phase space. 
We call the generator the \emph{classical eikonal}. 

The $S$-matrix in conventional treatments $(\hat{S} = \hat{I} + i\hat{T}/\hbar)$ has an operational definition given by the Dyson series. 
The counterpart of the Dyson series for the $\chi$-matrix is known as the \emph{Magnus series}~\cite{Magnus:1954zz}; see refs.~\cite{Blanes:2008xlr,ebrahimifard2023magnusexpansion} for a review. 
When the $\chi$-matrix was introduced in ref.~\cite{Damgaard:2021ipf}, it was written as a series expansion in the $T$-matrix. 
In this approach, the computation of the $\chi$-matrix elements required 
subtracting superclassical terms from the $T$-matrix. 
The Magnus expansion, consisting of nested commutators, allows us to compute the $\chi$-matrix without ever encountering a superclassical term.

In the bulk of this paper, through sections \ref{sec:eom}-\ref{sec:hopf}, 
in the simple context of non-relativistic scattering, 
we present three methods to compute the classical eikonal 
with varying degrees of mathematical abstraction and computational efficiency, where a worldline description is adopted for massive particles. 
The first approach (section~\ref{sec:eom}) is to solve the equations of motion (EOM) perturbatively 
and deduce the classical eikonal from the impulse (change of momentum), which provides the most straightforward definition of the eikonal.
The second approach (section~\ref{sec:magnus}) uses the Magnus expansion, which provides a shortcut to reach the same classical eikonal, albeit being more abstract than the former.
The third approach (section~\ref{sec:hopf}) relies on the Hopf algebra structure behind the Magnus expansion, 
which provides further efficiency at the price of further abstraction. 
It offers a novel and extremely fast way to compute the classical eikonal. 

In all approaches, the classical eikonal admits a tree-level Feynman diagram expansion, a viewpoint advocated by the worldline quantum field theory (WQFT)~\cite{Mogull:2020sak}. 
It is easy to enumerate all possible tree diagrams. Most of the computations are about how to determine the coefficient of each diagram. 
A subtlety, analyzed in great detail in this paper, is that the Magnus expansion assigns \emph{different causality prescription} of the propagators from the usual Feynman diagram calculations. The causality prescription is important for obtaining the correct eikonal. 
A na\"{i}ve use of Feynman propagators would lead to an infrared(IR)-divergent result for the eikonal.

In section~\ref{sec:PMdyn}, we carry over our methods to the relativistic setting and apply them to the post-Minkowskian (PM) expansion of gravitating binaries, adopting the Magnus expansion as the definition of the eikonal rather than that based on iterating EOM.
The key insight is that the Magnus expansion treats propagation of worldline degrees of freedom (DOF) and propagation of second-quantized field DOF on an equal footing.
Therefore, the tools developed for the dynamics of massive particles also apply to the dynamics of fields.
We demonstrate our methods by computing the 3PM eikonal and find agreement with previous results based on amplitude methods. 
Thus, the causality prescription from the Magnus expansion solves the long-standing puzzle about the IR-divergence of the 3PM eikonal. We expect that the same prescription will render a divergence-free eikonal to all orders.

It is convenient to separate the eikonal depending on the mass ratio according to the so-called self-force (SF) expansion. The 0SF terms treat one of the particles as a static source and the other as a probe. The best way to compute the 0SF eikonal is to use the equivalence between the eikonal and the radial action; see appendix~\ref{app:radial}. 
For the 1SF terms, the diagrammatic expansion developed in the main body of this paper is indispensable. We successfully reproduce the conservative and radiation-reaction parts of the 3PM 1SF eikonal. 

We do not cover the dissipative effects, since including a graviton as an external leg will require a further generalization of our methods. 
When applied to computing the impulse, we find that the eikonal computes the 3PM impulse up to dissipative effects due to radiated gravitational waves. We conjecture that the dissipative terms can be incorporated when we consider the enlarged phase space including the field DOF.

In section~\ref{sec:discussion}, we conclude the paper with discussions on possible generalizations and further applications. 
In appendix~\ref{app:radial}, we explain some aspects of the radial action and its relevance in the 0SF computation. 
In appendix~\ref{app:i0cuts}, we offer a diagrammatic representation of Poisson brackets which forms the basis of the EOM method of section~\ref{sec:eom}. 
In appendix~\ref{app:OPS_Hopf}, we 
provide an intuition for the Hopf algebra structure underlying the Magnus expansion.


We note that recent papers by Ajith et al.~\cite{Ajith:2024fna,Du:2024rkf} gave an in-depth analysis of the relationship between QFT, worldline, WQFT and eikonal exponentiation. 
Conceptually, these works have significant overlap with the current paper, but technically they are rather complementary to the current one. 

\section{Classical eikonal from equation of motion} \label{sec:eom}

\subsection{Classical limit of the exponential representation} 

We recall how the classical eikonal as a scattering generator was motivated by the classical limit of the quantum $S$-matrix~\cite{Kim:2024grz}.
We trade the unitary operator $\hat{S}$ for a hermitian operator $\hat{\chi}$ as 
$\hat{S} = \exp(i \hat{\chi}/\hbar)$~\cite{Lehmann:1957zz,Damgaard:2021ipf,Damgaard:2023ttc}. 
The change in a scattering observable, as in the KMOC formula~\cite{Kosower:2018adc}, can be expressed in terms of  $\hat{\chi}$ as~\cite{Damgaard:2023ttc}
\begin{align} \label{eq:KMOC-N}
    \hat{O}_{\text{out}} = \hat{S}^\dagger \hat{O}_{\text{in}} \hat{S} = e^{-i\hat{\chi}/\hbar} O_{\text{in}} e^{+i\hat{\chi}/\hbar} =: (e^{-i\hat{\chi}/\hbar})_{\text{adj}}[O_{\text{in}}] \,,
\end{align}
where the adjoint action of an operator is defined as an action through the commutator, 
\begin{align}
\begin{aligned}
    \hat{A}_{\text{adj}} [\hat{B}] &:= [\hat{A},\hat{B}] \,,
    \\ (e^{\hat{A}})_{\text{adj}}[\hat{B}] &= \sum_{n=0}^\infty \frac{1}{n!} \underbrace{[\hat{A},[\hat{A},\cdots,[\hat{A},}_{n \text{ times}} \hat{B}] \cdots ]] = \hat{B} + [\hat{A},\hat{B}] + \frac{1}{2!} [\hat{A},[\hat{A},\hat{B}]] + \cdots \,.
\end{aligned}
\end{align}
The equation \eqref{eq:KMOC-N} has an interpretation as a symmetry transformation; $\hat{O}_{\text{out}}$ is obtained from $\hat{O}_{\text{in}}$ by a symmetry transformation having $\hat{\chi}$ as the generator. Taking the classical limit of \eqref{eq:KMOC-N} turns the operator $\hat{\chi}$ to the classical eikonal $\chi$ and replaces the commutator by the Poisson bracket, 
\begin{align}
   \frac{1}{i \hbar}[ \hat{A} , \hat{B}] \;\; \to \;\; \{ A, B \} \,, 
   \label{Dirac-Poisson} 
\end{align}
we arrive at the interpretation of the classical eikonal as the 
scattering generator~\cite{Kim:2024grz},
\begin{align}
\begin{aligned} \label{eq:eik_scgen}
    O_{\text{out}} &= e^{\{ \chi, \bullet \}} [O_{\text{in}}]
    \\ &= O_{\text{in}} + \{ \chi , O_{\text{in}} \} + \frac{1}{2!} \{ \chi , \{ \chi , O_{\text{in}} \} \} + \frac{1}{3!} \{ \chi , \{ \chi , \{ \chi , O_{\text{in}} \} \} \} + \cdots \,,
\end{aligned}
\end{align}
where the final state scattering observable $O_{\text{out}}$ is obtained from the initial state scattering observable $O_{\text{in}}$ by a canonical transform $e^{\{ \chi, \bullet \}}$ generated by the eikonal $\chi$.

Note that this viewpoint of the eikonal is rather different from the typical viewpoint found in the literature, where the classical eikonal is understood as a saddle-point approximation of the scattering amplitude~\cite{DiVecchia:2023frv}. 
The latter viewpoint forces redefinition of the impact parameter due to longitudinal component of the impulse, leading to the distinction between the impact parameter $b^\m$ and the eikonal impact parameter $b_{\text{eik}}^\m$.
The viewpoint advocated here simply uses $b^\m$ and the ``frame rotation'' to $b_{\text{eik}}^\m$ is understood as a by-product of iterated brackets, which arises due to the ``noncommutative'' nature of the impact parameter space $\{ b^\m , b^\n \} \neq 0$~\cite{Kim:2023vgb,Kim:2024grz}.

The ``de-quantization" to Poisson brackets can be understood as Lie brackets of vector fields in the usual sense in Hamiltonian mechanics: 
\begin{align}
    X_f[h] = \{ f, h\} 
    \quad \Longrightarrow \quad 
    [X_f,X_g][h] = X_{\{f,g\}}[h] \,,  
\end{align}
where $X[h]$ denotes the derivative action of a vector field $X$ on a scalar function $h$. For a pair of vector fields, $[X,Y]$ denotes the Lie bracket. 
The identity relating Lie bracket and Poisson bracket holds thanks to the Jacobi identity.

\subsection{Equation of motion and iteration}

A straightforward (but inefficient) way to compute the classical eikonal is 
to solve the EOM perturbatively and use \eqref{eq:eik_scgen}. 
We compute $\chi_{(1)}$ and $\chi_{(2)}$ explicitly and develop a graphical method 
to reach higher orders.
For simplicity, we work in Newtonian mechanics with the Hamiltonian 
\begin{align}
    H = \frac{\vec{p}^2}{2m} + V(\vec{x}) \,.
\end{align}
The generalization to relativistic particles should be straightforward. 

The first order computation is almost trivial. 
\begin{align}
    \Delta_{(1)} \vec{p} = - \int_{-\infty}^\infty \partial_{\vec{x}} V(\vec{b}+\vec{v}t) dt 
    = \{ \chi_{(1)} , \vec{p} \} 
    \quad
    \Longrightarrow
    \quad 
    \chi_{(1)} = - \int_{-\infty}^{+\infty} V(\vec{b}+\vec{v}t) dt \,.
\end{align}
To avoid clutter, we will sometimes denote $ V(\vec{b}+\vec{v}t_a)$ by $V(t_a)$ or $V_a$ below. 
The integration range is to be understood as from $t=-\infty$ to $t= + \infty$ unless otherwise specified. 

\paragraph{2nd order}
At the second order, the EOM for $\vec{p}$ reads 
\begin{align}
     \frac{d}{dt} \vec{p}_{(2)} 
     = -  \partial_i \partial_{\vec{x}} V(\vec{b} + \vec{v} t)  x^i_{(1)}(t)  \,.
\label{2nd-EOM}
\end{align}
We need to feed in the first order solution $\vec{x}_{(1)}(t)$ at finite time. 
In terms of the retarded  Green's function,
\begin{align}
\begin{split}
    &R_{12} = (t_1 - t_2) \theta(t_1-t_2)  \,, 
\\
    &\left( \frac{d}{dt_1} \right)^2 R_{12} = \delta(t_1 -t_2) \,,
    \quad 
    R_{12} = 0 \quad \mbox{if} \quad t_1 < t_2 \,, 
\end{split}
    \label{retarded-propagator}
\end{align}
the first order solution is 
\begin{align}
    \vec{x}_{(1)} (t) = - \frac{1}{m} \int R_{ts}\, \partial_{\vec{x}} V(s) ds \,.
\end{align}
The second order impulse is then 
\begin{align}
    \Delta_{(2)}\vec{p} = \frac{1}{m}  \int  R_{ts} \left[ \partial_i \partial_{\vec{x}} V(t) \partial_i V(s) \right] ds dt \,. 
    \label{2nd-impulse}
\end{align}
The next step is to subtract the iteration term, 
\begin{align}
    \frac{1}{2} \{ \chi_{(1)} , \{ \chi_{(1)} , \vec{p} \} \} =  \frac{1}{2} \{ \chi_{(1)} , \Delta_{(1)} \vec{p} \} 
    = - \frac{1}{2m} \int (s-t) \partial_i V(s)  \partial_i \partial_{\vec{x}}V(t) \,.
    \label{1st-cut-a}
\end{align}
The $(s-t)$ factor is produced by the Poisson bracket 
\begin{align}
    \left\{ x_i + \frac{p_i}{m} s , x_j + \frac{p_j}{m} t \right\} = -\frac{1}{m} (s-t) \delta_{ij} 
    = -\frac{1}{m} (R_{st} - R_{ts}) \delta_{ij} \,. 
       \label{1st-cut-b}
\end{align}
We call the combination $(R_{st} - R_{ts})/m$ a \emph{causality cut}. 
The second order eikonal is 
\begin{align}
\begin{split}
\{\chi_{(2)} , \vec{p} \} = \D_{(2)}\vec{p} - \frac{1}{2} \{ \chi_{(1)} , \Delta_{(1)} \vec{p} \} 
\;\; \Longrightarrow \;\;
\chi_{(2)} =  \frac{1}{2m} \int R_{ts}[ \partial_{\vec{x}} V(t) \cdot \partial_{\vec{x}} V(s) ] ds dt\,.
\end{split}
\label{2nd-eiknonal}
\end{align}

\paragraph{Diagrams for EOM}

We introduce a diagrammatic representation of the computation so far. 
The basic ingredients are as follows:
\begin{align}
    \chi_{(1)} =  \; - \;\;
      \begin{fmffile}{dot}
        \parbox{10pt}{
        \begin{fmfgraph*}(5,10)\fmfkeep{dot}
            \fmfleft{i1}
            \fmfv{decor.shape=circle,decor.size=4}{i1}
        \end{fmfgraph*}
        }
    \end{fmffile} \hskip -1mm , 
    \qquad 
    \D_{(1)} \vec{p}\; = 
    \; - \;\; 
    \begin{fmffile}{penta}
        \parbox{10pt}{
        \begin{fmfgraph*}(10,10)\fmfkeep{penta} 
            \fmfleft{i1}
            \fmfv{decor.shape=pentagram,decor.size=6}{i1}
        \end{fmfgraph*}
        }
    \end{fmffile} \hskip -1mm , 
        \qquad 
    \frac{1}{m} R_{12} =  \;\;   
    \begin{fmffile}{retardedG}
        \parbox{30pt}{
        \begin{fmfgraph*}(30,10)
            \fmfleft{o1}
            \fmfright{i1}
            \fmfv{decor.shape=circle,decor.filled=empty,decor.size=6,label=$2$,label.angle=-90}{i1}
            \fmfv{decor.shape=circle,decor.filled=empty,decor.size=6,label=$1$,label.angle=-90}{o1}
            \fmf{fermion}{i1,o1}
        \end{fmfgraph*}
        }
    \end{fmffile} \;\; .
\end{align}
The pentagram vertex, corresponding to the impulse $\D_{(1)} \vec{p} = \{ \chi_{(1)} , \vec{p} \}$, can be interpreted as the derivative operator $\{ \bullet, \vec{p} \}$ acting on the vertex corresponding to $\chi_{(1)}$. 
The white circles at the two endpoints of the retarded propagator $R_{12}$ can be taken by a black dot or a pentagram from any subdiagram. 

The second order impulse \eqref{2nd-impulse} is mapped to 
\begin{align}
\begin{aligned}
    \D_{(2)} \vec{p} &= \;\;
    \begin{fmffile}{2ndEOM}
        \parbox{30pt}{
        \begin{fmfgraph*}(30,10)\fmfkeep{2ndEOM}
            \fmfleft{o1}
            \fmfright{i1}
            \fmfv{decor.shape=circle,decor.size=4}{i1}
            \fmfv{decor.shape=pentagram,decor.size=6}{o1}
            \fmf{fermion}{i1,o1}
        \end{fmfgraph*}
        } \;\; .
    \end{fmffile}
\end{aligned}
\end{align}
The cut in \eqref{1st-cut-a} and \eqref{1st-cut-b} can be represented by 
\begin{align}
\begin{aligned}
\{ \chi_{(1)} , \Delta_{(1)} \vec{p} \, \} &= 
    \left\{ \;\;
    \parbox{10pt}{\fmfreuse{dot}}
    \hskip -1mm , \hskip 2mm
     \parbox{10pt}{\fmfreuse{penta}}
    \hskip -1mm
    \right\} 
    \\
    &= \;\;    
    \begin{fmffile}{cut-1st}
        \parbox{30pt}{
        \begin{fmfgraph*}(30,10)
            \fmfleft{i1}
            \fmfright{o1}
            \fmfv{decor.shape=circle,decor.size=4}{i1}
            \fmfv{decor.shape=pentagram,decor.size=6}{o1}
            \fmf{phantom}{i1,c,o1}
            \fmfv{decor.shape=cross,decor.size=10}{c}
            \fmf{fermion}{i1,o1}
        \end{fmfgraph*}
        }
    \end{fmffile} 
    \;\; = \;\; 
       \begin{fmffile}{cut-1st-1}
        \parbox{30pt}{
        \begin{fmfgraph*}(30,10)
            \fmfleft{i1}
            \fmfright{o1}
            \fmfv{decor.shape=circle,decor.size=4}{i1}
            \fmfv{decor.shape=pentagram,decor.size=6}{o1}
            \fmf{fermion}{i1,o1}
        \end{fmfgraph*}
        }
    \end{fmffile} 
    \;\; - \;\; 
       \begin{fmffile}{cut-1st-2}
        \parbox{30pt}{
        \begin{fmfgraph*}(30,10)
            \fmfleft{i1}
            \fmfright{o1}
            \fmfv{decor.shape=circle,decor.size=4}{i1}
            \fmfv{decor.shape=pentagram,decor.size=6}{o1}
            \fmf{fermion}{o1,i1}
        \end{fmfgraph*}
        }
    \end{fmffile} \;\; .
\end{aligned}
\end{align}
The last step in determining $\chi_{(2)}$ in \eqref{2nd-eiknonal} can be depicted as 
\begin{align}
\begin{aligned}
 \{ \chi_{(2)} , \vec{p} \} &= \Delta_{(2)} \vec{p} - \frac{1}{2} \{ \chi_{(1)} , \Delta_{(1)} \vec{p} \} 
 \\
 &= \frac{1}{2} \left( \;\;
 \parbox{30pt}{\fmfreuse{2ndEOM}}
    \;\; + \;\; 
       \begin{fmffile}{cut-1st-4}
        \parbox{30pt}{
        \begin{fmfgraph*}(30,10)
            \fmfleft{i1}
            \fmfright{o1}
            \fmfv{decor.shape=circle,decor.size=4}{i1}
            \fmfv{decor.shape=pentagram,decor.size=6}{o1}
            \fmf{fermion}{o1,i1}
        \end{fmfgraph*}
        }
    \end{fmffile}
    \;\; \right) 
    = \left\{ \frac{1}{2} \;\; 
 \begin{fmffile}{chi2}
        \parbox{30pt}{
        \begin{fmfgraph*}(30,10)\fmfkeep{chi2}
            \fmfleft{i1}
            \fmfright{o1}
            \fmfv{decor.shape=circle,decor.size=4}{i1}
            \fmfv{decor.shape=circle,decor.size=4}{o1}
            \fmf{fermion}{o1,i1}
        \end{fmfgraph*}
        }
    \end{fmffile}
    \;\; , 
    \vec{p}
    \, \right\} \,.
\end{aligned}
\end{align}
Similar to $\D_{(1)} \vec{p}$, the pentagram vertex can be interpreted as the derivative operator $\{ \bullet , \vec{p} \}$ acting on the vertex. The sum over all possible single pentagram vertex substitutions corresponds to the Leibniz rule for derivatives.

\paragraph{3rd order}

The third order EOM, 
\begin{align}
     \frac{d}{dt} \vec{p}_{(3)} 
     = -   \partial_i \partial_{\vec{x}} V(t) \,x^i_{(2)}(t) - \frac{1}{2} \,\partial_i \partial_j \partial_{\vec{x}} V(t) \,x^i_{(1)}(t) \,x^j_{(1)}(t) \,,
\end{align}
leads to the following diagrammatic representation of the impulse, 
\begin{align}
\begin{aligned}
    \D_{(3)} \vec{p} &= \;\; - \;\; 
    \begin{fmffile}{3pmimp1}
        \parbox{50pt}{
        \begin{fmfgraph*}(50,10)\fmfkeep{3pmimp1}
            \fmfleft{o1}
            \fmfright{i1}
            \fmfv{decor.shape=circle,decor.size=4}{i1}
            \fmfv{decor.shape=pentagram,decor.size=6}{o1}
            \fmf{fermion}{i1,c,o1}
            \fmfv{decor.shape=circle,decor.size=5}{c}
        \end{fmfgraph*}
        }
    \end{fmffile}
    \;\; - \frac{1}{2} \;\;\;
    \begin{fmffile}{3pmimp2}
        \parbox{25pt}{
        \begin{fmfgraph*}(25,20)\fmfkeep{3pmimp2}
            \fmfleft{o1}
            \fmfright{i1,i2}
            \fmfv{decor.shape=circle,decor.size=4}{i1}
            \fmfv{decor.shape=circle,decor.size=4}{i2}
            \fmfv{decor.shape=pentagram,decor.size=6}{o1}
            \fmf{fermion}{i1,o1}
            \fmf{fermion}{i2,o1}
        \end{fmfgraph*}
        }
    \end{fmffile} \;\; .
\end{aligned}
\end{align}
The diagrammatic computation of $\chi_{(3)}$ can be summarized as 
\begin{align}
\begin{aligned}
    &\{ \chi_{(3)} , \vec{p} \} 
    \\
    &= 
    \D_{(3)} \vec{p} - \frac{1}{2} \{ \chi_{(1)} , \{ \chi_{(2)} , \vec{p} \} \} - \frac{1}{2} \{ \chi_{(2)} , \D_{(1)} \vec{p} \} - \frac{1}{6} \{ \chi_{(1)} , \{ \chi_{(1)} , \D_{(1)} \vec{p} \} \}
    \\
     &= \D_{(3)} \vec{p} - \frac{1}{2} \{ \chi_{(1)} , \D_{(2)} \vec{p} \} - \frac{1}{2} \{ \chi_{(2)} , \D_{(1)} \vec{p} \} + \frac{1}{12} \{ \chi_{(1)} , \{ \chi_{(1)} , \D_{(1)} \vec{p} \} \}
    \\ &= - \frac{1}{3} \left( \;\;
    \parbox{50pt}{\fmfreuse{3pmimp1}}
    \;\; + \;\;
    \begin{fmffile}{3pmimp10}
        \parbox{50pt}{
        \begin{fmfgraph*}(50,10)
            \fmfleft{o1}
            \fmfright{i1}
            \fmfv{decor.shape=circle,decor.size=4}{i1}
            \fmfv{decor.shape=circle,decor.size=4}{o1}
            \fmf{fermion}{i1,c,o1}
            \fmfv{decor.shape=pentagram,decor.size=6}{c}
        \end{fmfgraph*}
        }
    \end{fmffile}
    \;\; + \;\;
    \begin{fmffile}{3pmimp11}
        \parbox{50pt}{
        \begin{fmfgraph*}(50,10)
            \fmfleft{o1}
            \fmfright{i1}
            \fmfv{decor.shape=pentagram,decor.size=6}{i1}
            \fmfv{decor.shape=circle,decor.size=4}{o1}
            \fmf{fermion}{i1,c,o1}
            \fmfv{decor.shape=circle,decor.size=4}{c}
        \end{fmfgraph*}
        }
    \end{fmffile}
    \;\; \right)
    \\ &\qquad - \frac{1}{12} \left(\;\;
    \parbox{25pt}{\fmfreuse{3pmimp2}}
    \;\; + 2 \;\;\;
    \begin{fmffile}{3pmimp21}
        \parbox{25pt}{
        \begin{fmfgraph*}(25,20)
            \fmfleft{o1}
            \fmfright{i1,i2}
            \fmfv{decor.shape=circle,decor.size=4}{i1}
            \fmfv{decor.shape=pentagram,decor.size=6}{i2}
            \fmfv{decor.shape=circle,decor.size=4}{o1}
            \fmf{fermion}{i1,o1}
            \fmf{fermion}{i2,o1}
        \end{fmfgraph*}
        }
    \end{fmffile}
    \;\; \right) - \frac{1}{12} \left(\;\;
    \begin{fmffile}{3pmchi_rev1}
        \parbox{25pt}{
        \begin{fmfgraph*}(25,20)
            \fmfright{o1}
            \fmfleft{i1,i2}
            \fmfv{decor.shape=circle,decor.size=4}{i1}
            \fmfv{decor.shape=circle,decor.size=4}{i2}
            \fmfv{decor.shape=pentagram,decor.size=6}{o1}
            \fmf{fermion}{o1,i1}
            \fmf{fermion}{o1,i2}
        \end{fmfgraph*}
        }
    \end{fmffile}
    \;\; + 2 \;\;
    \begin{fmffile}{3pmchi_rev2}
        \parbox{25pt}{
        \begin{fmfgraph*}(25,20)
            \fmfright{o1}
            \fmfleft{i1,i2}
            \fmfv{decor.shape=circle,decor.size=4}{i1}
            \fmfv{decor.shape=pentagram,decor.size=6}{i2}
            \fmfv{decor.shape=circle,decor.size=4}{o1}
            \fmf{fermion}{o1,i1}
            \fmf{fermion}{o1,i2}
        \end{fmfgraph*}
        }
    \end{fmffile}
    \;\; \right)
    \\ &= - \left\{ \frac{1}{3} \;\;
    \begin{fmffile}{3pmchi1}
        \parbox{50pt}{
        \begin{fmfgraph*}(50,10)
            \fmfleft{o1}
            \fmfright{i1}
            \fmfv{decor.shape=circle,decor.size=4}{i1}
            \fmfv{decor.shape=circle,decor.size=4}{o1}
            \fmf{fermion}{i1,c,o1}
            \fmfv{decor.shape=circle,decor.size=4}{c}
        \end{fmfgraph*}
        }
    \end{fmffile}
    \;\; + \frac{1}{12} \left( \;\;
    \begin{fmffile}{3pmchi2}
        \parbox{25pt}{
        \begin{fmfgraph*}(25,20)
            \fmfleft{o1}
            \fmfright{i1,i2}
            \fmfv{decor.shape=circle,decor.size=4}{i1}
            \fmfv{decor.shape=circle,decor.size=4}{i2}
            \fmfv{decor.shape=circle,decor.size=4}{o1}
            \fmf{fermion}{i1,o1}
            \fmf{fermion}{i2,o1}
        \end{fmfgraph*}
        }
    \end{fmffile}
    \;\; + \;\;
    \begin{fmffile}{3pmchi3}
        \parbox{25pt}{
        \begin{fmfgraph*}(25,20)
            \fmfright{o1}
            \fmfleft{i1,i2}
            \fmfv{decor.shape=circle,decor.size=4}{i1}
            \fmfv{decor.shape=circle,decor.size=4}{i2}
            \fmfv{decor.shape=circle,decor.size=4}{o1}
            \fmf{fermion}{o1,i1}
            \fmf{fermion}{o1,i2}
        \end{fmfgraph*}
        }
    \end{fmffile}
    \;\; \right)
    , \; \vec{p} \right\} \,.
\end{aligned}
\label{chi3-EOM-final}
\end{align}
Each term on the RHS can be computed diagrammatically. 
For example, 
\begin{align}
\begin{aligned}
- \{  \chi_{(1)} , \D_{(2)} \vec{p} \}
    &= \left\{ \;\;
    \parbox{10pt}{\fmfreuse{dot}}
   , \;\;
    \begin{fmffile}{pb12-b}
        \parbox{10pt}{
        \begin{fmfgraph*}(10,25)
            \fmfleft{o1,o2}
            \fmfv{decor.shape=circle,decor.size=4}{o1}
            \fmfv{decor.shape=pentagram,decor.size=6}{o2}
            \fmf{fermion}{o1,o2}
        \end{fmfgraph*}
        }
    \end{fmffile}
    \right\}
    = \;\;
    \begin{fmffile}{pb12_1}
        \parbox{30pt}{
        \begin{fmfgraph*}(30,25)
            \fmfleft{i1,i2}
            \fmfright{o1,o2}
            \fmfv{decor.shape=circle,decor.size=4}{i2}
            \fmfv{decor.shape=circle,decor.size=4}{o1}
            \fmfv{decor.shape=pentagram,decor.size=6}{o2}
            \fmf{fermion}{o1,o2}
            \fmf{phantom}{i2,c,o2}
            \fmfv{decor.shape=cross,decor.size=10}{c}
            \fmf{fermion}{i2,o2}
        \end{fmfgraph*}
        }
    \end{fmffile}
    \;\; + \;\;
    \begin{fmffile}{pb12_2}
        \parbox{30pt}{
        \begin{fmfgraph*}(30,25)
            \fmfleft{i1,i2}
            \fmfright{o1,o2}
            \fmfv{decor.shape=circle,decor.size=4}{i1}
            \fmfv{decor.shape=circle,decor.size=4}{o1}
            \fmfv{decor.shape=pentagram,decor.size=6}{o2}
            \fmf{fermion}{o1,o2}
            \fmf{phantom}{i1,c,o1}
            \fmfv{decor.shape=cross,decor.size=10}{c}
            \fmf{fermion}{i1,o1}
        \end{fmfgraph*}
        }
    \end{fmffile}
    \\
    \\ &= \left( \;\;
    \begin{fmffile}{pb12_11}
        \parbox{30pt}{
        \begin{fmfgraph*}(30,25)
            \fmfleft{i1,i2}
            \fmfright{o1,o2}
            \fmfv{decor.shape=circle,decor.size=4}{i2}
            \fmfv{decor.shape=circle,decor.size=4}{o1}
            \fmfv{decor.shape=pentagram,decor.size=6}{o2}
            \fmf{fermion}{o1,o2}
            \fmf{fermion}{i2,o2}
        \end{fmfgraph*}
        }
    \end{fmffile}
    \;\; - \;\;
    \begin{fmffile}{pb12_12}
        \parbox{30pt}{
        \begin{fmfgraph*}(30,25)
            \fmfleft{i1,i2}
            \fmfright{o1,o2}
            \fmfv{decor.shape=circle,decor.size=4}{i2}
            \fmfv{decor.shape=circle,decor.size=4}{o1}
            \fmfv{decor.shape=pentagram,decor.size=6}{o2}
            \fmf{fermion}{o1,o2}
            \fmf{fermion}{o2,i2}
        \end{fmfgraph*}
        }
    \end{fmffile}
    \;\; \right) + \left( \;\;
    \begin{fmffile}{pb12_21}
        \parbox{35pt}{
        \begin{fmfgraph*}(30,25)
            \fmfleft{i1,i2}
            \fmfright{o1,o2}
            \fmfv{decor.shape=circle,decor.size=4}{i1}
            \fmfv{decor.shape=circle,decor.size=4}{o1}
            \fmfv{decor.shape=pentagram,decor.size=6}{o2}
            \fmf{fermion}{o1,o2}
            \fmf{fermion}{i1,o1}
        \end{fmfgraph*}
        }
    \end{fmffile}
    \;\; - \;\;
    \begin{fmffile}{pb12_22}
        \parbox{30pt}{
        \begin{fmfgraph*}(30,25)
            \fmfleft{i1,i2}
            \fmfright{o1,o2}
            \fmfv{decor.shape=circle,decor.size=4}{i1}
            \fmfv{decor.shape=circle,decor.size=4}{o1}
            \fmfv{decor.shape=pentagram,decor.size=6}{o2}
            \fmf{fermion}{o1,o2}
            \fmf{fermion}{o1,i1}
        \end{fmfgraph*}
        }
    \end{fmffile}
    \;\; \right) \,.
\end{aligned}
\end{align}
Similarly, 
\begin{align}
\begin{aligned}
- 2 \{  \chi_{(2)} , \D_{(1)} \vec{p} \}
    &= \left\{ \;\;
        \begin{fmffile}{pb21-a}
        \parbox{10pt}{
        \begin{fmfgraph*}(10,25)
            \fmfleft{o1,o2}
            \fmfv{decor.shape=circle,decor.size=4}{o1}
            \fmfv{decor.shape=circle,decor.size=4}{o2}
            \fmf{fermion}{o1,o2}
        \end{fmfgraph*}
        }
    \end{fmffile}
     , \;\;
     \parbox{10pt}{\fmfreuse{penta}}
    \right\}
    = \;\;
    \begin{fmffile}{pb21_1}
        \parbox{30pt}{
        \begin{fmfgraph*}(30,25)
            \fmfleft{i1,i2}
            \fmfright{o1,o2}
            \fmfv{decor.shape=circle,decor.size=4}{i1}
            \fmfv{decor.shape=circle,decor.size=4}{i2}
            \fmfv{decor.shape=pentagram,decor.size=6}{o2}
            \fmf{fermion}{i1,i2}
            \fmf{phantom}{i2,c,o2}
            \fmfv{decor.shape=cross,decor.size=10}{c}
            \fmf{fermion}{i2,o2}
        \end{fmfgraph*}
        }
    \end{fmffile}
    \;\; + \;\;
    \begin{fmffile}{pb21_2}
        \parbox{30pt}{
        \begin{fmfgraph*}(30,25)
            \fmfleft{i1,i2}
            \fmfright{o1,o2}
            \fmfv{decor.shape=circle,decor.size=4}{i1}
            \fmfv{decor.shape=circle,decor.size=4}{i2}
            \fmfv{decor.shape=pentagram,decor.size=6}{o1}
            \fmf{fermion}{i1,i2}
            \fmf{phantom}{i1,c,o1}
            \fmfv{decor.shape=cross,decor.size=10}{c}
            \fmf{fermion}{i1,o1}
        \end{fmfgraph*}
        }
    \end{fmffile}
    \\
    \\ &= \left( \;\;
    \begin{fmffile}{pb21_11}
        \parbox{30pt}{
        \begin{fmfgraph*}(30,25)
            \fmfleft{i1,i2}
            \fmfright{o1,o2}
            \fmfv{decor.shape=circle,decor.size=4}{i1}
            \fmfv{decor.shape=circle,decor.size=4}{i2}
            \fmfv{decor.shape=pentagram,decor.size=6}{o2}
            \fmf{fermion}{i1,i2}
            \fmf{fermion}{i2,o2}
        \end{fmfgraph*}
        }
    \end{fmffile}
    \;\; - \;\;
    \begin{fmffile}{pb21_12}
        \parbox{30pt}{
        \begin{fmfgraph*}(30,25)
            \fmfleft{i1,i2}
            \fmfright{o1,o2}
            \fmfv{decor.shape=circle,decor.size=4}{i1}
            \fmfv{decor.shape=circle,decor.size=4}{i2}
            \fmfv{decor.shape=pentagram,decor.size=6}{o2}
            \fmf{fermion}{i1,i2}
            \fmf{fermion}{o2,i2}
        \end{fmfgraph*}
        }
    \end{fmffile}
    \;\; \right) + \left( \;\;
    \begin{fmffile}{pb21_21}
        \parbox{30pt}{
        \begin{fmfgraph*}(30,25)
            \fmfleft{i1,i2}
            \fmfright{o1,o2}
            \fmfv{decor.shape=circle,decor.size=4}{i1}
            \fmfv{decor.shape=circle,decor.size=4}{i2}
            \fmfv{decor.shape=pentagram,decor.size=6}{o1}
            \fmf{fermion}{i1,i2}
            \fmf{fermion}{i1,o1}
        \end{fmfgraph*}
        }
    \end{fmffile}
    \;\; - \;\;
    \begin{fmffile}{pb21_22}
        \parbox{30pt}{
        \begin{fmfgraph*}(30,25)
            \fmfleft{i1,i2}
            \fmfright{o1,o2}
            \fmfv{decor.shape=circle,decor.size=4}{i1}
            \fmfv{decor.shape=circle,decor.size=4}{i2}
            \fmfv{decor.shape=pentagram,decor.size=6}{o1}
            \fmf{fermion}{i1,i2}
            \fmf{fermion}{o1,i1}
        \end{fmfgraph*}
        }
    \end{fmffile}
    \;\; \right) \,.
\end{aligned}
\end{align}
In both of the examples, the diagrams were organized so that horizontally oriented propagators originate from Poisson brackets computed as causality cuts.
We leave the last term, $\{ \chi_{(1)} , \{ \chi_{(1)}, \D_{(1)} \vec{p} \} \}$, as an exercise to the readers. 

\paragraph{Higher orders} 

\begin{figure}[htbp]
    \centering
    \includegraphics[width=0.85\linewidth]{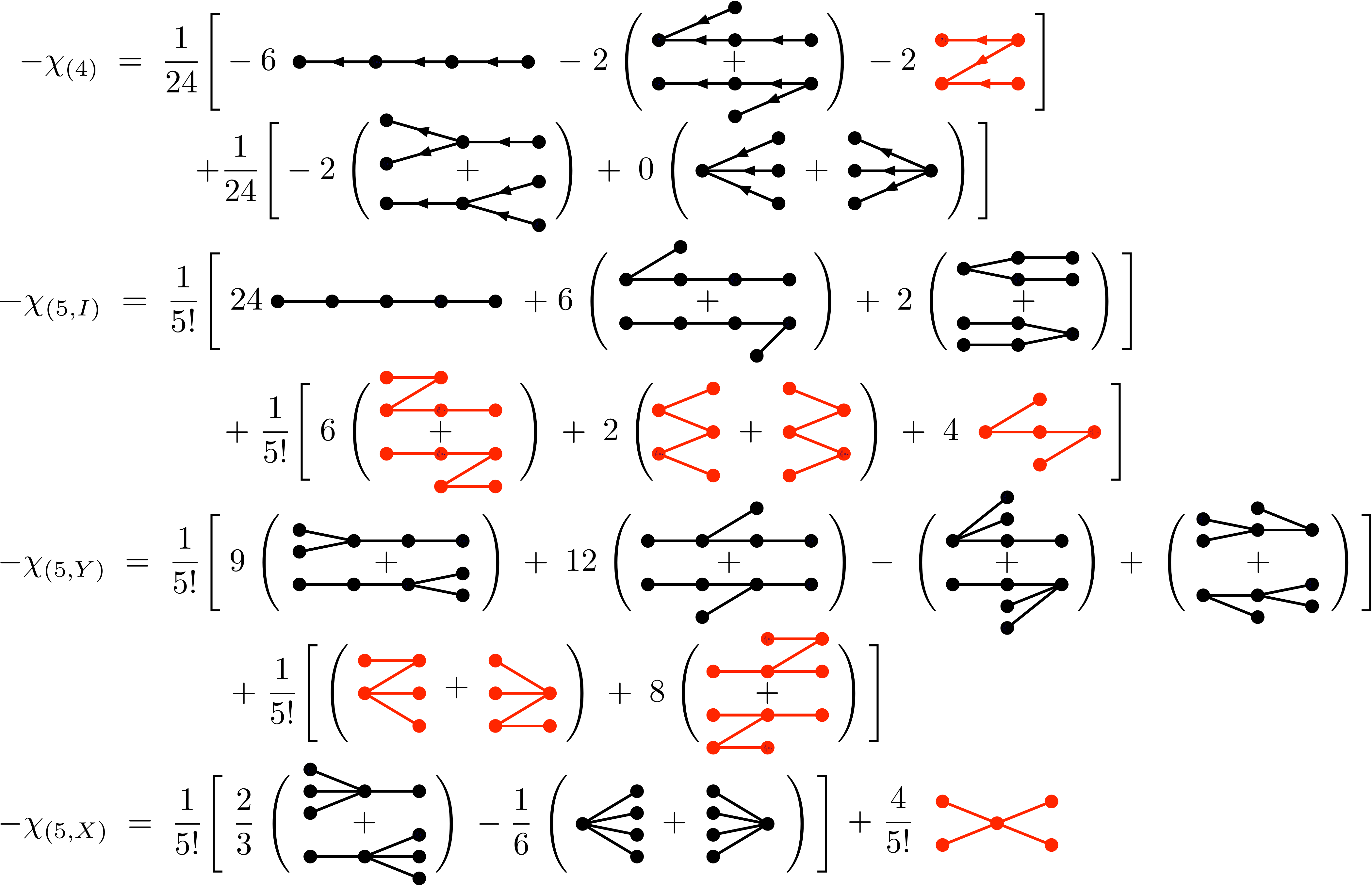}
    \caption{$\chi_{(4)}$ and $\chi_{(5)}$.
    The black trees are rooted, while the red ones are non-rooted. For $\chi_{(5)}$, it is understood that all edges are oriented with an arrow from right to left.}
    \label{fig:Magnus-5}
\end{figure}

The diagrammatic computation generalizes straightforwardly to higher orders, 
as shown in detail in appendix~\ref{app:i0cuts}, 
but the computational cost grows quite quickly.
We automated the procedure using \texttt{Mathematica}, which produces $\chi_{(n)}$ up to $n=7$ in a few minutes. 
For the last illustration before we move on to more advanced methods, 
we present the results for $n=4,5$ in Figure~\ref{fig:Magnus-5}. 
%

\subsection{Tree graphs, functions and integrals} \label{tree functions}

\paragraph{WQFT}

Our diagrammatic notation arose from an attempt to refine a similar one introduced by the worldline quantum field theory (WQFT) initiated in ref.~\cite{Mogull:2020sak}. There it was suggested that the classical eikonal $\chi$ can be computed from the logarithm of the partition function of the WQFT in the background of the free (straight line) worldline. The classical limit collects tree diagrams only. The logarithm collects connected diagrams only. 

\begin{figure}[htbp]
    \centering
    \includegraphics[width=0.95\linewidth]{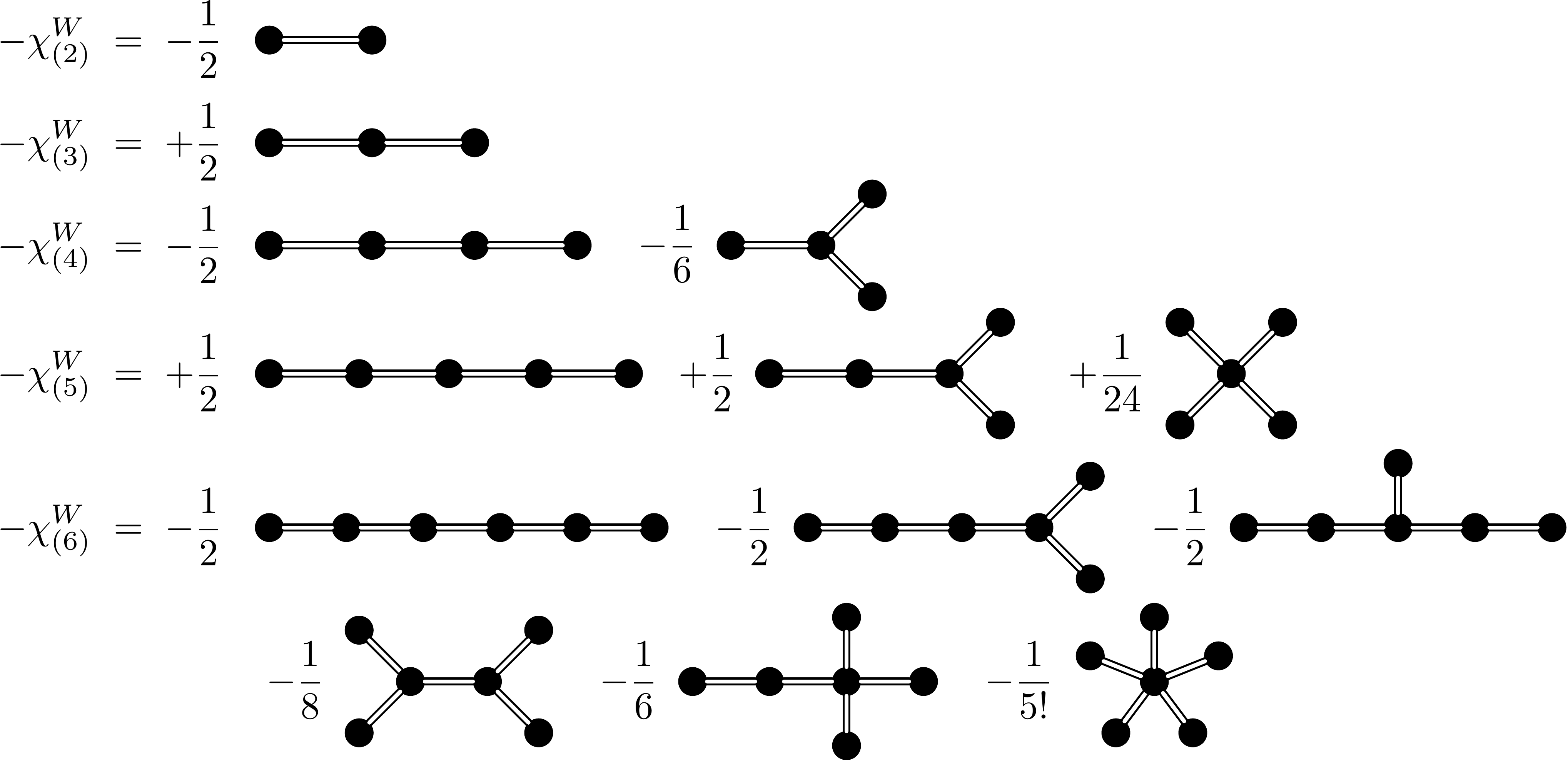}
    \caption{Feynman rules for $\chi$ in the formal WQFT. \\
    The coefficients are $(-1)^{n-1}$ times the symmetry factors of the diagrams.}
    \label{fig:WQFT-formal}
\end{figure}

Figure~\ref{fig:WQFT-formal} shows the WQFT diagrams up to $n=6$. We call it a \emph{formal} expansion, to emphasize the fact that we have not specified the choice of the propagator. 
The coefficients of the diagrams are the standard symmetry factors (up to signs).

For $\chi_{(2)}$, the choice of propagator is immaterial. 
But, beginning from $\chi_{(3)}$, the causal flow in the diagrams becomes important. 
We observe in \eqref{chi3-EOM-final} that $\chi_{(3)}$ carries three sub-digrams (oriented graphs) with coefficients $(1/3,1/12,1/12)$. 
The sum, $1/3+2(1/12) = 1/2$, matches the coefficient in the formal WQFT. 
As we will see shortly, the phenomenon continues to $\chi_{(n\ge 4)}$. 
The main body of this paper is devoted to an efficient algorithm to compute the coefficient of every oriented tree graph for all $n$.

\paragraph{Tree graphs}

Three types of tree graphs are relevant: unoriented, rooted, and oriented. 
Unoriented trees form the basis of the formal WQFT expansion. 
Oriented trees span the eikonal diagrams.
Rooted trees are oriented tress which contains a unique node (called root), 
for which all arrows are incoming. 
For example, in \eqref{chi3-EOM-final}, the first two diagrams are rooted, while the third is not. 
When we discuss the Magnus expansion and Hopf algebra in later sections, 
the distinction between rooted and non-rooted trees will be useful. 
Non-rooted trees carry two or more nodes for which all arrows are incoming. 
We call such nodes \emph{semi-roots}.

\begin{figure}[htbp]
    \centering
    \includegraphics[width=0.5\linewidth]{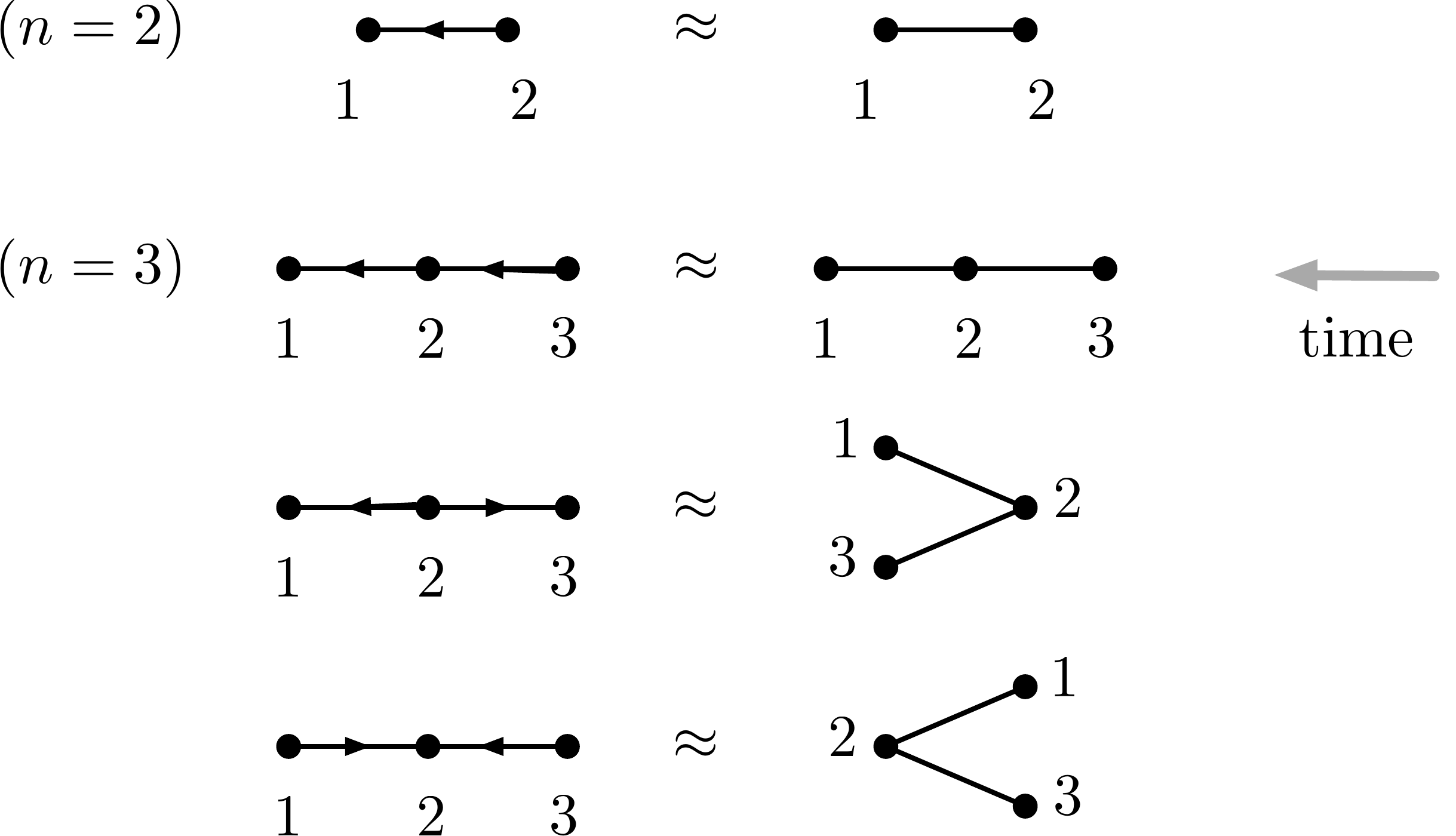}
    \caption{Two graphical notations: time-ordering by arrows or by a time flow.}
    \label{fig:retarded-tree}
\end{figure}

Sometimes an alternative notation for oriented graph is useful. Instead of arrows, we can use a ``time flow" to indicate the orientation of the edge connecting two nodes. In the latter, the relative horizontal position of connected nodes specifies the orientation of the edge. Our convention is that time flows from right to left. See Figure~\ref{fig:retarded-tree} for an illustration.

Given the number of (unlabeled) nodes, $n$, 
the number of possible trees can be counted recursively. 
We denote the number of unoriented/rooted/oriented trees by 
$u_n$/$r_n$/$o_n$, respectively. 
Their values for small $n$ are given in 
OEIS A000055/A000081/A000238, respectively, 
which we quote here for the readers' convenience:
\begin{align}
    \begin{array}{c|rrrrrrrrrrrr}
    n \;& \;1\; & 2\; & 3\; & 4 & 5 & 6 & 7 & 8 & 9 & 10 & 11 & 12
    \\
    \hline
    u_n \;& \;1\; & 1\; & 1\; & 2 & 3 & 6 & 11 & 23 & 47 & 106 & 235 & 551
    \\
    r_n \;& \;1\; & 1\; & 2\; & 4 & 9 & 20 & 48 & 115 & 286 & 719 & 1842 & 4766
    \\
    o_n \;& \;1\; & 1\; & 3\; & 8 & 27 & 91 & 350 & 1376 & 5743 & 24635 & 108968 & 492180 
    \end{array}
\end{align}
These sequences satisfy some interesting properties. 
For instance, the generating function for rooted trees, $R(x) = r_1 x + r_2 x^2 + \cdots$, satisfies 
\begin{align}
    R(x) = x \exp\left[ R(x) + \frac{1}{2} R(x^2) + \frac{1}{3} R(x^3) + \frac{1}{4} R(x^4) + \cdots \right] \,.
\end{align}
It is related to the generating function for unoriented trees, $U(x) = u_1 x + u_2 x^2 + \cdots$, as 
\begin{align}
    U(x) = R(x) - \frac{1}{2} R(x)^2 + \frac{1}{2} R(x^2) \,. 
\end{align}

\paragraph{Tree functions and integrals}

In the remainder of this paper, a \emph{tree} (denoted generically by $\tau$) will always refer to a connected, oriented tree graph. 
A tree function is a map from the set of all trees (including the empty tree) to another set ($\mathbb{Z}$, $\mathbb{Q}$, $\mathbb{R}$, etc.). 
A basic example is $|\tau|$, the number of vertices of $\tau$. 
Another simple example is the symmetry factor $\sigma(\tau)$, 
the number of permutations of the vertices that leave $\tau$ invariant (with the orientation of edges taken into account). 
A \emph{forest} is a formal product of trees. 
The tree function evaluated for a forest is defined as the product of the function evaluated for each connected component:
\begin{align}
    f(\tau_1 \tau_2 \tau_3 \cdots \tau_N) = f(\tau_1) f(\tau_2) \cdots f(\tau_N)\,.
\end{align}

\begin{figure}[htbp]
    \centering
    \includegraphics[width=0.8\linewidth]{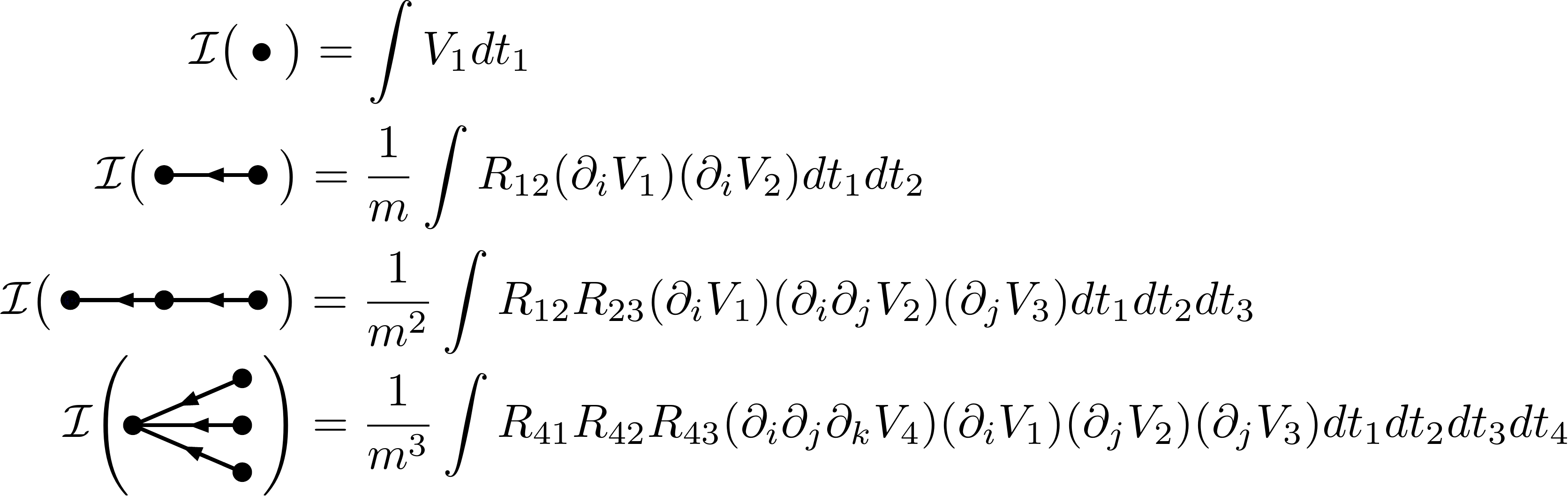}
    \caption{Examples of tree integrals. The generalization to arbitrary trees is clear.}
    \label{fig:tree-integral}
\end{figure}

One of the main goals of this paper is to compute $\chi_{(n)}$. 
At each order $n$, in the tree notation, we can write the result as 
\begin{align}
    - \chi_{(n)} = \sum_{|\tau| = n} \frac{\omega(\tau)}{\sigma(\tau)} \mathcal{I}(\tau) \,.
    \label{eikonal-coefficient-overview} 
\end{align}
We have already defined $|\tau|$ and $\sigma(\tau)$. The other tree function $\omega(\tau)$ is our target; we will explain a few algorithms to compute it efficiently. 
Normalizing each coefficient with $\s(\t)$ in the denominator is similar to attaching symmetry factors of Feynman diagrams. 
We will see the precise connection shortly. 
$\mathcal{I}(\tau)$ is the integral expression associated to $\tau$, 
which generalize what we have seen up to $n=4$ earlier; see Figure~\ref{fig:tree-integral} for further illustration. 

\begin{figure}[htbp]
    \centering
    \includegraphics[width=0.8\linewidth]{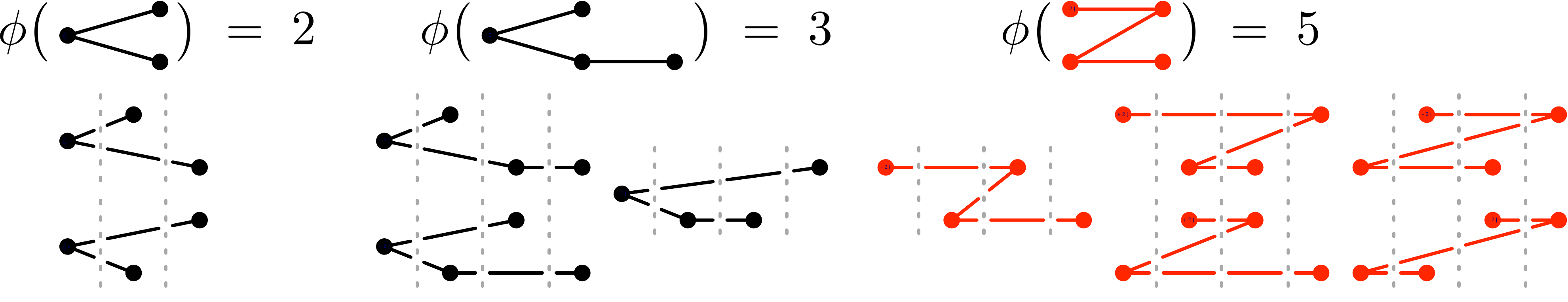}
    \caption{How to count $\phi(\tau)$.}
    \label{fig:free-factor}
\end{figure}

In our search for $\omega(\tau)$ in later sections, another tree function $e(\tau)$ will play a crucial role. We define $e(\tau)$ in a few steps. 
We first introduce $\gamma(\tau)$ known as the \emph{tree factorial}. 
Its definition begins with $\gamma(\emptyset)=\gamma(\bullet)=1$. When multiple rooted trees $(\tau_1, \cdots, \tau_N)$ are ``grafted" by a new root to form a tree $\tau$ (denoted by $\tau = [\tau_1, \cdots, \tau_N]$) such that $|\tau| = \sum_a |\tau_a| + 1$, $\gamma(\tau)$ is defined recursively as 
\begin{align}
    \gamma(\tau) = |\tau| \prod_a \gamma(\tau_a) \,.
    \label{tree-factorial-recursion}
\end{align}
Next, we introduce $\phi(\tau)$ which counts the number of the \emph{linear extensions of a poset}. Put simply, it counts the number of all possible time orderings of labeled vertices that are compatible with the oriented edges of a tree. 
See Figure~\ref{fig:free-factor} for an illustration. 

\noindent 
For rooted trees, it is well known that the two functions satisfy 
\begin{align}
    \phi(\tau) \gamma(\tau) = |\tau|! \,.
    \label{phi-gamma-n-factorial}
\end{align}

The tree function $e(\tau)$ is originally defined for rooted trees simply as $e(\tau)= 1/\gamma(\tau)$. 
It may not look obvious how to generalize the tree factorial $\gamma(\tau)$ 
to non-rooted trees.
Fortunately, $\phi(\tau)$ is equally well-defined for 
rooted and non-rooted trees alike, 
as shown in Figure~\ref{fig:free-factor}, 
so we can turn the relation \eqref{phi-gamma-n-factorial} into a definition of $e(\tau)$ for non-rooted trees:
\begin{align}
    e(\tau) := \frac{1}{\gamma(\tau)} := \frac{\phi(\tau)}{|\tau|!} \,. 
    \label{e-tree-extended}
\end{align}
We note that for a non-rooted tree, $\gamma(\tau)$ is not necessarily an integer.

\begin{figure}[htbp]
    \centering
    \includegraphics[width=0.6\linewidth]{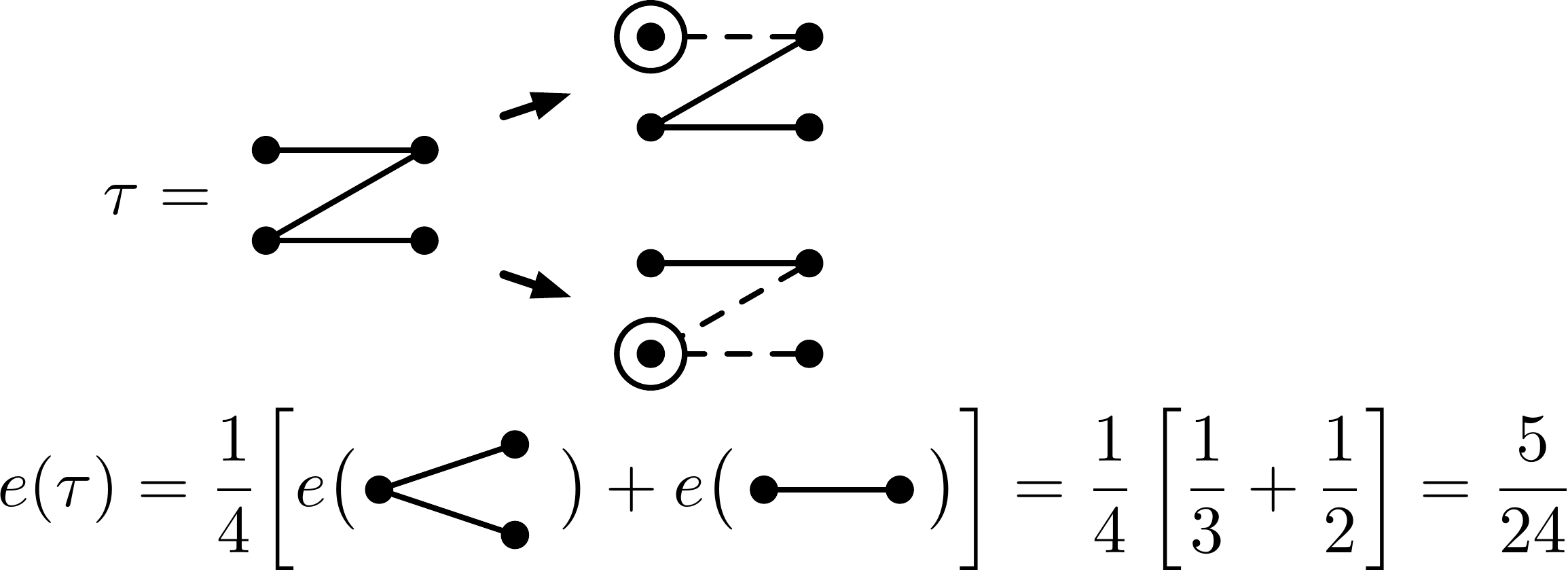}
    \caption{Recursion relation for $e(\tau)$ for non-rooted trees.}
    \label{fig:e-recursion}
\end{figure}

It is possible to generalize the recursive definition of $\gamma(\tau)$ to build an efficient algorithm to compute $e(\tau)$ even for non-rooted trees. 
For a rooted tree $\tau = [\tau_1, \cdots, \tau_N]$, 
\eqref{tree-factorial-recursion} means 
\begin{align}
    e(\tau) = \frac{1}{|\tau|} \prod_{a=1}^N e(\tau_a) \,.
\end{align}
For a non-rooted tree, we use essentially the same grafting formula except that we sum over the semi-roots; 
see Figure~\ref{fig:e-recursion} for an illustration. 

\section{Magnus expansion} \label{sec:magnus}

The Magnus expansion takes the log of a time-ordered exponential integral. 
It has a wide range of applications not only in physics but also in many areas of applied mathematics. See, {\it e.g.}, ref.~\cite{Blanes:2008xlr} for a comprehensive review for physicists and ref.~\cite{ebrahimifard2023magnusexpansion} for a more abstract mathematical treatment. 
See ref.~\cite{Argeri:2014qva} for applications of Magnus series for Feynman integrals.

\subsection{Overview}

We recall some facts about the Magnus expansion following ref.~\cite{ebrahimifard2023magnusexpansion}. 
Consider a matrix-valued differential equation, 
\begin{align}
    \dot{Y}(t) = A(t) Y(t) \,,
    \quad 
    Y(t_0) = Y_0
    \quad 
    \Longrightarrow
    \quad 
    Y(t)=\exp\left[ \Omega^{(A)}(t) \right] Y_0 \,.
\end{align}
Magnus~\cite{Magnus:1954zz} showed that $\Omega^{(A)}(t)$ is the unique solution of the differential equation, 
\begin{align}
    \dot{\Omega}(t) = A(t) + \sum_{k>1} \frac{B_k}{k!} \operatorname{ad}^k_{\Omega^{(A)}}(A)(t) = \frac{\operatorname{ad}_{\Omega^{(A)}}}{e^{\operatorname{ad}_{\Omega^{(A)}}}-1}(A)(t) \,, 
    \quad 
    \Omega^{(A)}(t_0) = 0 \,.
\end{align}
Here, $\operatorname{ad}^k$ is the $k$-th iteration of $\operatorname{ad}_X(Y) := [X,Y]$ and $B_n = B_n^-$ are the Bernoulli numbers. 
With a formal parameter $h$, the Magnus expansion is given by 
\begin{align}
    \Omega^{(h A)}(t)=\sum_{n \geq 1} h^n \Omega_n^{(A)}(t) \,,
    \quad 
    \Omega_1^{(A)} (t)=\int_{t_0}^t A(s) d s \,.
\end{align}
We will omit the superscript $(A)$ from here on. 
For $n\ge 2$, $\Omega_{n}$ can be computed recursively by integrating 
\begin{align}
\begin{split}
\dot{\Omega}_n(t)& = \sum_{k > 0} \frac{B_k}{k!} \sum_{\{r\}_k} [\Omega_{r_k}(t), \cdots [\Omega_{r_2}(t), [\Omega_{r_1}(t), A(t) ]] \cdots ]\,,
\end{split}
\label{Magnus-recursion}
\end{align}
where $\{r\}_k$ is the set of all $k$-tuples of integers satisfying 
\begin{align}
    r_1 + r_2 + \cdots r_k = n-1 \,,
    \quad 
    r_i > 0 \,.
\end{align} 

\paragraph{Quantum scattering}

In a full-fledged QFT, the analytic structure of the $S$-matrix can be extremely complicated, plagued by a plethora of poles and cuts.
Here, we pursue a much more modest goal; we focus on a single particle quantum mechanics where the $S$-matrix can be constructed rigorously in the interaction picture. 
Let us quickly review  the construction, following ref.~\cite{taylor2012scattering}.

Assume that the Hamiltonian takes the form 
\begin{align}
    \hat{H} = \hat{H}_0 + \hat{V} \,.
\end{align}
The time-evolution operator in the interaction picture is given by
\begin{align}
    \hat{U}_I(t,t_0) = e^{- \hat{H}_0 t /i\hbar} e^{ \hat{H} (t - t_0)/i\hbar} e^{ \hat{H}_0 t_0 /i\hbar} \,.
    \label{T-ordered-exp}
\end{align}
It can be expressed as a time-ordered integral: 
\begin{align}
    \hat{U}_I(t,t_0)  = \mathcal{T} \exp\left( \frac{1}{i\hbar} \int^{t}_{t_0} \hat{V}_I(t) dt \right) \,, 
    \quad 
    \hat{V}_I(t) = e^{- \hat{H}_0 t /i\hbar}\hat{V} e^{ \hat{H}_0 t /i\hbar} \,.
\end{align}
The $S$-matrix is defined by the limit, 
\begin{align}
    \hat{S} = \lim_{t \rightarrow +\infty} \hat{U}_I(+t,-t) \,.
\end{align}

One could expand the integral in \eqref{T-ordered-exp} to obtain the Dyson series, 
\begin{align}
    \label{Dyson-series}
    \hat{S} \,=\,
        \hat{1} 
        + \frac{1}{i\hbar} \int^\infty_{-\infty} dt_1\,\,
            \hat{V}_I(t_1) 
        + \frac{1}{(i\hbar)^2} \int_{t_1 > t_2} dt_1 dt_2\,\,
            \hat{V}_I(t_1) \hat{V}_I(t_2) 
        + \cdots
    \,.
\end{align}
But, since we aim for $\hat{\chi} = - i\hbar \log \hat{S}$, a better choice is the Magnus expansion~\cite{Magnus:1954zz}.
The first few terms of the expansion are 
\begin{align}
\begin{split}
      - \hat{\chi}_{(1)} &= \int^\infty_{-\infty} \hat{V}_I(t_1) dt_1   \,,
      \\
      - \hat{\chi}_{(2)} &= \frac{1}{i\hbar} \frac{1}{2} \int_{t_1 > t_2} [ \hat{V}_I(t_1) , \hat{V}_I(t_2) ] dt_1 dt_2 \,,
      \\
      - \hat{\chi}_{(3)} &= \frac{1}{(i\hbar)^2} \frac{1}{6} \int_{t_1 > t_2>t_3} \left( [ \hat{V}_1 , [ \hat{V}_2, \hat{V}_3 ]] + [ \hat{V}_3 , [ \hat{V}_2, \hat{V}_1 ]] \right) dt_1 dt_2 dt_3 \,, 
      \\
      - \hat{\chi}_{(4)} &= \frac{1}{(i\hbar)^3} \int_{t_1 > t_2>t_3 >t_4} \hat{\Upsilon}_4(t_1,t_2,t_3,t_4) \,  dt_1 dt_2 dt_3 dt_4\,, 
      \\ 
      \hat{\Upsilon}_4 &= \frac{1}{12} \left( [ \hat{V}_1 , [ \hat{V}_2, [\hat{V}_3 ,\hat{V}_4 ]]] +  [ \hat{V}_2 , [ \hat{V}_3, [\hat{V}_4 ,\hat{V}_1 ]]] \right.
      \\
      &\qquad \qquad
      \left. + [[ \hat{V}_1, \hat{V}_2], \hat{V}_3], \hat{V}_4] + [\hat{V}_1, [[ \hat{V}_2, \hat{V}_3], \hat{V}_4]\right) \,.
      \label{Magnus3-quantum}
\end{split}
\end{align}
From the third line on, we abbreviated $\hat{V}_I(t_k)$ by $\hat{V}_k$. 
Several ways to spell out the $n$-th order term of the Magnus expansion are known. 
For now, it suffices to note that
\begin{align}
    - \hat{\chi}_{(n)} &= \frac{1}{(i\hbar)^{n-1}} \int_{t_1 > \ldots > t_n} \hat{\Upsilon}_n(t_1,\cdots , t_n) \,dt_1 \cdots dt_n \,, 
\end{align}
where the operator $\hat{\Upsilon}_n(t_1,\cdots , t_n)$ is a sum of $(n-1)$-fold nested commutators with constant coefficients, where each term is multi-linear in $\hat{V}_k$ ($k=1,\ldots,n$).

\paragraph{Classical limit}

A notable feature of the Magnus expansion is that the classical limit is almost trivial; we simply replace commutators with Poisson brackets as in \eqref{Dirac-Poisson}. The result is the Magnus expansion for the classical eikonal:
\begin{align}
\begin{split}
      - \chi_{(1)} &= \int V_1 \, dt_1   \,,
      \\
      - \chi_{(2)} &= \frac{1}{2} \int_{t_1 > t_2} \{ V_1 , V_2 \} \, dt_1 dt_2 \,,
      \\
      - \chi_{(3)} &= \frac{1}{6} \int_{t_1 > t_2>t_3} \left( \{ V_1 , \{  V_2, V_3 \}\} + \{ V_3, \{  V_2 , V_1 \}\} \right) dt_1 dt_2 dt_3 \,,
      \\
       - \chi_{(n)} &= \int_{t_1 > \ldots > t_n} \Upsilon_n(t_1,\cdots , t_n) \,dt_1 \cdots dt_n \,.
      \label{Magnus3-classical}
\end{split}
\end{align}
Here, $V_k$ is the classical limit of the operator $\hat{V}_I(t_k)$. 
An advantage of this expansion over the Dyson series \eqref{Dyson-series} is that the latter is prone to the so-called 
``super-classical" terms, which are not well-behaved in the $\hbar \rightarrow 0$ limit. 
In contrast, each term in the Magnus expansion carries $1/(i\hbar)^{n-1}$ to turn the $(n-1)$-fold commutators to Poisson brackets, 
resulting in a perfectly well-defined classical limit.

To be concrete, from here on, we choose to work in non-relativistic quantum mechanics with the free Hamiltonian for a particle, 
\begin{align}
    H_0 = \frac{\vec{p}^2}{2m} \,.
\end{align}
Then a simple algebra shows that 
\begin{align}
    V_k = V\left(\vec{x} + \frac{\vec{p}}{m} t_k \right) \,.
    \label{V-int-classical}
\end{align}

\paragraph{From brackets to worldline propagators} 

The expansion \eqref{Magnus3-classical} is completely classical, but the brackets are still not commutative. 
To gain further insights, let us compute the brackets explicitly with \eqref{V-int-classical}. 

For $\chi_{(2)}$, we get 
\begin{align}
    -\chi_{(2)} = 
    -\frac{1}{2m}
    \int_{t_1>t_2} t_{12} \left[ (\partial_{\vec{x}} V_1) \cdot ( \partial_{\vec{x}}  V_2) \right] dt_1 dt_2 \,, 
    \quad 
    t_{kl} = t_k - t_l \,.
\end{align}
The final result for $\chi_{(2)}$ is then 
\begin{align}
     -\chi_{(2)} = -\frac{1}{2m} \int R_{12} \left[ (\partial_{\vec{x}} V_1) \cdot ( \partial_{\vec{x}}  V_2) \right] dt_1 dt_2 \,.
\end{align}
It agrees (rather trivially) with $\chi_{(2)}$ obtained from the EOM \eqref{2nd-eiknonal}. 

For $\chi_{(3)}$, after some reshuffling, we get 
\begin{align}
\begin{split}
      - \chi_{(3)} &= \frac{1}{m^2} \int P_3(t_1, t_2, t_3) 
    \left[ (\partial_i V_1) ( \partial_i \partial_j V_2) (\partial_j V_3)\right] dt_1 dt_2 dt_3 \,,
    \\
    &\quad P_3(t_1,t_2,t_3) = \frac{1}{6} \left( 2R_{12} R_{23} + \frac{1}{2} R_{12} R_{32} + \frac{1}{2} R_{21} R_{23}  \right) \,.
\end{split}
\end{align}
The partial derivative $\partial_i$ is taken with respect to $\vec{x} = (x^1, x^2, x^3)$. Again, it agrees perfectly with the result from the EOM iteration.

\paragraph{Notations}

To connect two similar sets of notations, we provide a dictionary, 
\begin{align}
    \Omega_n \; \leftrightarrow \; - \chi_{n} \,,
    \quad 
    A(t)  \; \leftrightarrow \; V(t) \,,
    \quad 
    [\;\; , \;\; ] \; \leftrightarrow \;  \{\;\; , \;\; \} \,.
\end{align}
We also recall from \eqref{1st-cut-b} that 
\begin{align}
    \{ V_2 , V_1 \} \;\; \propto \;\;   + (R_{12} - R_{21}) 
    \;\; \propto \;\;    
    \;\;    
    \begin{fmffile}{cut-x}
        \parbox{30pt}{
        \begin{fmfgraph*}(30,10)
            \fmfleft{o1}
            \fmfright{i1}
            \fmfv{decor.shape=circle,decor.size=4,label=\small{2},label.angle=90}{i1}
            \fmfv{decor.shape=circle,decor.size=4,label=\small{1},label.angle=90}{o1}
            \fmf{phantom}{i1,c,o1}
            \fmfv{decor.shape=cross,decor.size=10}{c}
            \fmf{fermion}{i1,o1}
        \end{fmfgraph*}
        }
    \end{fmffile} 
    \;\; = \;\; 
       \begin{fmffile}{cut-x-1}
        \parbox{30pt}{
        \begin{fmfgraph*}(30,10)
            \fmfleft{o1}
            \fmfright{i1}
            \fmfv{decor.shape=circle,decor.size=4,label=\small{2},label.angle=90}{i1}
            \fmfv{decor.shape=circle,decor.size=4,label=\small{1},label.angle=90}{o1}
            \fmf{fermion}{i1,o1}
        \end{fmfgraph*}
        }
    \end{fmffile} 
    \;\; - \;\; 
       \begin{fmffile}{cut-x-2}
        \parbox{30pt}{
        \begin{fmfgraph*}(30,10)
            \fmfleft{o1}
            \fmfright{i1}
            \fmfv{decor.shape=circle,decor.size=4,label=\small{2},label.angle=90}{i1}
            \fmfv{decor.shape=circle,decor.size=4,label=\small{1},label.angle=90}{o1}
            \fmf{fermion}{o1,i1}
        \end{fmfgraph*}
        }
    \end{fmffile} \;\; .
    \label{cut-x}
\end{align}

\subsection{Murua's formula for rooted trees} \label{sec:Murua_root}

We recall from \eqref{eikonal-coefficient-overview} that 
our goal is to compute the tree function $\omega(\tau)$ in the diagrammatic expansion, 
\begin{align}
    - \chi_{(n)} = \sum_{|\tau| = n} \frac{\omega(\tau)}{\sigma(\tau)} \mathcal{I}(\tau) \,.
    \nonumber
\end{align}
For rooted trees, Murua~\cite{Murua_2006} defined $\omega(\tau)$ in a different (but related) context, and gave a recursive formula to compute $\omega(\tau)$: 
\begin{align}
    \omega(\tau) = \sum_{p \in P_r(\tau)} B_{|p|-1}  e(p') \omega(\tau\backslash p) \,.
    \label{Murua-original}
\end{align}
Here, $P_r(\tau)$ is the set of partitions of $\tau$ that necessarily contain all edges connected to the root. 
A \emph{partition} $p$ of a tree $\tau$ is an arbitrary collection of edges of $\tau$; 
there are $2^{|\tau|-1}$ partitions when there is no restriction. 
It is understood that $p$ is made connected by shrinking each connected components of the remainder $(\tau\backslash p)$.
$p'$ is what is left of $p$ after amputating all the edges connected to the root. 
The function $e(\tau)$ is the inverse of the tree factorial, as explained in section~\ref{tree functions}.

Here is a simple example of \eqref{Murua-original} taken from Murua's original paper:
\begin{align}
    \adjustbox{valign=c}{\includegraphics[width=0.75\linewidth]{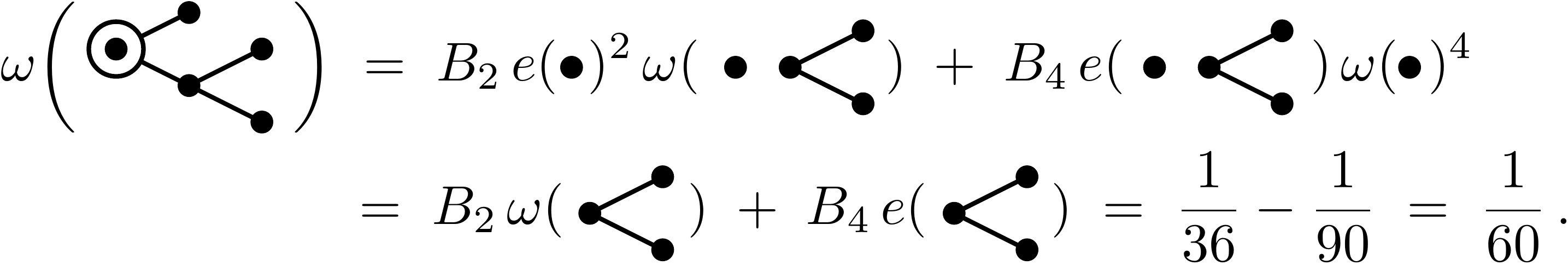}}
    \label{Murua-ex}
\end{align}

\paragraph{Sketch of the derivation} 
The formula \eqref{Murua-original}  can be derived from 
the Magnus recursion relation \eqref{Magnus-recursion}. 
We do not attempt a rigorous proof, for which the readers can consult ref.~\cite{Murua_2006}. Instead, we give a down-to-earth sketch of the derivation,  
where each element of the formula is explained intuitively based on examples. 
In this subsection, we focus on rooted trees as in ref.~\cite{Murua_2006}, for which it suffices to take only the ``positive branch" from the ``cut" \eqref{cut-x}. 

In the tree integral $\mathcal{I}(\tau)$ for the eikonal, all the vertices $V(t_a)$ are integrated from $t_a = -\infty$ to $t_a = +\infty$. 
The time ordering, denoted by the oriented edges of the tree, is enforced by the retarded Green's function, 
so we need not impose restrictions on the integration range. 
To take advantage of the Magnus recursion \eqref{Magnus-recursion}, we single out 
the largest time variable consistent with the time ordering and write the integral as 
\begin{align}
    \mathcal{I}(\tau) = \int_{-\infty}^\infty \dot{\mathcal{I}}(\tau, t) dt 
    \label{largest time rooted}
\end{align}
Other time variables are hidden inside the integrand. 
For a rooted tree, $t$ is simply the time variable for the root vertex. 

On the LHS of \eqref{Magnus-recursion}, we expect  
\begin{align}
    \dot{\Omega}_n(t) = \sum_{|\tau|=n} \frac{\omega(\tau)}{\sigma(\tau)} \dot{\mathcal{I}}(\tau;t) \,,
    \label{dot-Omega-tree-sum}
\end{align}
On the RHS, for each $\Omega$ factor, we write 
\begin{align}
    \Omega_{r_a}(t) = \int_{-\infty}^t \dot{\Omega}_{r_a}(s_a) ds_a \,.
\end{align}
It is understood that $\dot{\Omega}$ in the integrand can again be decomposed as in \eqref{dot-Omega-tree-sum}.

The partition, a key ingredient in Murua's formula \eqref{Murua-original}, 
naturally arises from \eqref{Magnus-recursion}. 
The ``new" edges produced by the brackets constitute the partition $p$, 
whereas the ``old" edges included in $\dot{\Omega}_{r_a}$ contribute to $(\tau\backslash p)$. 
From this perspective, the reason why $\omega(\tau\backslash p)$ appear on the RHS is obvious. 

\begin{figure}[htbp]
    \centering
    \includegraphics[width=0.6\linewidth]{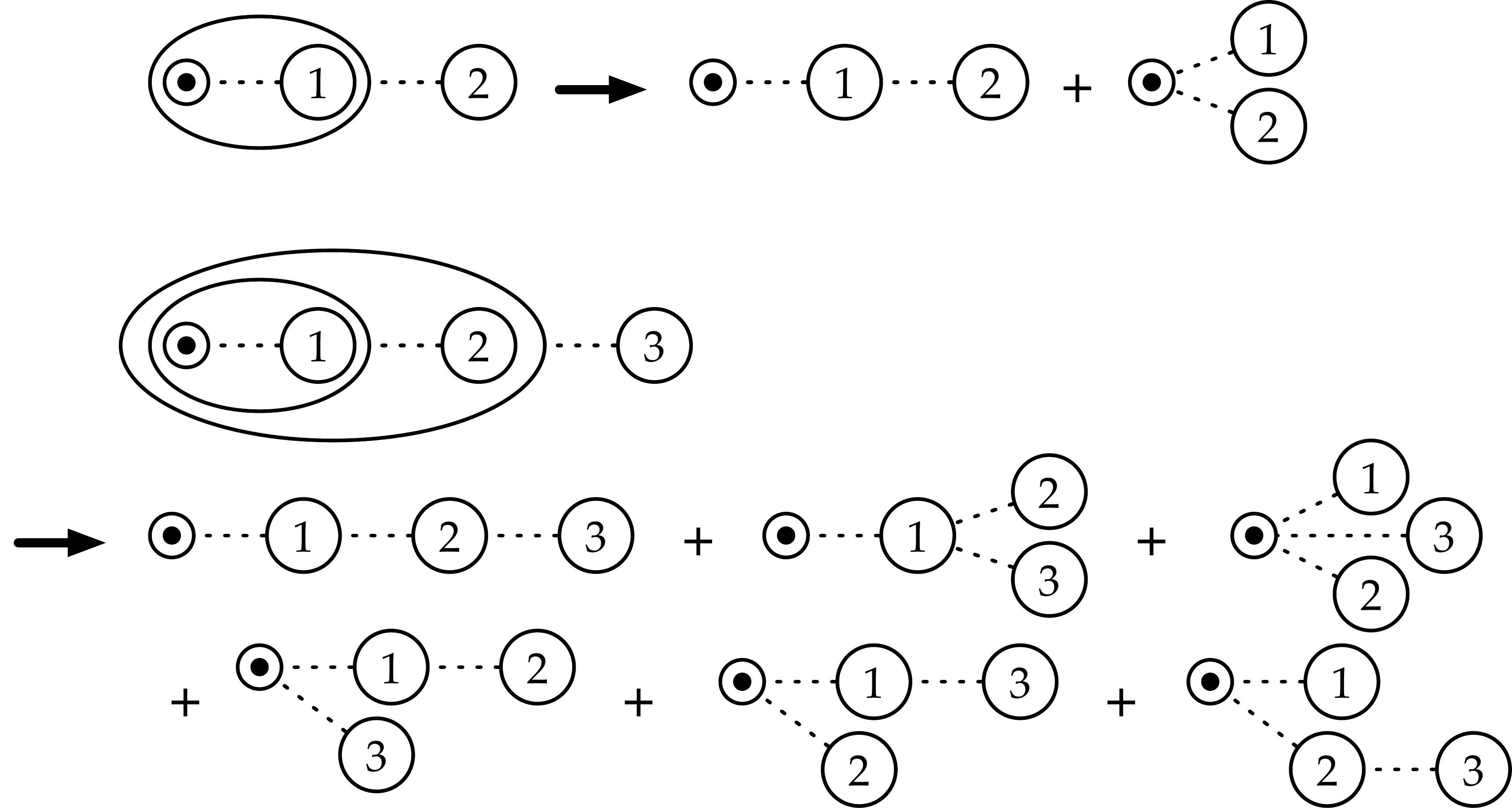}
    \caption{Popping bubbles. The numbers $a=1,2,3$ in circles represent $\Omega_{r_a}$. \\ Non-rooted trees are suppressed temporarily.}
    \label{fig:bubble-rooted}
\end{figure}

At this stage, it is useful to introduce a graphical notation 
(which traces back at least to Penrose~\cite{penrose2007road});
see Figure~\ref{fig:bubble-rooted}.
We evaluate the nested brackets one at a time. 
When a bracket is inserted into another bracket, we denote the inner one by a bubble. 
Each bracket producing the cut \eqref{cut-x} is represented by a dotted line. 
When a bracket acts on a bubble, it ``pops" the bubble to produce a sum, where the dotted line 
gets connected to components inside the bubble one by one (Leibniz rule). 
The operation is associative; the result does not depend on which bubble we pop first. 

We now explain how $e(p')$ arise. 
When the partition, regarded as a tree on its own, has no symmetry, the $e(p')$ factor 
simply follows from a sequence of equalities, with $|p| = k+1$, 
\begin{align}
    \frac{\phi(p)}{k!} = \frac{\phi(p)}{(|p|-1)!} = |p| \frac{\phi(p)}{|p|!} = |p|e(p) = e(p') \,.
    \label{gamma-p-prime}
\end{align}
The third step used \eqref{phi-gamma-n-factorial},
so the only non-trivial point in \eqref{gamma-p-prime} is how the sum over $\{r\}_k$ in \eqref{Magnus-recursion} 
produces multiplicity $\phi(p)$. 
Let us illustrate the idea with the trees at the bottom of the $k=3$ example in Figure~\ref{fig:bubble-rooted}. 
When the three circles are all distinct, we should sum over all 6 permutations of $(1,2,3)$, 
so that we get a total of 18 trees of the same shape. They split into 6 groups of multiplicity 3, 
which is the correct value for $\phi(p)$. When all three circles are indistinguishable, we have no permutations to sum over, 
but we still get multiplicity 3 from the three diagrams in the Figure with the labels removed.  

\begin{figure}[htbp]
    \centering
    \includegraphics[width=0.75\linewidth]{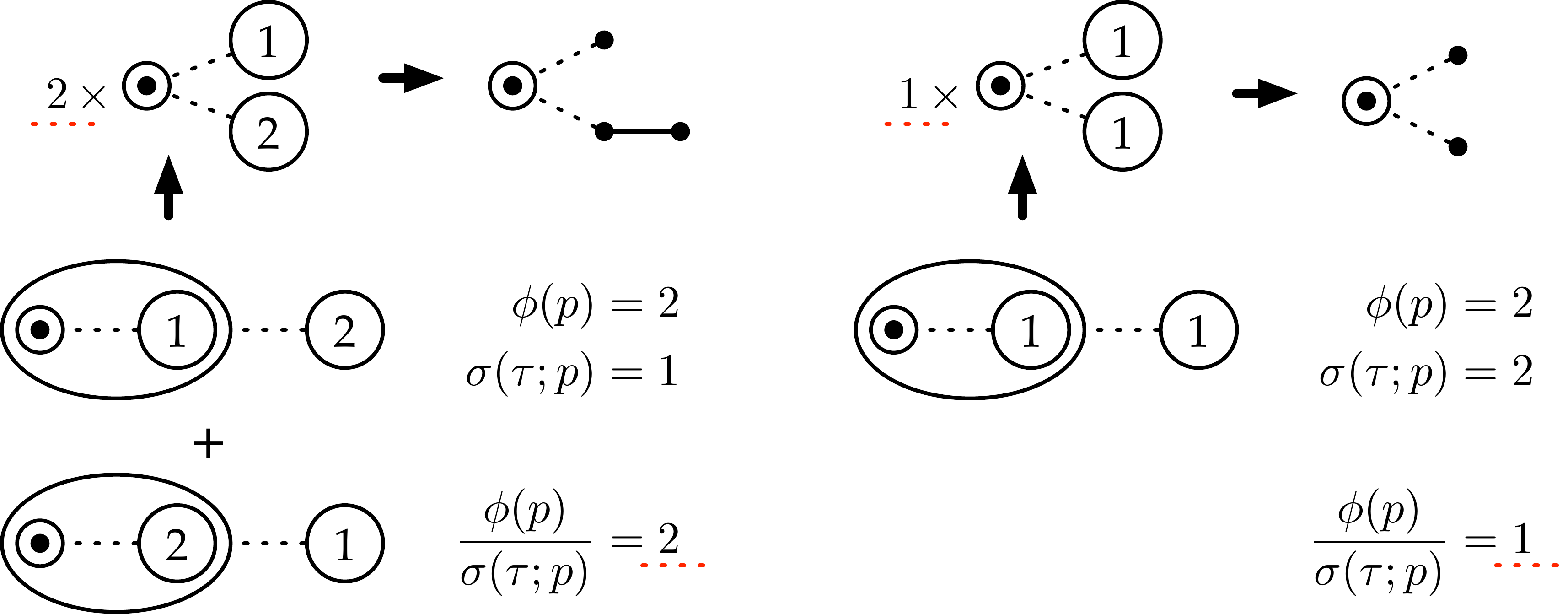}
    \caption{Symmetry factor of the two-colored tree formed by a partition.}
    \label{fig:bubble-symmetry1}
\end{figure}

We now turn to the most delicate aspect of Murua's formula: why the symmetry factor $\sigma(\tau)$ present in \eqref{dot-Omega-tree-sum} disappears completely from \eqref{Murua-original}.  
Given a partition $p$, suppose we color the edges participating in $p$ differently from those belonging to $(\tau\backslash p)$. 
We denote the symmetry factor of the two-colored tree by $\sigma(\tau;p)$. 
When $\sigma(\tau;p) > 1$, the $\phi(p)$ factor in \eqref{gamma-p-prime} gets modified as 
\begin{align}
    \phi(p) \;\; \rightarrow \;\; \frac{\phi(p)}{\sigma(\tau ; p)} \,.
\end{align}
The $\sigma(\tau;p)$ factor in the denominator can be traced back to how a term in the Magnus recursion 
is weighted differently depending on identical connected components of $(\tau\backslash p)$; 
see Figure~\ref{fig:bubble-symmetry1} for an illustration. 

\begin{figure}[htbp]
    \centering
    \includegraphics[width=0.65\linewidth]{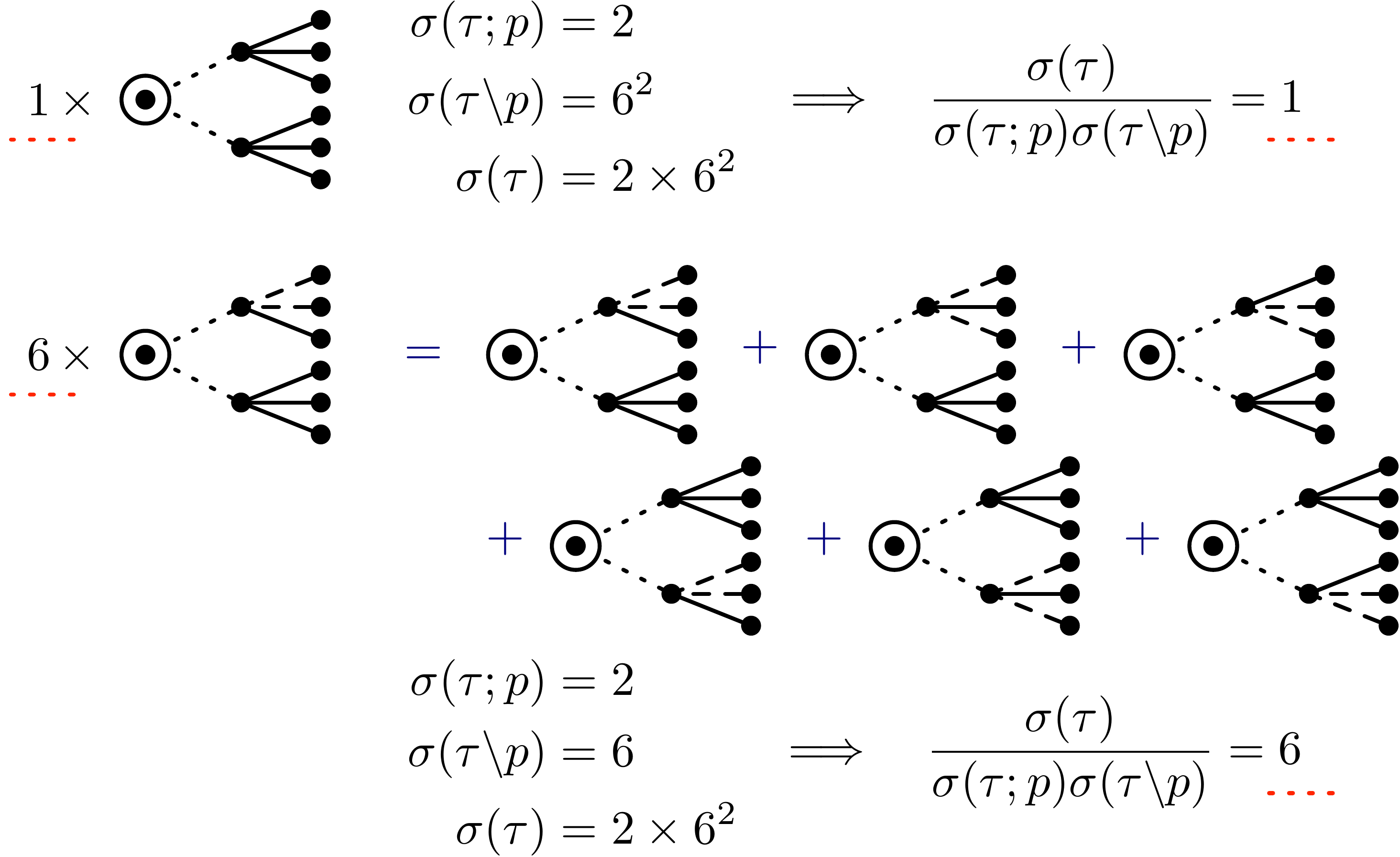}
    \caption{Mismatch in symmetry factors compensated by the multiplicity of equivalent partitions.}
    \label{fig:bubble-symmetry2}
\end{figure}

\noindent 
In some cases, as in the top example of Figure~\ref{fig:bubble-symmetry2}, the product $\sigma(\tau;p) \sigma(\sigma\backslash p)$ equals $\sigma(\tau)$. In other cases, as in the bottom example of Figure~\ref{fig:bubble-symmetry2}, 
$\sigma(\tau;p) \sigma(\sigma\backslash p)$ is an integer divisor of $\sigma(\tau)$, and the quotient $\sigma(\tau)/(\sigma(\tau;p) \sigma(\sigma\backslash p))$ counts the multiplicity of ``equivalent" partitions related to each other by the symmetry of the whole tree $\tau$. The multiplicity is included in the partition sum in \eqref{Murua-original}.  
In all cases, we observe that the symmetry factors and the multiplicity factor conspire to 
remove all explicit dependencies on $\sigma(\tau)$ and $\sigma(\tau\backslash p)$ from Murua's formula. 

\subsection{Extension to non-rooted trees} 

Having understood how Murua's formula \eqref{Murua-original} can be derived from the Magnus recursion \eqref{Magnus-recursion}, it is not too difficult to extend the formula to non-rooted trees. 

Recall that a non-rooted tree carries two or more root-like vertices where all edges are incoming, 
and that we call them semi-roots. 
With an arbitrary choice $s$ among the semi-roots, our extended formula reads 
\begin{align}
\begin{split}
    \omega(\tau) = \sum_{p \in P_s(\tau)}  (-1)^{\ell(p)} B_{|p|-1} e(p')   \omega(\tau\backslash p) \,.
\end{split}
\label{Murua-extended}
\end{align}
It appears to involve the choice $s$, 
but the resulting $\omega(\tau)$ turns out to be independent of $s$. 
$P_s(\tau)$ is the set of partitions of $\tau$ that contain all edges connected to the semi-root $s$.
Given a partition $p$, $p'$ is defined in two steps. First, we flip the orientation of all the ``wrong" edges of $p$ to obtain a new tree $\bar{p}$ \emph{rooted} with respect to $s$. Then $p'$ is what is left of $\bar{p}$ after amputating all edges connected to $s$. The function $\ell(p)$ counts how many edges of $p$ should be flipped to reach $\bar{p}$. 
Since $\bar{p}$ and $p'$ are rooted trees, this formula does not require $e(\tau)$ for non-rooted trees.

\begin{figure}[htbp]
    \centering
    \includegraphics[width=0.65\linewidth]{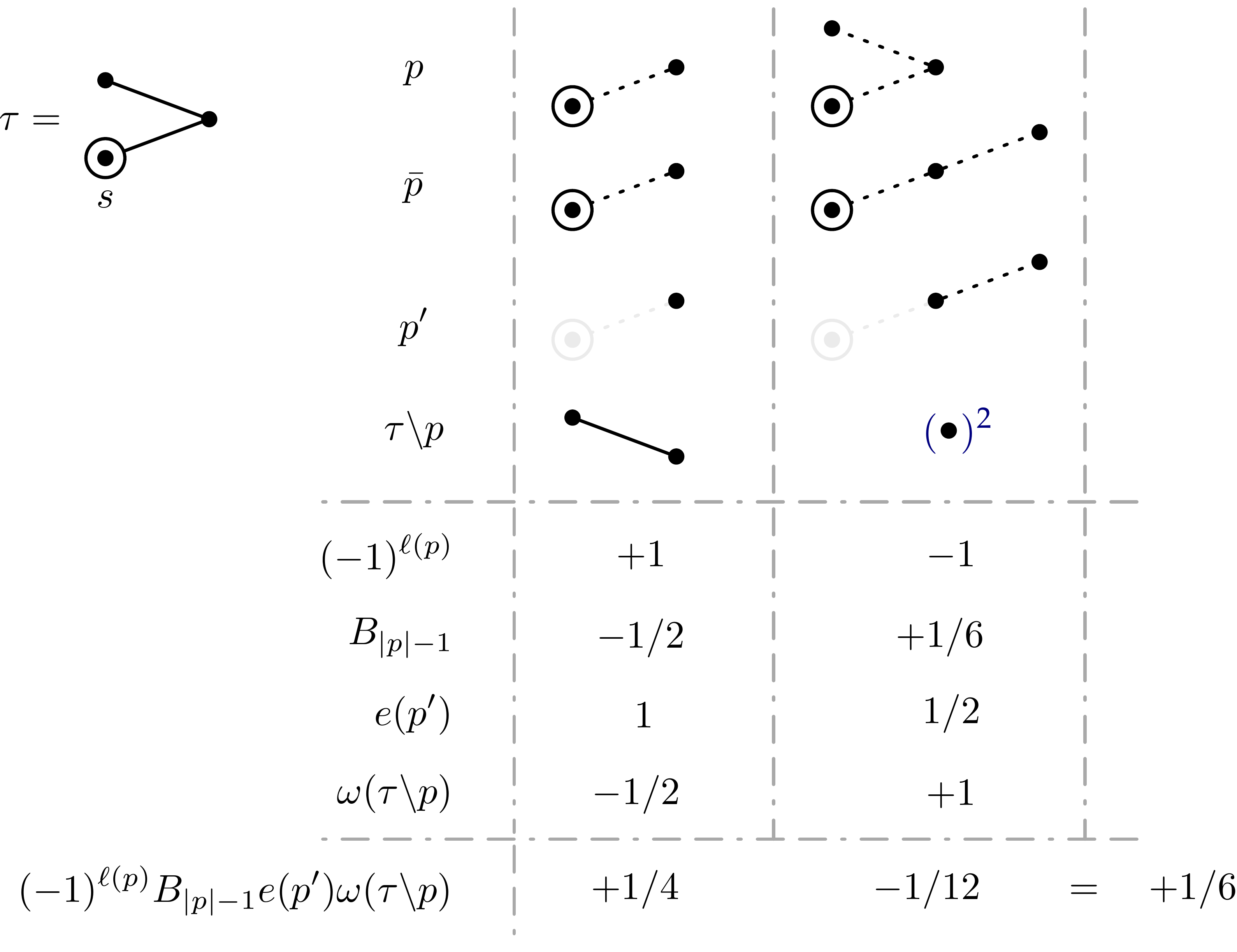}
    \caption{A rooted tree turned upside-down, regarded as a non-rooted tree.}
    \label{fig:Murua-Sungsoo-ex1}
\end{figure}

\begin{figure}[htbp]
    \centering
    \includegraphics[width=0.78\linewidth]{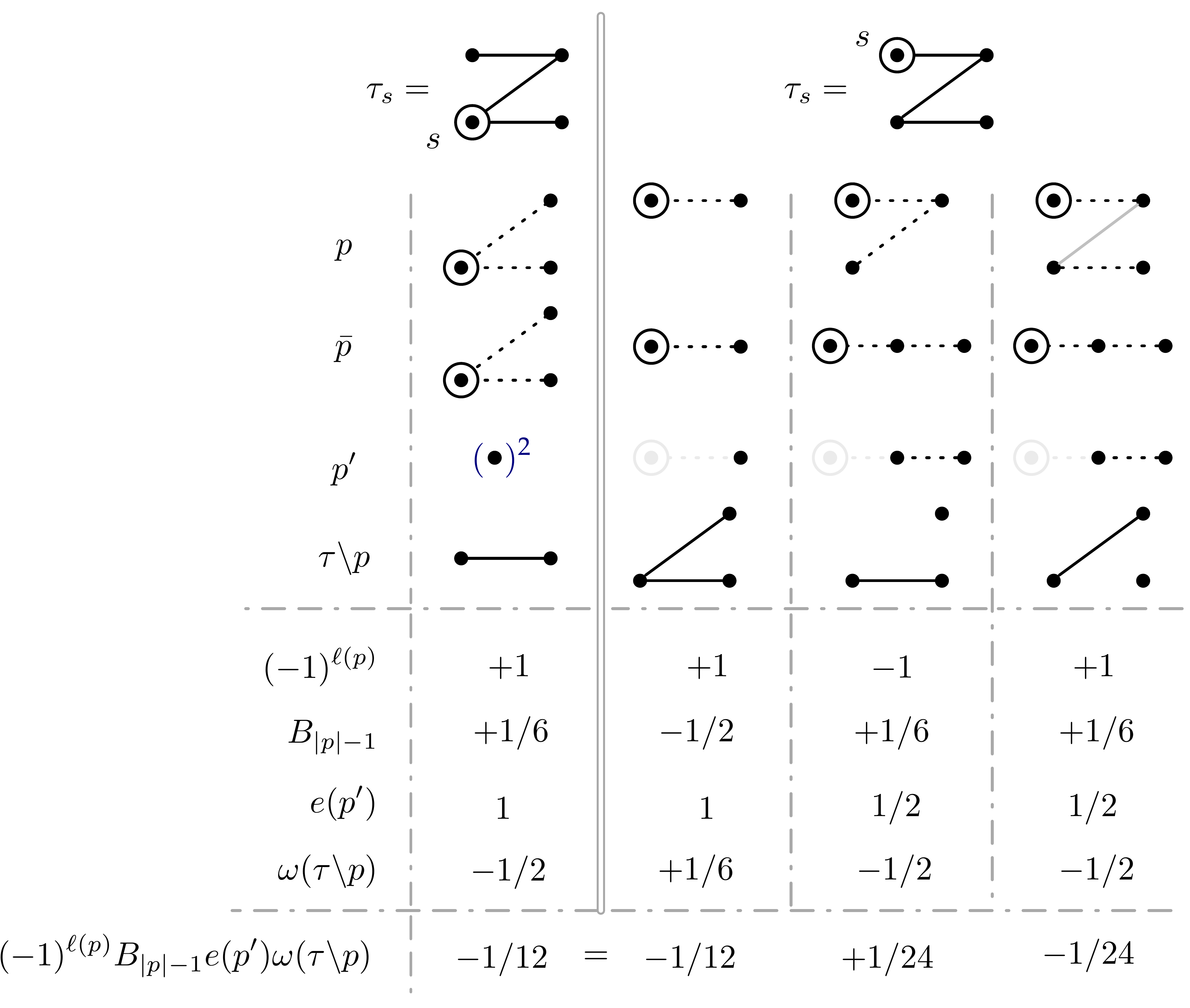}
    \caption{The simplest genuine non-rooted tree with two distinct choices of the semi-root.}
    \label{fig:Murua-Sungsoo-ex2}
\end{figure}

\paragraph{Examples} 
The formula is best explained by examples. The simplest one is the ``upside-down rooted tree" in Figure~\ref{fig:Murua-Sungsoo-ex1}. The two semi-roots are on an equal footing. 
Two partitions contribute to the sum and give the same $\omega(\tau)$ as the original ``right way up" rooted tree. 
The first genuine non-rooted tree, shown in Figure~\ref{fig:Murua-Sungsoo-ex2}, admits two different choices of the semi-root, but both yield the same $\omega(\tau)$. 


\paragraph{Derivation}

The derivation of the extended Murua formula \eqref{Murua-extended} 
from the Magnus recursion \eqref{Magnus-recursion} 
is similar to the one for the original formula, 
but we should discuss a few novel features.

For a non-rooted tree, two or more semi-roots can compete for the largest time. 
The full integral should include all possible semi-roots, so 
the integral \eqref{largest time rooted} is generalized to
\begin{align}
  \dot{\mathcal{I}}(\tau, t) =  \sum_{s} \dot{\mathcal{I}}(\tau_s, t) \,,
  \label{largest time non-rooted}
\end{align}
where the label $s$ runs over the semi-roots and $\tau_s$ denotes the tree $\tau$ with the semi-root $s$ marked distinctly 
from other vertices.

\begin{figure}[htbp]
    \centering
    \includegraphics[width=0.65\linewidth]{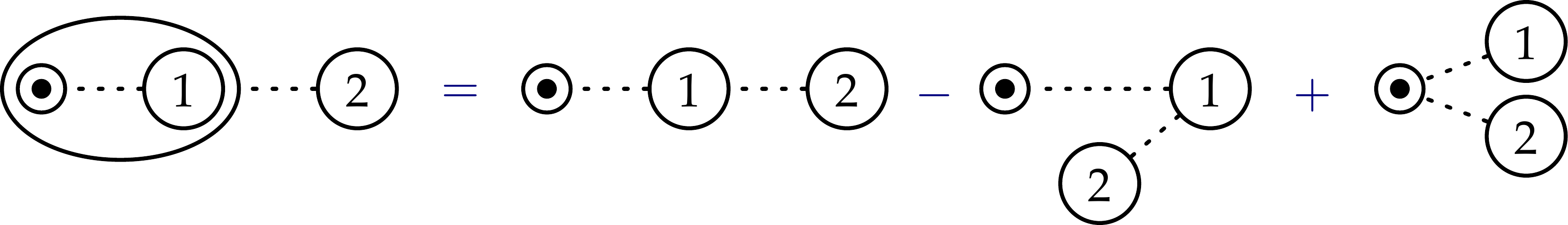}
    \caption{Popping bubbles to produce non-rooted trees.}
    \label{fig:bubble-non-rooted}
\end{figure}

The sign factor $(-1)^{\ell(p)}$ follows from the relative minus sign in \eqref{cut-x}. 
A simple example is given in Figure~\ref{fig:bubble-non-rooted}. 
The edges connected to the root vertex can only go forward in time. Other edges in the partition receive both forward and backward branches from \eqref{cut-x}. 
The generalization to arbitrary partitions is straightforward. 
The discussion on $e(p')$ and the symmetry factors go through with little modification. 
The fact that $\omega(\tau)$ for non-rooted trees can be computed with an arbitrary choice of semi-root means that the tree integral for non-rooted trees can be written as a single integral without restrictions on the integration range other than those required by the retarded Green's functions.

\section{Hopf algebra} \label{sec:hopf}

After the EOM iteration and the Magnus expansion, 
we reach another level of abstraction and discuss the Hopf algebra structure behind the eikonal expansion. 
Our discussion follows early papers on the subject, such as refs.~\cite{Chartier_2010,Calaque_2011}.\footnote{Perhaps the most well known Hopf algebra for physicists is the one by Connes and Kreimer \cite{Connes:1998qv} in the context of renormalization of QFT. The Hopf algebra to be discussed in this section is different from the Connes-Kreimer Hopf algebra. The relation between the two Hopf algebras is clearly explained in ref.~\cite{Chartier_2010}.}
For a recent update on the mathematics of Magnus expansion, see ref.~\cite{ebrahimifard2023magnusexpansion} 
and references therein.

Our interest lies not in the Hopf algebra for its own sake, but in the practical application arising from it. 
After reviewing the Hopf algebra originally developed for rooted trees, we propose an extension to non-rooted trees compatible with the non-rooted extension of the previous section. 
The Hopf algebra allows us to prove non-trivial relations satisfied by $e(\tau)$ and $\omega(\tau)$, which we turn to an efficient algorithm to compute them. 

\subsection{CEM Hopf algebra for rooted trees} \label{sec:CEM_Hopf}

We give an informal introduction to the Hopf algebra of Calaque, Ebrahimi-Fard, Manchon (CEM)~\cite{Calaque_2011}, mainly for physicists not familiar with Hopf algebra. The authors' interest in the Hopf algebra began with the realization that the ``eikonal coefficients" in \eqref{eikonal-coefficient-overview}, when restricted to rooted trees, agree perfectly with $\omega(\tau)/\sigma(\tau)$ tabulated up to $n=5$ in ref.~\cite{Calaque_2011}. 

\paragraph{Composition of power series}

Consider two functions defined by infinite series:
\begin{align}
    \begin{split}
        f(x) = f_0 + f_1 x + f_2 x^2 + f_3 x^3 +\cdots \,,
        \\
        g(x) = g_0 + g_1 x + g_2 x^2 + g_3 x^3 +\cdots \,.
    \end{split}
    \label{f-g-power}
\end{align}
We impose two mild simplifying assumptions:
\begin{align}
    f_0 = 0 \,, \quad g_0 = 1 \,.
    \label{f-g-condition}
\end{align}
It is straightforward to compute the coefficients of the composite series order by order, 
\begin{align}
\begin{split}
     (g\circ f)(x) = g(f(x)) &= 1 + (g\circ f)_1 x + (g\circ f)_2 x^2 +  (g\circ f)_3 x^3 + \cdots \,,
    \\
    (g\circ f)_1 &= g_1 f_1 \,,
    \\
      (g\circ f)_2 &= g_1 f_2 + g_2 f_1^2 \,,
      \\  
      (g\circ f)_3 &= g_1 f_3 + 2 g_2 f_1 f_2 + g_3 f_1^3 \,. 
\end{split}
\label{g-f-composition}
\end{align}
To all orders, every monomial on the RHS is linear in $g_k$ and 
the orders of $f_k$ add up to $n$.

\paragraph{Composition of tree series} 

To each power series in \eqref{f-g-power}, we associate a \emph{tree series}:
\begin{align}
\begin{aligned}
    F &= \sum_{\tau} f(\tau) \tau 
    = 
    f(\emptyset) + 
    f( \;\,
    \parbox{5pt}{\fmfreuse{dot}}
        ) \;\,
        \parbox{10pt}{\fmfreuse{dot}}
        \hskip -4pt 
        + \, f( \;\,
       \begin{fmffile}{dot2}
        \parbox{25pt}{
        \begin{fmfgraph*}(20,20)\fmfkeep{dot2}
            \fmfright{i1}
        \fmfleft{o1}
            \fmfv{decor.shape=circle,decor.size=4}{i1}
            \fmfv{decor.shape=circle,decor.size=4}{o1}
            \fmf{fermion}{i1,o1}
        \end{fmfgraph*}
        }
       \end{fmffile}
       ) \;\, 
       \parbox{25pt}{\fmfreuse{dot2}} \, 
       + \, f( \;\,
       \begin{fmffile}{dot3}
        \parbox{25pt}{
        \begin{fmfgraph*}(20,20)\fmfkeep{dot3}
            \fmfright{i1}
            \fmfleft{o1,o2}
            \fmfv{decor.shape=circle,decor.size=4}{i1}
            \fmfv{decor.shape=circle,decor.size=4}{o1}
            \fmfv{decor.shape=circle,decor.size=4}{o2}
            \fmf{fermion}{i1,o1}
            \fmf{fermion}{i1,o2}
        \end{fmfgraph*}
        }
       \end{fmffile}
       ) \;\, 
       \parbox{25pt}{\fmfreuse{dot3}}
       \, + \, \cdots \,,
    \\
        G &= \sum_{\tau} g(\tau) \tau 
        = g(\emptyset) + g( \;\;
        \parbox{10pt}{\fmfreuse{penta}} \hskip -5pt
        ) \;\; 
        \parbox{10pt}{\fmfreuse{penta}}
        \, + \, g( \;\,
       \begin{fmffile}{penta2}
        \parbox{25pt}{
        \begin{fmfgraph*}(20,20)\fmfkeep{penta2}
            \fmfright{i1}
            \fmfleft{o1}
            \fmfv{decor.shape=pentagram,decor.size=5}{i1}
            \fmfv{decor.shape=pentagram,decor.size=5}{o1}
            \fmf{fermion}{i1,o1}
        \end{fmfgraph*}
        }
       \end{fmffile}
       ) \;\, 
       \parbox{25pt}{\fmfreuse{penta2}}
       \, + \, g( \;\,
       \begin{fmffile}{penta3}
        \parbox{25pt}{
        \begin{fmfgraph*}(20,20)\fmfkeep{penta3}
            \fmfright{i1}
            \fmfleft{o1,o2}
            \fmfv{decor.shape=pentagram,decor.size=6}{i1}
            \fmfv{decor.shape=pentagram,decor.size=6}{o1}
            \fmfv{decor.shape=pentagram,decor.size=6}{o2}
            \fmf{fermion}{i1,o1}
            \fmf{fermion}{i1,o2}
        \end{fmfgraph*}
        }
       \end{fmffile}
       ) \;\, 
       \parbox{25pt}{\fmfreuse{penta3}}
       \, + \, \cdots \,, 
\end{aligned}    
\end{align}
where $f(\tau)$, $g(\tau)$ are numerical coefficients, and the sum over trees is a formal sum. 
We assume that for linear trees  $\ell_n$ with $n$ vertices, the 
tree coefficients are determined by those in the power series, 
\begin{align}
    g(\emptyset) = g_0 \,,
    \quad 
    g(\bullet) = g_1 \,,
    \quad 
    g(\bullet \hskip -1.3mm -  \hskip -1.3mm\bullet) = g_2 \,,
    \quad 
    g(\ell_n) = g_n \,,
    \label{g-linear}
\end{align}
and that there exists a canonical extension to non-linear trees. 

The composition of two tree series follows from the replacement rule, 
\begin{align}
    \begin{aligned}
        \parbox{10pt}{\fmfreuse{penta}}
        &\quad \rightarrow \quad 
        f( \;\;
        \parbox{5pt}{\fmfreuse{dot}}
       ) \;\; 
        \parbox{10pt}{\fmfreuse{dot}}
        \hskip -4pt 
        + \, f( \;\,
        \parbox{25pt}{\fmfreuse{dot2}}
       ) \;\, 
       \parbox{25pt}{\fmfreuse{dot2}}
       \, + \, f( \;\,
       \parbox{25pt}{\fmfreuse{dot3}}
       ) \;\, 
       \parbox{25pt}{\fmfreuse{dot3}}
       \, + \, \cdots \,,
\end{aligned}
\label{replacement-rule}
\end{align}
and the result is best summarized in terms of partitions,\footnote{The ordering of $G$ and $F$ in $(G \star F)$ is reversed from the definition of ref.~\cite{Chartier_2010}.}
\begin{align}
G\star F = \sum_\tau (g\star f) (\tau) \tau \,,
\quad 
 (g \star f)(\tau) = \sum_{p \in P(\tau)} g(p) f(\tau\backslash p) \,.
        \label{g-f-star}
\end{align}
The role of partitions here is similar to the one in the previous section. 
When restricted to linear trees, the result agrees with the ordinary composition of power series \eqref{g-f-composition}. 

\begin{figure}[htbp]
    \centering
    \includegraphics[width=0.75\linewidth]{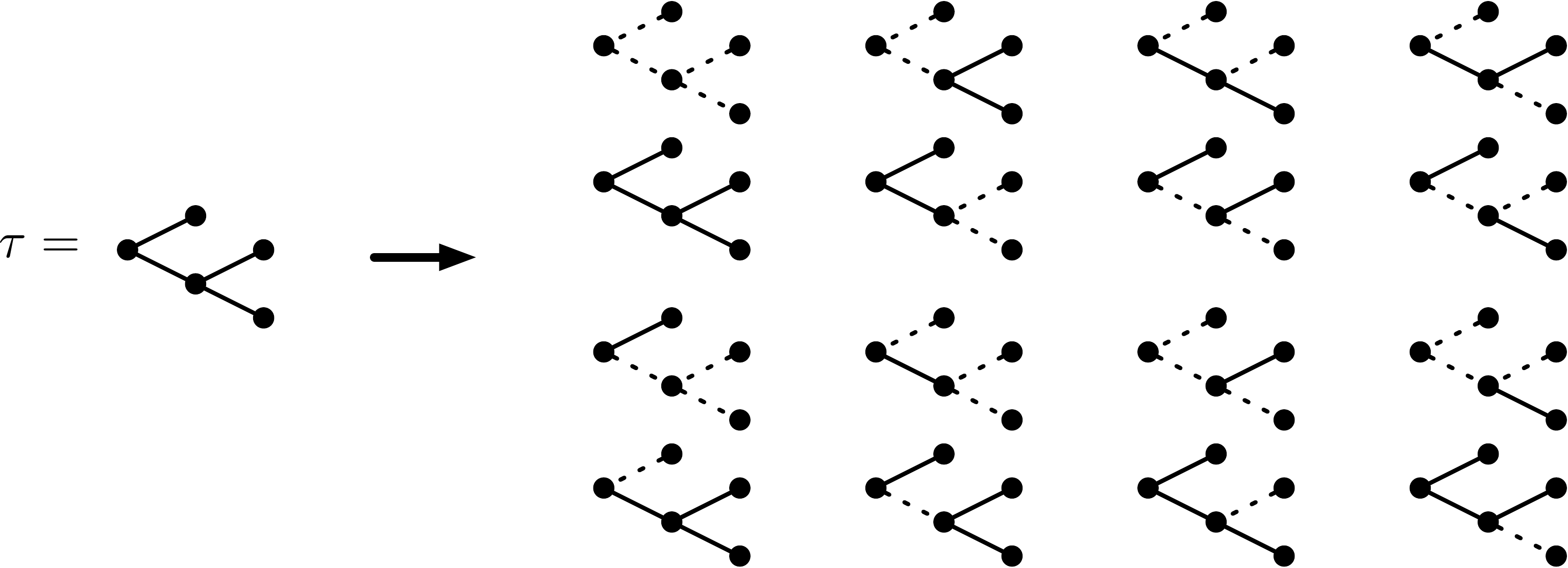}
    \caption{A rooted tree and its partitions.}
    \label{fig:partitions-16}
\end{figure}

Figure~\ref{fig:partitions-16} lists all 16 partitions of a tree with four edges (five vertices). 
The edges belonging to the partition are denoted by dotted lines; in the composition $[G\star F]$, they come from the $G$ side. 
The solid lines come from $F$ through the replacement rule \eqref{replacement-rule}. 

\paragraph{Hopf algebra} 

A key ingredient of the CEM Hopf algebra is the co-product, 
\begin{align}
    \Delta(\tau) = \sum_{p \in P(\tau)} p \otimes (\tau\backslash p) \,, 
\end{align}
which is coassociative, 
\begin{align}
    (\operatorname{id} \otimes \Delta) \circ \Delta = ( \Delta \otimes \operatorname{id} ) \circ \Delta\,. 
\end{align}
We can rewrite the composition rule for the tree series \eqref{g-f-star} slightly abstractly as 
\begin{align}
    (g \star f)(\tau) = \left(\mu \circ (g\otimes f) \circ \Delta \right) (\tau) \,,
\end{align}
where $\mu$ takes the product of two functions. 

Another key element of the CEM Hopf algebra is the antipode $S$ satisfying 
\begin{align}
    \mu \circ (S \otimes \operatorname{id} ) \circ \Delta = \mu \circ ( \operatorname{id} \otimes S) \circ \Delta 
    = \bullet \, \delta_{\bullet} \,,
\end{align}
where the single-vertex tree $(\bullet)$ is considered as an identity element among trees and $\delta_{\tau_1}(\tau_2)$ 
is the Kronecker delta on trees:
\begin{align}
    \delta_{\tau_1} (\tau_2) = 1 \quad (\tau_1 = \tau_2) \,,
    \quad 
     \delta_{\tau_1} (\tau_2) = 0 \quad (\tau_1 \neq \tau_2) \,. 
\end{align}
Intuitively, the antipode allows us to compute the ``inverse" of a tree series with $f(\bullet)=1$.

\paragraph{Exp-Log pair}

A prototypical example of a pair of tree series satisfying the ``inverse" relation is the pair ($e(\tau)$, $\omega(\tau)$), 
which play a central role throughout this paper: 
\begin{align}
\begin{split}
        e(\tau) \quad &\leftrightarrow \quad e^x = \sum_{n=0}^\infty \frac{1}{n!} x^n \,, 
        \\
        \omega(\tau) \quad &\leftrightarrow \quad \log(1+x) = 
        \sum_{n=1}^\infty \frac{(-1)^{n-1}}{n} x^n \,.
\end{split}
\end{align}
As explained in ref.~\cite{Chartier_2010}, the precise relation between the two tree series is 
\begin{align}
    (e \star \omega)(\tau) = \delta_\emptyset(\tau) + \delta_\bullet(\tau)  
    \quad &\leftrightarrow \quad 
    e^y|_{y=\log(1+x)} = 1 + x \,.
    \label{e star omega}
\end{align}
We may exchange the role of the two functions 
and write another composition rule, 
\begin{align}
 (\omega_+ \star e_-)(\tau) =  \delta_\emptyset(\tau) + \delta_\bullet(\tau) 
   \quad &\leftrightarrow \quad 
   \left[ 1 + \log (1+y) \right]_{y=e^x-1} = 1 + x \,.
   \label{omega star e} 
\end{align}
The shift by $(\pm 1)$ is to ensure the conditions \eqref{f-g-condition}. 

\paragraph{Computing $\omega$}

As shown in ref.~\cite{Chartier_2010}, given $e(\tau)$, we can compute $\omega(\tau)$ by solving \eqref{e star omega}. Similarly, we can compute $\omega(\tau)$ by solving \eqref{omega star e}. 
We are interested in trees with $|\tau|\ge 2$, which sets the RHS to zero. The LHS can be expanded using \eqref{g-f-star}:
\begin{align}
\begin{split}
  (e \star \omega) (\tau) &=  e(\bullet) \omega(\tau) + \mbox{(many smaller trees)}  +  e(\tau) \omega(\bullet)^n 
    \\
    &= \omega(\tau)   + \mbox{(many smaller trees)}  +   e(\tau)  \,.   
\end{split}
\label{omega-hopf}
\end{align}
When we work recursively, we assume that $\omega(\tau')$ for all smaller trees $\tau'$ are known, so it is clear that $\omega(\tau)$ can be determined uniquely. 

\begin{figure}[htbp]
    \centering
    \includegraphics[width=0.5\linewidth]{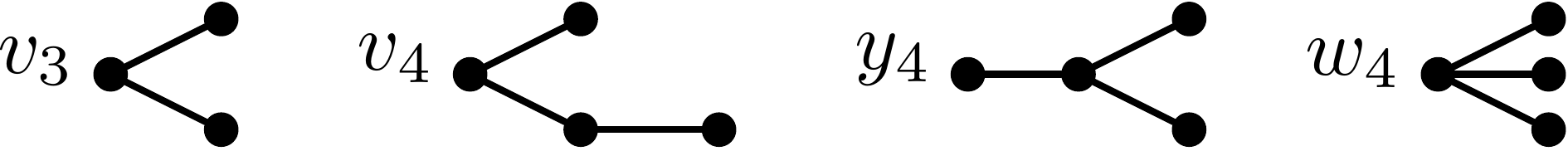}
    \caption{Some nicknames for trees.}
    \label{fig:tree-nicknames}
\end{figure}

Let us apply \eqref{e star omega} to the example in Figure~\ref{fig:partitions-16}, which admits 16 partitions. After grouping similar terms, we obtain an identity:
\begin{align}
    \begin{split}
        0 &= \omega(\tau) + e(\tau) 
        + 2 e(v_3) \omega(v_3) 
        + 2 e(v_3) \omega(\ell_3) + 2e(\ell_3) \omega(\ell_2)^2 
        \\
        &\quad + e(y_4) \omega(\ell_2) + e(\ell_2) \omega(y_4) + e(w_4) \omega(\ell_2) + e(\ell_2) \omega(\ell_2)\omega(v_3) 
        \\
        &\quad + 2 e(v_4) \omega(\ell_2) + 2 e(\ell_2) \omega(v_4) \,, 
    \end{split}
    \label{e star omega example}
\end{align}
where $\ell_n$ denotes the linear tree with $n$ vertices, and other nicknames for trees are given in Figure~\ref{fig:tree-nicknames}. Upon inserting $e(\tau)=1/15$ and the $e$, $\omega$ values for smaller trees, we obtain $\omega(\tau) = 1/60$, in agreement with \eqref{Murua-ex}. 
If we instead use \eqref{omega star e}, we obtain a similar identity as \eqref{e star omega example}, with the replacement $e \leftrightarrow \omega$, 
which still gives the same value for $\omega(\tau)$.

\subsection{Extension to non-rooted trees} 

The original CEM Hopf algebra is defined on rooted trees. The tree functions $e$ and $\omega$ are ``inverses" of each other (they satisfy the antipode relation), so extending $e(\tau)$ to non-rooted trees will automatically extend $\omega(\tau)$ and vice versa. 

The extended Murua formula \eqref{Murua-extended}, derived from the Magnus expansion, suggests a way to extend $\omega(\tau)$ before figuring out how to extend $e(\tau)$. 
Here, we take an alternative approach; we extend $e(\tau)$ first based on the relation $e(\tau) = \phi(\tau)/|\tau|!$ as explained in section~\ref{tree functions}, and then use the Hopf algebra to extend $\omega(\tau)$. 
Happily, the two extensions agree perfectly. Using some relations among the tree functions, we will be able to \emph{prove} that the two approaches lead to the same extension of $e(\tau)$ and $\omega(\tau)$.

\paragraph{Computing $\omega$ of non-rooted trees} 

\begin{figure}[htbp]
    \centering
    \includegraphics[width=0.54\linewidth]{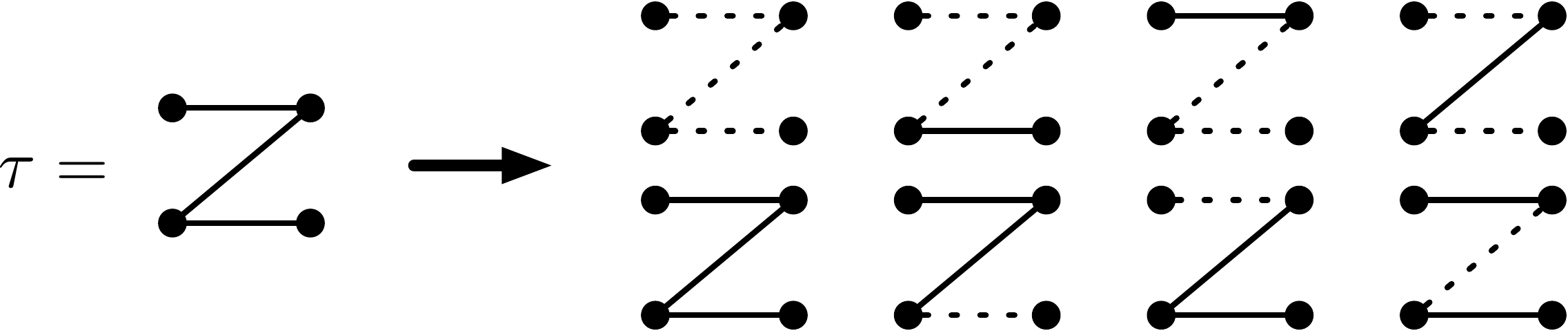}
    \caption{A non-rooted tree and its partitions.}
    \label{fig:partitions-8}
\end{figure}

The computation of $\omega(\tau)$ using either \eqref{e star omega} or \eqref{omega star e} works 
as well for non-rooted trees as for rooted ones. 
Applying the rules to the example in Figure~\ref{fig:partitions-8}, we obtain an equation,
\begin{align}
    \begin{split}
        0 &= \omega(\tau) + e(\tau) 
        + 2 e(v_3) \omega(\ell_2) 
        + 2 e(\ell_2) \omega(v_3) + e(\ell_3)\omega(\ell_2) + e(\ell_2)\omega(\ell_2)^2 \,, 
    \end{split}
    \label{e star omega non-rooted}
\end{align}
where we use the same nicknames as in Figure~\ref{fig:tree-nicknames}. 
Inserting $e(\tau)=5/24$ and the $e$, $\omega$ values for smaller trees, we obtain $\omega(\tau) = -1/12$, in agreement with the result of the extended Murua formula shown in Figure~\ref{fig:Murua-Sungsoo-ex2}.

The tree functions $e(\tau)$ and $\omega(\tau)$ satisfy a few interesting properties. We list them and discuss their derivation, interpretation and applications.

\paragraph{Sum rules} 

Let $\hat{\tau}$ be an unoriented tree with $n$ vertices.
We associate two sets with $\hat{\tau}$. 
Let $A(\hat{\tau})$ be the set of $2^{n-1}$ oriented trees obtained by assigning all possible orientations to the edges of $\hat{\tau}$. Some elements of $A(\hat{\tau})$ are isomorphic to each other as oriented trees. 
Let $B(\hat{\tau})$ be a subset of $A(\hat{\tau})$, where we count isomorphic oriented trees as one element. 
In other words, $B(\hat{\tau})$ is an equivalence class among all oriented trees with $n$ vertices. 
The sum rules are stated as follows. 

\emph{Sum rules for $e(\tau)$}:
\begin{align}
    \sum_{\tau \in A(\hat{\tau})} e(\tau) = 1 \,,
    \qquad 
    \sum_{\tau \in B(\hat{\tau})} \frac{e(\tau)}{\sigma(\tau)} = \frac{1}{S(\hat{\tau})}  \,.
    \label{eq:e_sum_rules}
\end{align}

\emph{Sum rules for $\omega(\tau)$}: 
\begin{align}
    \sum_{\tau \in A(\hat{\tau})} \omega(\tau) = (-1)^{|\hat{\tau}|-1} \,,
    \quad 
    \sum_{\tau \in B(\hat{\tau})} \frac{\omega(\tau)}{\sigma(\tau)} = \frac{(-1)^{|\hat{\tau}|-1}}{S(\hat{\tau})}  \,. \label{eq:omega_sum_rules}
\end{align}
For both sum rules $S(\hat{\tau})$ is the symmetry factor of the unoriented tree. 
The second $\omega$ sum rule may be called ``Feynman reduction," 
as it relates the eikonal expansion to the formal WQFT expansion.
For both functions, the second variant of the sum rule follows from the first one by the orbit-stabilizer theorem in group theory; for any tree function $f(\tau)$, 
\begin{align}
    \sum_{\tau \in B(\hat{\tau})} \frac{S(\hat{\tau})}{\sigma(\tau)} f(\tau) =    \sum_{\tau \in A(\hat{\tau})} f(\tau) \,.
\end{align}

Feynman reduction is equivalent to the following statement. Ignoring causality prescriptions of all propagators and summing over the coefficients of all equivalent diagrams, we obtain the coefficient (inverse of the symmetry factor) that would have been associated to the corresponding Feynman diagram. The physical motivation is that the effective action computed from the usual diagrammatic rules should be equal to the vacuum expectation value of the $\hat{\chi}$ operator computed using the Magnus expansion, if all propagators can never go on-shell. In other words, the Magnus expansion captures the subtleties in Wick-rotating Euclidean computations to Lorentzian signatures in the form of causality prescriptions. 
This property is important since tools developed for scattering amplitudes, such as generalized unitarity, can be used to construct the integrands. The $i0^+$ prescriptions can be determined after reduction to master integrals; see section~\ref{sec:3PM_1SF_WQFT_eik}.

\begin{figure}[htbp]
    \centering
    \includegraphics[width=0.66\linewidth]{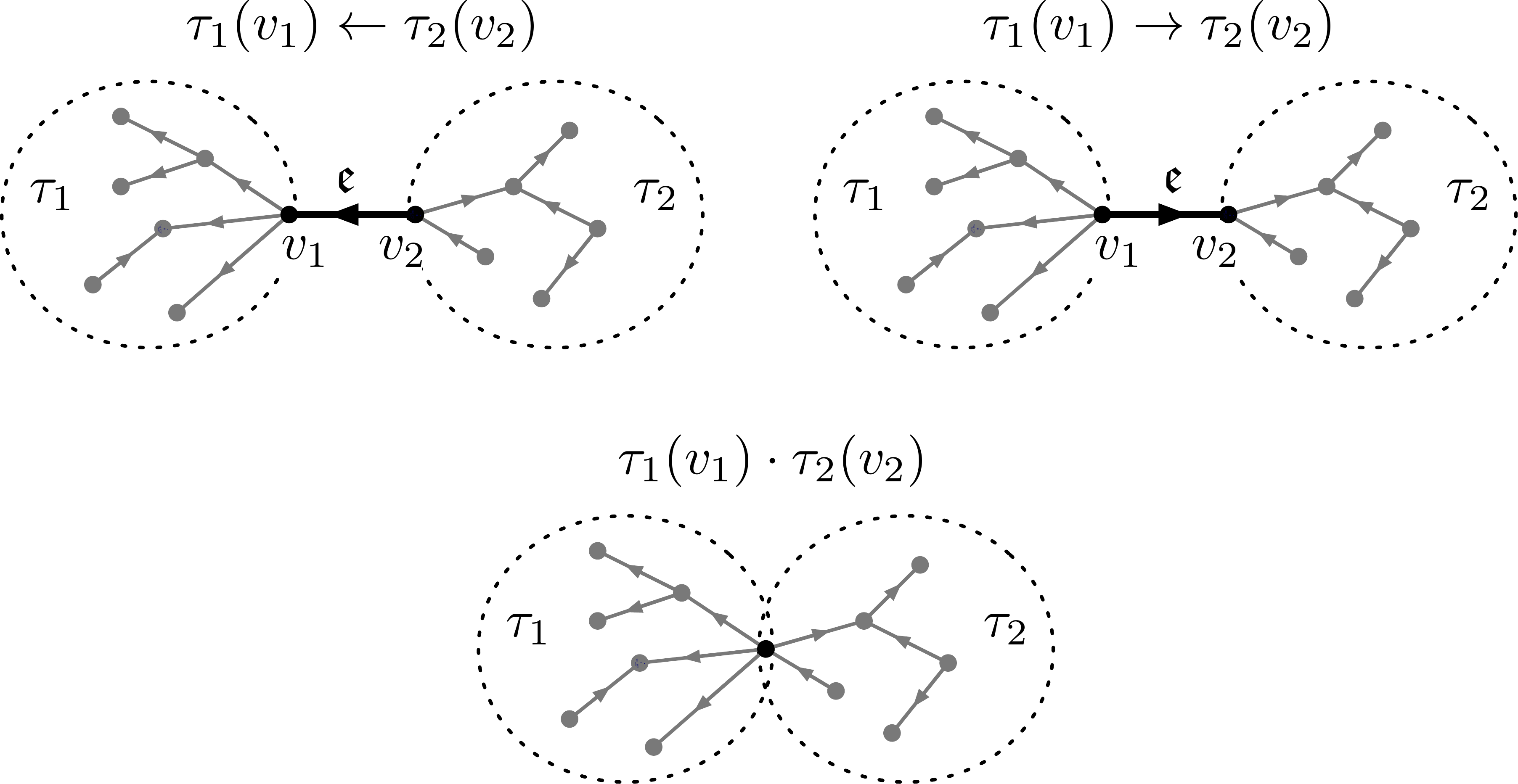}
    \caption{Three ways to combine two trees.}
    \label{fig:contraction-rules}
\end{figure}

\paragraph{Contraction rules}

To state the contraction rules, we introduce three ways to combine two trees, $\tau_1$ and $\tau_2$, to form a bigger tree. The definitions are pictorially explained in Figure~\ref{fig:contraction-rules}. 
Although the definitions depend crucially on the choice of vertices, for brevity, we may sometimes omit the reference to the vertices and write $\tau_{[1 \leftarrow 2]}$, $\tau_{[1 \rightarrow 2]}$, $\tau_1 \cdot \tau_2$.

\emph{Contraction rule for $e(\tau)$}: 
\begin{align}
    e(\tau_{[1 \leftarrow 2]}) + e(\tau_{[1 \rightarrow 2]}) = e(\tau_1) e(\tau_2) \,.
     \label{eq:e_contraction}
\end{align}

\emph{Contraction rule for $\omega(\tau)$}: 
\begin{align}
    \omega(\tau_{[1 \leftarrow 2]}) + \omega(\tau_{[1 \rightarrow 2]}) = - \omega(\tau_1 \cdot \tau_2) \,.
    \label{eq:omega_contraction}
\end{align}
We may state the latter slightly differently. 
 For a graph $\tau$, the graph $\t|_{\bar{\mathfrak{e}}}$ is obtained by selecting an edge $\mathfrak{e} \in \tau$ and reversing its orientation. Define the graph $\t_{\mathfrak{e}_c}$ as the graph obtained from $\t$ by shrinking the edge $\mathfrak{e}$ to a vertex. 
Then, the contraction rule reads 
    \begin{align}
          \w (\t) + \w (\t|_{\bar{\mathfrak{e}}})  = - \w(\t_{\mathfrak{e}_c}) 
         \,. \label{eq:omega_contraction-2}
    \end{align}

The edge contraction can be physically motivated as follows. Consider a UV theory with the interaction Hamiltonian $H_{\text{UV}}$. Integrating out heavy degrees of freedom, we get an effective theory with the interaction Hamiltonian $H_{\text{eff}}$.
Given a graph $\t$, assume that the edge $\mathfrak{e} \in \t$ corresponds to the propagator of the heavy degrees of freedom in the UV theory. In the effective theory, the heavy degrees of freedom cannot become on-shell, so its causality prescription is irrelevant. This means the graph $\t$ and the graph $\t|_{\bar{\mathfrak{e}}}$ obtained by reversing the causality of edge $\mathfrak{e}$ results in the same integral, and they can be considered the same. Moreover, the two $H_{\text{UV}}$ operators joined by the edge $\mathfrak{e}$ can be shrunk to the effective operator $H_{\text{eff}}$, leading to the graph $\t_{\mathfrak{e}_c}$ obtained by shrinking $\mathfrak{e} \in \t$ to a vertex. Requiring the sum of the first two integrals solely composed of $H_{\text{UV}}$ to be equal to the second integral with the edge $\mathfrak{e}$ replaced by $H_{\text{eff}}$ yields the edge contraction relation.

\paragraph{Combinatorics of the rules for $e$}

For $e(\tau)$, both the sum rule and the contraction rule admit simple combinatorial interpretation. 
In terms of the ordering counting function $\phi(\tau)$, the first sum rule reads 
\begin{align}
 \sum_{\tau \in A(\hat{\tau})} \phi(\tau) = n! \,, 
\end{align}
which splits $n!$ possible orderings of $n$ vertices into subsets depending on the orientation of the edges of $\tau$. 
The second sum rule is equivalent to the first one, except that we group trees related by symmetry. 

Similarly, the contraction rule can be understood as 
\begin{align}
    \phi(\tau_{[1 \leftarrow 2]}) + \phi(\tau_{[1 \rightarrow 2]}) = \phi(\tau_1) \phi(\tau_2) \frac{n!}{n_1! n_2!} \,.
\end{align}
The LHS counts the ordering of vertices subject to the oriented tree, except that the orientation of the edge connecting $\tau_1$ and $\tau_2$ is ignored. 
The RHS counts the ordering within $\tau_1$ and $\tau_2$ separately, and then take account of the relative ordering 
between the $n_1$ vertices of $\tau_1$ and the $n_2$ vertices of $\tau_2$. 

\paragraph{Contraction rule for $\omega$}

When we regard an unoriented tree $\hat{\tau}$ as an equivalence class among oriented trees, 
the contraction rule allows us to compute $\omega$ for all trees in a class from that of any one tree in the same class. 
In a situation where we have two or more candidate functions for $\omega$, 
if all of them satisfy the contraction rule and agree on the $\omega$ value for at least one tree in each class, 
then they should all be the same. 

We have three candidates for $\omega$: the extended Murua formula \eqref{Murua-extended}, 
and two variants of the Hopf algebra relation \eqref{e star omega}, \eqref{omega star e}. 
We will show that all three satisfy the contraction rule. 
Assuming that the original Murua formula \eqref{Murua-original} agrees with Hopf algebra for rooted trees, 
which include at least one oriented tree in each class, we conclude that the extended Murua formula also agrees with the extended Hopf algebra. 

The verification of the contraction rule works recursively. It can be checked explicitly for $|\tau|=2$. 
Assuming that the rule holds for all trees up to $|\tau| = n-1$, we try to prove that it should hold for $|\tau|=n$.

\begin{figure}[htbp]
    \centering
    \includegraphics[width=0.7\linewidth]{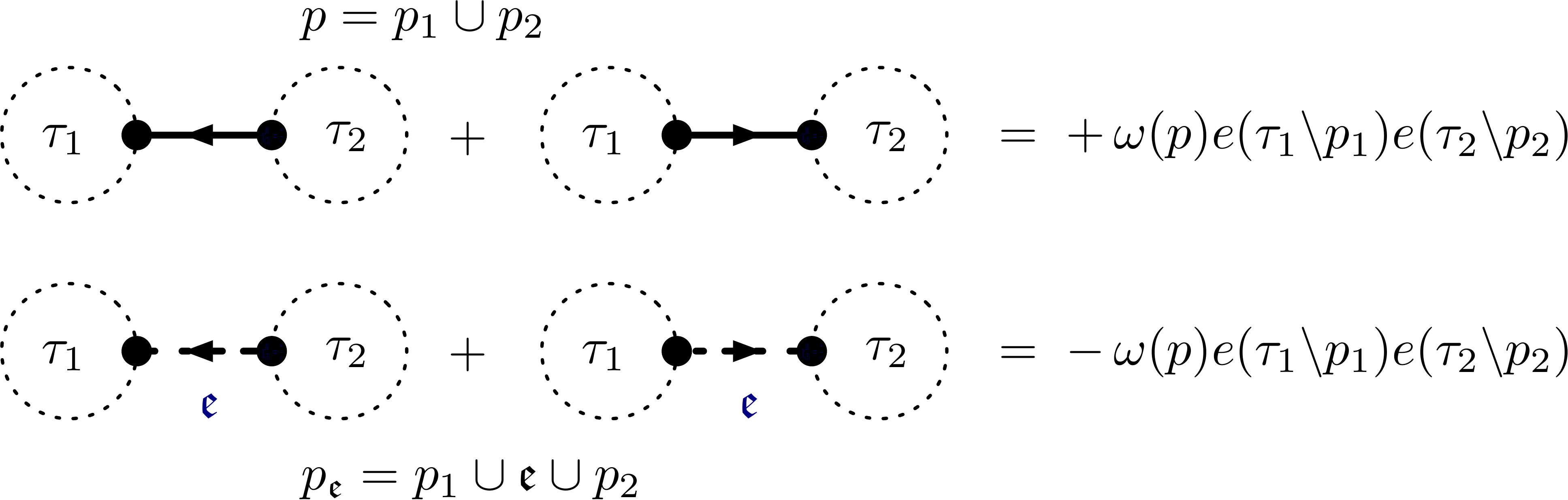}
    \caption{Cancellation based on contraction rule for smaller trees.}
    \label{fig:contraction-check1}
\end{figure}

Consider the Hopf algebra relation \eqref{omega star e}. 
For any tree $\tau$ with $|\tau| \ge 2$, we have
\begin{align}
J(\tau) =   \sum_{p\in P(\tau)} \omega(p) e(\tau\backslash p) = 0 \,.
\end{align}
The sum contains $\omega(\tau)$ from the case $p=\tau$.
For a proof of the contraction rule, consider
\begin{align}
   J(\tau_{[1 \leftarrow 2]}) +  J(\tau_{[1 \rightarrow 2]}) = 0 \,.
\end{align}
For almost all partitions, two terms from $\tau_{[1 \leftarrow 2]}$ and two from $\tau_{[1 \rightarrow 2]}$ 
cancel out, as shown in Figure~\ref{fig:contraction-check1}. It is assumed that 
the contraction rules for $e$ and $\omega$ hold for smaller trees. 
The only terms not cancelled in this manner are those with $p_1 = \tau_1$, $p_2 = \tau_2$, 
which gives 
\begin{align}
     \omega(\tau_{[1\leftarrow 2]}) +  \omega(\tau_{[1\rightarrow 2]}) + \omega(\tau_1\cdot \tau_2) = 0 \,,
\end{align}
which proves the contraction rule for $|\tau|=n$. 

\begin{figure}[htbp]
    \centering
    \includegraphics[width=0.7\linewidth]{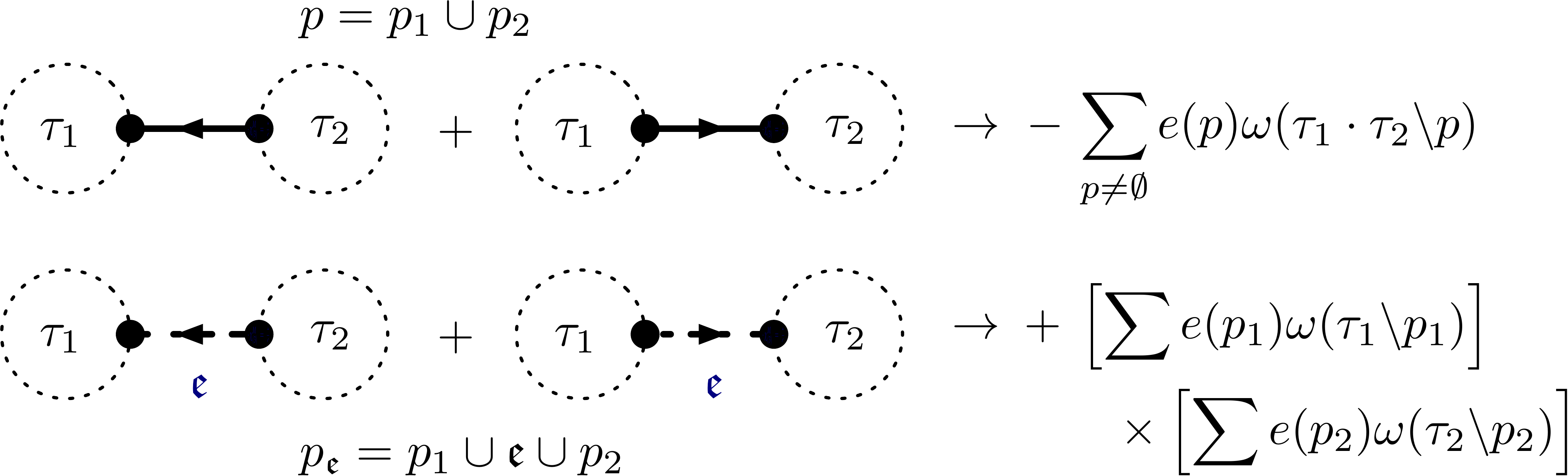}
    \caption{The partitions may or may not contain the edge to be contracted.}
    \label{fig:contraction-check2}
\end{figure}

For the other Hopf algebra relation \eqref{e star omega}, the proof proceeds similarly. 
We begin with 
\begin{align}
K(\tau) =   \sum_{p\in P(\tau)} e(p) \omega(\tau\backslash p) = 0 \,, 
\label{K-vanishing}
\end{align}
and consider the sum 
\begin{align}
    K(\tau_{[1 \leftarrow 2]}) +  K(\tau_{[1 \rightarrow 2]}) = 0 \,. 
\end{align}
For the sum over partitions containing the edge $\mathfrak{e}$ to be contracted, 
the contraction rule for $e(\tau)$ leads to the factorization 
\begin{align}
    \left[ \sum e(p_1) \omega (\tau_1\backslash p_1) \right] \left[ \sum e(p_2) \omega (\tau_2\backslash p_2) \right] \,,
\end{align}
which vanishes by \eqref{K-vanishing}. 
For the sum over partitions not containing the edge $\mathfrak{e}$, 
we get 
\begin{align}
    \omega(\tau_{[1\leftarrow 2]}) +  \omega(\tau_{[1\rightarrow 2]}) - \sum_{p\neq \emptyset} e(p) \omega(\tau_1\cdot \tau_2/p) = 
     \omega(\tau_{[1\leftarrow 2]}) +  \omega(\tau_{[1\rightarrow 2]}) + \omega(\tau_1 \cdot \tau_2) \,,
\end{align}
where we used the $\omega$ contraction rule for smaller trees and \eqref{K-vanishing}. 
Again, we conclude that the $\omega$ contraction rule holds for $|\tau|=n$. 

Finally, we show that the extended Murua formula \eqref{Murua-extended} satisfies the contraction rule. 
Let us assume that the contracted edge $\mathfrak{e}$ is not attached to the semi-root. Among all partitions, those with $\mathfrak{e} \notin p'$ produces the Murua formula for $-\omega(\tau_1\cdot \tau_2)$ (like the top line in Figure~\ref{fig:contraction-check2}). The partitions with $\mathfrak{e} \in p'$ cancel out, as the two terms carry the same $B_{|p|-1} e(p')\omega(\tau\backslash p)$ but opposite signs from $(-1)^{\ell(p)}$. 

The assumption that the contracted edge is not connected to the semi-root does not hold when both $\tau_1$ and $\tau_2$ are rooted and their roots are connected by $\mathfrak{e}$. But, we can turn the trees upside-down to keep $\mathfrak{e}$ away from the semi-root.

\paragraph{Sum rule for $\omega$}

\begin{figure}[htbp]
    \centering
    \includegraphics[width=0.65\linewidth]{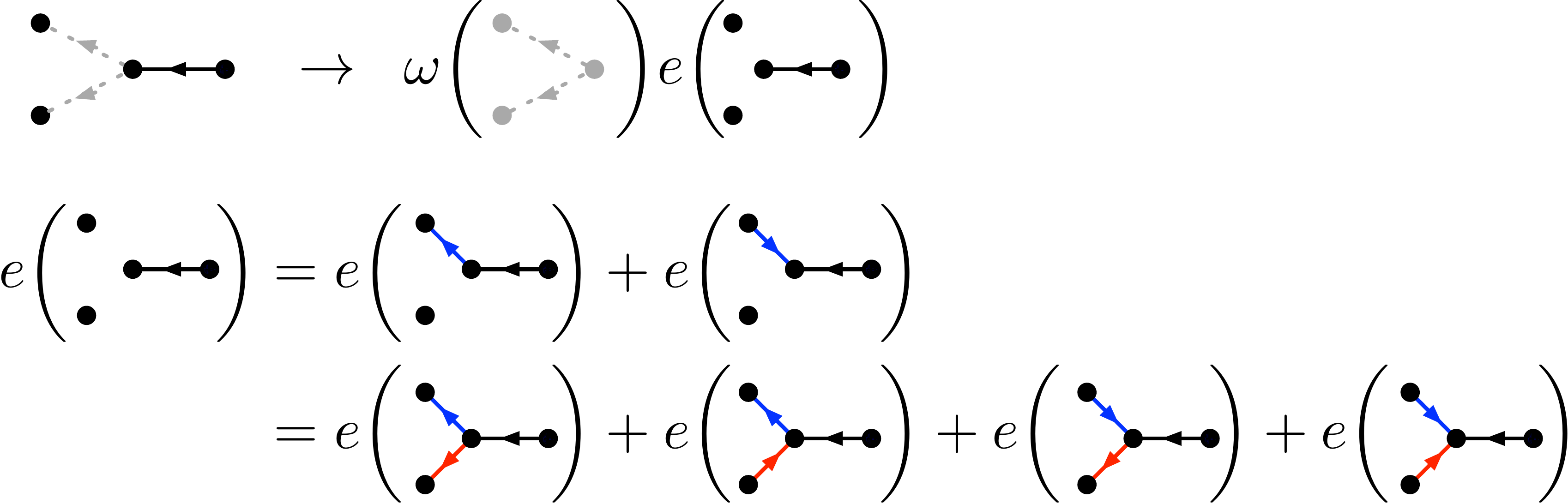}
    \caption{Using the contraction rule in the reverse order to turn $e(\tau\backslash p)$ to $e(\tau')$.}
    \label{fig:contraction-reversed}
\end{figure}

The sum rule for $\omega$ is also proved recursively. It holds trivially for $|\hat{\tau}|=1$. 
We assume that the rule holds for all sets $A(\hat{\tau})$ up to $|\hat{\tau}|=n-1$,
and examine the sum for $|\hat{\tau}|=n$. Consider a chain of equalities, 
\begin{align}
        \sum_{\tau \in A(\hat{\tau})} \omega(\tau) &=  - \sum_{\tau \in A(\hat{\tau})} \sum_{p\neq \tau} \omega(p) e(\tau\backslash p) 
        =  - \sum_{\tau \in A(\hat{\tau})} \sum_{p\neq \tau} \omega(p) 2^{|p|-|\tau|} \,.
\end{align}
The first equality is the Hopf algebra relation \eqref{omega star e} for $|\tau|\ge 2$. 
The second equality requires two steps. 
First, we use the contraction rule for $e$ in the reverse order to recover $e(\tau')$ for several $\tau' \in A(\hat{\tau})$ 
from disconnected components in $(\tau\backslash p)$; 
see Figure~\ref{fig:contraction-reversed}. 
Second, after rearranging the $e(\tau')$ within the sum over $\tau \in A(\hat{\tau})$, 
we use the sum rule for $e$ to replace $e$ by its average value, $2^{|p|-|\tau|}$. 

Similarly, we apply the sum rules for $\omega$ for smaller trees to replace $\omega(p)$ by its average value, $(-1)^{|p|-1} \cdot 2^{1-|p|}$. 
Counting the number of distinct partitions at a fixed $|p|$, we find 
\begin{align}
\begin{split}
        \sum_{\tau \in A(\hat{\tau})} \omega(\tau) 
        &= - 2^{|\hat{\tau}|-1}   \sum_{|p| = 1}^{|\hat{\tau}|-1} \binom{|\hat{\tau}|-1}{|p|-1} (-1)^{|p|-1} \cdot 2^{1-|p|} \cdot 2^{|p|-|\hat{\tau}|} 
        \\
        &= - \sum_{|p| = 1}^{|\hat{\tau}|-1} \binom{|\hat{\tau}|-1}{|p|-1} (-1)^{|p|-1} = (-1)^{|\hat{\tau}|-1} \,.
\end{split}
\end{align}
Hence, the proof is complete. 

\paragraph{Fast algorithm for computing $e$ and $\omega$}

\begin{figure}[htbp]
    \centering
     \includegraphics[width=0.45\linewidth]{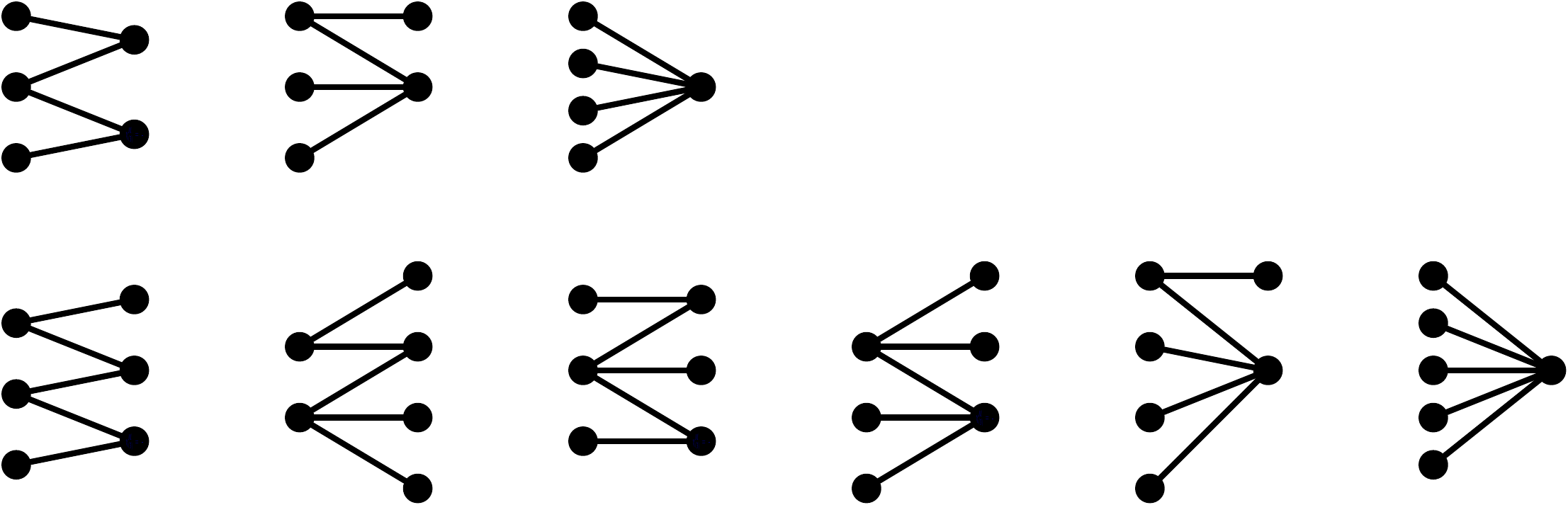}
    \caption{Maximally folded trees for $n=5$ (top) and $n=6$ (bottom). 
    For each unoriented tree, the maximally folded tree is unique up to an overall flip of orientation.}
    \label{fig:max-fold}
\end{figure}

The contraction rules can be used to devise a novel algorithm to compute $e(\tau)$ and $\omega(\tau)$. 
For each equivalence class $\hat{\tau}$, we take the ``maximally folded tree" in the sense shown in Figure~\ref{fig:max-fold}. We apply the extended Murua formula \eqref{Murua-extended} to compute $\omega(\tau)$ and the recursion formula to compute $e(\tau)$. For all the other trees in the same class, we use the contraction rules to compute $e$ and $\omega$.  

The \texttt{Mathematica} notebook, \texttt{contraction-omega.nb}, included in the set of ancillary files to this paper, computes $e(\tau)$ and $\omega(\tau)$ for all (rooted and non-rooted) trees up to $n=12$ in less than an hour in a state-of-the-art laptop computer, which is substantially faster than the EOM-based method and the Hopf-algebra-based method explained earlier.
Empirically, the computing time for the fast algorithm scales as $c^n$ ($c\sim 6$), 
a notable reduction from the $\sim (n!)^2$ scaling of the EOM-based method and the 
$\sim (n!)$ scaling of the Hopf-algebra-based method.

\section{Application to post-Minkowskian dynamics} \label{sec:PMdyn}

Generalization to relativistic scattering (including dynamical background field degrees of freedom) becomes almost trivial when we repackage the classical computation as a quantum field theory computation through the worldline quantum field theory (WQFT) formalism~\cite{Mogull:2020sak}, where propagation of worldline DOFs and graviton DOFs are treated on an equal footing as quantum field theory (QFT) two-point functions. 

The necessary small leap is a simple shift of view on the eikonal. Up to this point, we used the scattering generator equation \eqref{eq:eik_scgen} as the definition for the eikonal and determined propagator causality prescriptions based on diagrammatic representation of Poisson brackets (causality cuts), and then showed that the same causality prescription is given by the Magnus expansion. We now reverse the logic: We \emph{define} the eikonal by the Magnus expansion, and argue that the diagrams corresponding to scattering observables are obtained from the scattering generator equation \eqref{eq:eik_scgen} where Poisson brackets are computed by causality cuts. The nontrivial part of the leap is arguing that the causality cuts for graviton propagators will also compute Poisson brackets in the extended phase space where both worldline DOFs and graviton DOFs are present.

This section is organized as follows. We first recall the properties of two-point functions in QFT, and provide evidence that the causality cut for graviton propagators also compute Poisson brackets. Next, we compute the 3PM 1SF eikonal from WQFT as a demonstration that the causality prescription given by the Magnus expansion also applies to graviton propagators. We conclude the section by checking the 3PM impulse computed from the eikonal against full impulse calculations and commenting on the missing bremsstrahlung loss, which may be recovered when ``radiative eikonal'' is included in the scattering generator equation \eqref{eq:eik_scgen}.

\subsection{Wightman, Pauli and Jordan, and all that} \label{sec:Wightman}
The elementary building blocks for perturbative QFT are the Wightman functions,\footnote{We mostly follow ref.~\cite{Birrell:1982ix}, with some factors of $i$'s removed to simplify the diagrammatic rules. Although we only work with bosonic fields, generalization to fermionic fields should be straightforward.}
\begin{align}
    G^+ (x,y) := \langle 0 | \phi (x) \phi (y) | 0 \rangle \,,\quad G^- (x,y) := \langle 0 | \phi (y) \phi (x) | 0 \rangle \,,
\end{align}
where $\phi (x)$ is an abstract (fluctuation) field; it stands for any (dynamical) field content of the theory. The Wightman functions satisfy the free field equations for both arguments,
\begin{align}
    (\square_x - m^2) \, G^\pm (x,y) = (\square_y - m^2) \, G^\pm (x,y) = 0 \,,
\end{align}
where $\square_x$ is the d'Alembertian in the variable $x$ and $m$ is the mass of the field $\phi$.\footnote{We use mostly positive metric signature. For worldline fields, $\square_\s = - \frac{\partial^2}{\partial \s^2}$ is the double time derivative.}

The Pauli-Jordan function is the vacuum expectation value of the commutator,
\begin{align}
    G (x,y) &:= \langle 0 | [\phi (x), \phi (y)] | 0 \rangle = G^+ (x,y) - G^- (x,y) = [\phi (x), \phi (y)] \,,
\end{align}
where the last equation follows from the fact that the commutator of fields is a $c$-number. 

The retarded/advanced Green's functions are 
\begin{align}
\begin{aligned}
    G_R (x,y) &:= + \th (x^0 - y^0) \, G(x,y) \,,
    \\ G_A (x,y) &:= - \th (y^0 - x^0) \, G(x,y) = G_R (y,x) \,,
\end{aligned}
\end{align}
where $\th(x)$ is the Heaviside step function. The Green's functions satisfy the inhomogeneous wave equations,
\begin{align} \label{eq:IHWE_GRA}
    (\square_x - m^2) \, G_{R,A} (x,y) = - \delta( x^0 - y^0 ) \, [ \partial_0 \phi (x) , \phi (y) ] = i \delta^D (x - y) \,,
\end{align}
where we have used equal time commutation relations between the field $\phi(y)$ and its conjugate momentum field $\pi (x) = \partial_0 \phi(x)$. 

The Feynman propagator is
\begin{align}
\begin{aligned}
    G_F (x,y) &:= \langle 0 | T \phi(x) \phi(y) | 0 \rangle = \th (x^0 - y^0) \, G^+ (x,y) + \th (y^0 - x^0) \, G^- (x,y)
    \\ &= \frac{1}{2} \left[ G_R (x,y) + G_A (x,y) \right] + \frac{1}{2} \left[ G^+ (x,y) + G^- (x,y) \right] \,,
\end{aligned}
\end{align}
which also satisfies the inhomogeneous wave equation \eqref{eq:IHWE_GRA}. Note that the Feynman propagator is equivalent to the average of retarded and advanced Green's functions for worldline fields, since the average over Wightman functions vanish.

The \emph{causality cut} is defined as the difference between retarded and advanced Green's functions extend to general fields,\footnote{This definition is slightly different from that of ref.~\cite{Kim:2024grz}, which was restricted to worldline fields. This combination of Green's functions is also known as the Peierls bracket~\cite{Peierls:1952cb}.}
\begin{align}
    G_{R-A} (x,y) := G_R (x,y) - G_A (x,y) = G (x,y) = [\phi(x) , \phi(y)] \,,
\end{align}
and corresponds to the Pauli-Jordan function. ``De-quantizing'' the causality cut yields the Poisson bracket between fields,
\begin{align} \label{eq:cc_dq}
    G_{R-A} (x,y) = [\phi(x) , \phi(y)] \qquad \stackrel{\text{de-quantise}}{\longrightarrow} \qquad G_{R-A} (x,y) = - i \{ \phi(x) , \phi(y) \} \,.
\end{align}
When attached to vertices, the causality cuts of worldline fields compute the Poisson bracket between the amputated vertices; see appendix~\ref{app:i0cuts} for a detailed proof. 

We also expect that causality cuts of bulk graviton fields compute the Poisson bracket between amputated vertices. 
This is because the key ingredient of the proof is \eqref{eq:cc_dq} with the Poisson bracket between fluctuation fields substituted by that of background fields. 
The vertex amputation rules only computed derivatives of vertex rules, and it was necessary to pair them with causality cuts to complete them into a Poisson bracket between vertices.
However, there is a loophole that needs to be closed; the Lagrangian (2nd order) action is implicitly assumed in this section, while the Hamiltonian (1st order) action is assumed in appendix~\ref{app:i0cuts}. 
Nonetheless, we expect that the gap is not a serious obstacle for extending results in one action formulation to the other.

\subsection{3PM 1SF eikonal from WQFT} \label{sec:3PM_1SF_WQFT_eik}
The expansion in terms of the (symmetric) mass ratio $\n = \frac{m_1 m_2}{(m_1 + m_2)^2}$ is colloquially known as the self-force (SF) expansion. 
The leading order (0SF) corresponds to the background-probe limit, 
where the radial action coincides with the classical eikonal 
and can be most efficiently computed from the Schwarzschild solution in the isotropic coordinates;
see appendix~\ref{app:radial} and references therein. 
The leading order eikonal where causality prescription matters for propagators of massless particles\footnote{More precisely, where propagators of massless particles can become on-shell in the integration domain of loop momenta and therefore the integral depends on their causality prescriptions.} is the 1SF sector of the 3PM eikonal. As a demonstration that the causality prescription from the Magnus expansion extends to propagators of massless particles, we compute the 3PM 1SF eikonal using WQFT.

We use Feynman rules for WQFT presented in ref.~\cite{Mogull:2020sak}, therefore adopt mostly negative metric signature in this section.\footnote{The bulk graviton Feynman rules were provided by Gustav Jakobsen to one of the authors (JWK).} The \texttt{xTras} bundle of \texttt{xAct} package~\cite{xAct} was used for evaluating the integrands, and \texttt{LiteRed2} package~\cite{Lee:2012cn} was used for integration-by-parts (IBP) reduction to master integrals. 

We use notations of ref.~\cite{Jakobsen:2023oow} for the master integrals whose values are taken from ref.~\cite{Jakobsen:2022psy}.\footnote{The tabulated canonical basis in (4.59) of ref.~\cite{Jakobsen:2023oow} is missing a factor of $\zeta$; a factor of $\zeta$ should be present in all canonical integrals, not only $I^{\text{C},+\pm}_{(4)}$. JWK would like to thank Gustav Jakobsen for clarification.} For the reader's convenience, the notation for the master integrals are reproduced below.
\begin{gather}
    I_{n_1 \cdots n_7}^{\s_1 \s_2} = \int_{l_1,l_2} \frac{\deltabar(l_1 \cdot v_2) \deltabar(l_2 \cdot v_1)}{\prod_{i=1}^7 D_i^{n_i}} \,, \label{eq:MIdef} \phantom{asdfasdf} \\
    \begin{aligned}
        D_1 &= l_1 \cdot v_1 + \s_1 i 0^+ \,,\quad & D_2 &= l_2 \cdot v_2 + \s_2 i 0^+ \,,\quad & D_3 &= (k^0 + i0^+)^2 - \vec{k}^2 \,,
        \\ D_4 &= l_1^2 \,,\quad & D_5 &= l_2^2 \,, & &
        \\ D_6 &= (l_1 + q)^2 \,,\quad & D_7 &= (l_2 + q)^2 \,. & &
    \end{aligned}
\end{gather}
The $v_i^\m = p_i^\m / m_i$ are the velocity vectors.
We have fixed the causality prescription for $D_3$, where $k^\m = l_1^\m + l_2^\m + q^\m$. The alternative causality prescription for $D_3$ can be mapped to this set of integrals with $\s_{1,2} \to - \s_{1,2}$. The relevant master integrals are
\begin{align}
\begin{aligned}
    I_{(1)} &= I_{0,0,1,1,1,0,0} \,,\quad & I_{(2)} &= I_{0,0,2,1,1,0,0} \,,\quad & I_{(3)} &= I_{-1,-1,3,1,1,0,0} \,, 
    \\ I_{(4)} &= I_{1,1,1,1,1,0,0} \,,\quad & I_{(5)} &= I_{0,0,1,1,0,1,0} \,,\quad & I_{(6)} &= I_{0,0,1,1,1,1,1} \,, 
    \\ I_{(7)} &= I_{-1,-1,1,1,1,1,1} \,,\quad & I_{(8)} &= I_{0,0,0,1,1,1,1} \,. 
\end{aligned}
\end{align}
The causality prescription only matters for the master integral $I_{(4)}$, where the two distinct values are given by $I_{(4)}^{++} = I_{(4)}^{--}$ and $I_{(4)}^{+-} = I_{(4)}^{-+}$. The listed master integrals were classified as $b$-type integrals based on the observation that the integrals are relevant for the impulse in the direction of $b^\m$~\cite{Jakobsen:2023oow}. Note that the other master integrals, referred to as the $v$-type integrals based on the observation that the integrals are relevant for the impulse in the direction of $v_i^\m$, do not appear in the calculations. This is consistent with the expectation that $\{ \chi_{(3)} , p_1^\m \}$ can only contribute to the impulse in the $b^\m$ direction, and that the ``longitudinal impulse'' arises from iterations. The observation implies that we need to solve for less number of master integrals when we compute the eikonal, when compared to computing the impulse.

\begin{figure}
    \centering
    \begin{subfigure}[t]{0.25\linewidth}
        \centering
        \includegraphics[scale=0.75]{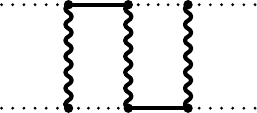}
        \caption{}
        \label{fig:3PMa}
    \end{subfigure}
    \quad
    \begin{subfigure}[t]{0.25\linewidth}
        \centering
        \includegraphics[scale=0.75]{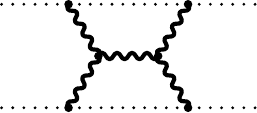}
        \caption{}
        \label{fig:3PMb}
    \end{subfigure}
    \quad
    \begin{subfigure}[t]{0.25\linewidth}
        \centering
        \includegraphics[scale=0.75]{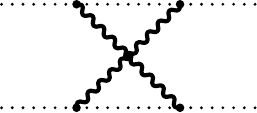}
        \caption{}
        \label{fig:3PMc}
    \end{subfigure}
    \\ \vskip 10pt
    \begin{subfigure}[t]{0.25\linewidth}
        \centering
        \includegraphics[scale=0.75]{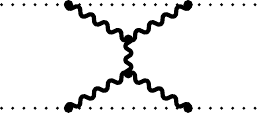}
        \caption{}
        \label{fig:3PMd}
    \end{subfigure}
    \quad
    \begin{subfigure}[t]{0.25\linewidth}
        \centering
        \includegraphics[scale=0.75]{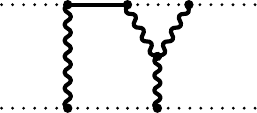}
        \caption{}
        \label{fig:3PMe}
    \end{subfigure}
    \quad
    \begin{subfigure}[t]{0.25\linewidth}
        \centering
        \includegraphics[scale=0.75]{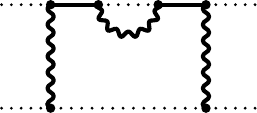}
        \caption{}
        \label{fig:3PMf}
    \end{subfigure}
    \caption{WQFT diagrams for 3PM 1SF eikonal.\\ Mirror diagrams ($1 \leftrightarrow 2$) of (e) and (f) should be added for full results.}
    \label{fig:3PMeik}
\end{figure}

The relevant diagrams for the eikonal are given in Figure~\ref{fig:3PMeik}. For the last two diagrams, \ref{fig:3PMe} and \ref{fig:3PMf}, their mirror images obtained by exchanging worldlines $1 \leftrightarrow 2$ should also be added for full results. We use Feynman reduction property of the tree function $\w(\t)$ \eqref{eq:omega_sum_rules} and construct the integrand through the usual Feynman rules, perform IBP reduction of the integrand to master integrals, substitute the master integrals whose weights are determined by the tree function $\w(\t)$, and then perform the impact parameter space Fourier transform. The evaluated diagrams (before weighting the master integrals) are presented in the ancillary file \texttt{diagram\_data.dat.m}. We comment on evaluation of the diagrams before presenting the results. 

The diagram \ref{fig:3PMa} is the most interesting in that it is the only diagram where causality prescription of worldline propagators matters. The relevant master integral is $I_{(4)}$, which we weight by $I_{(4)} = w_{++} I_{(4)}^{++} + w_{+-} I_{(4)}^{+-}$. The weights $w_{+\pm}$ can be determined from the tree function $\w(\t)$, where we use edge contraction rule \eqref{eq:omega_contraction} to amputate the outermost graviton propagators since their causality prescriptions are irrelevant.\footnote{In principle each $\w(\t)$ needs to be rescaled by the symmetry factor, $\w(\t)/\s(\t)$, but since all relevant propagators are distinct symmetry factors can be dropped in this calculation.}
\begingroup
\allowdisplaybreaks
\begin{subequations}
\begin{align}
    w_{++} &= - \left[ \w \left( \;
    \begin{fmffile}{z1}
        \parbox{25pt}{
        \begin{fmfgraph*}(25,20)
            \fmfright{r1,r2}
            \fmfleft{l1,l2}
            \fmfv{decor.shape=circle,decor.size=4}{r1}
            \fmfv{decor.shape=circle,decor.size=4}{r2}
            \fmfv{decor.shape=circle,decor.size=4}{l1}
            \fmfv{decor.shape=circle,decor.size=4}{l2}
            \fmf{fermion}{r1,l1,r2,l2}
        \end{fmfgraph*}
        }
    \end{fmffile}
    \;
    \right) + \w \left( \;
    \begin{fmffile}{z2}
        \parbox{25pt}{
        \begin{fmfgraph*}(25,20)
            \fmfright{r1,r2}
            \fmfleft{l1,l2}
            \fmfv{decor.shape=circle,decor.size=4}{r1}
            \fmfv{decor.shape=circle,decor.size=4}{r2}
            \fmfv{decor.shape=circle,decor.size=4}{l1}
            \fmfv{decor.shape=circle,decor.size=4}{l2}
            \fmf{fermion}{r1,l1}
            \fmf{fermion}{r2,l1}
            \fmf{fermion}{r2,l2}
        \end{fmfgraph*}
        }
    \end{fmffile}
    \;
    \right) + \w \left( \;
    \begin{fmffile}{z3}
        \parbox{25pt}{
        \begin{fmfgraph*}(25,20)
            \fmfright{r1,r2}
            \fmfleft{l1,l2}
            \fmfv{decor.shape=circle,decor.size=4}{r1}
            \fmfv{decor.shape=circle,decor.size=4}{r2}
            \fmfv{decor.shape=circle,decor.size=4}{l1}
            \fmfv{decor.shape=circle,decor.size=4}{l2}
            \fmf{fermion}{l1,r1}
            \fmf{fermion}{l1,r2}
            \fmf{fermion}{l2,r2}
        \end{fmfgraph*}
        }
    \end{fmffile}
    \;
    \right) + \w \left( \;
    \begin{fmffile}{z4}
        \parbox{25pt}{
        \begin{fmfgraph*}(25,20)
            \fmfright{r1,r2}
            \fmfleft{l1,l2}
            \fmfv{decor.shape=circle,decor.size=4}{r1}
            \fmfv{decor.shape=circle,decor.size=4}{r2}
            \fmfv{decor.shape=circle,decor.size=4}{l1}
            \fmfv{decor.shape=circle,decor.size=4}{l2}
            \fmf{fermion}{l2,r2,l1,r1}
        \end{fmfgraph*}
        }
    \end{fmffile}
    \;
    \right) \right] \nn
    \\[5pt] &= - \left[ \left( - \frac{1}{4} \right) + \left( - \frac{1}{12} \right) + \left( - \frac{1}{12} \right) + \left( - \frac{1}{4} \right) \right] = \frac{2}{3} \,,
    \\[5pt] w_{+-} &=- \left[ \w \left( \;
    \begin{fmffile}{z5}
        \parbox{25pt}{
        \begin{fmfgraph*}(25,20)
            \fmfright{r1,r2}
            \fmfleft{l1,l2}
            \fmfv{decor.shape=circle,decor.size=4}{r1}
            \fmfv{decor.shape=circle,decor.size=4}{r2}
            \fmfv{decor.shape=circle,decor.size=4}{l1}
            \fmfv{decor.shape=circle,decor.size=4}{l2}
            \fmf{fermion}{r1,l1,r2}
            \fmf{fermion}{l2,r2}
        \end{fmfgraph*}
        }
    \end{fmffile}
    \;
    \right) + \w \left( \;
    \begin{fmffile}{z6}
        \parbox{25pt}{
        \begin{fmfgraph*}(25,20)
            \fmfright{r1,r2}
            \fmfleft{l1,l2}
            \fmfv{decor.shape=circle,decor.size=4}{r1}
            \fmfv{decor.shape=circle,decor.size=4}{r2}
            \fmfv{decor.shape=circle,decor.size=4}{l1}
            \fmfv{decor.shape=circle,decor.size=4}{l2}
            \fmf{fermion}{r1,l1}
            \fmf{fermion}{r2,l1}
            \fmf{fermion}{l2,r2}
        \end{fmfgraph*}
        }
    \end{fmffile}
    \;
    \right) + \w \left( \;
    \begin{fmffile}{z7}
        \parbox{25pt}{
        \begin{fmfgraph*}(25,20)
            \fmfright{r1,r2}
            \fmfleft{l1,l2}
            \fmfv{decor.shape=circle,decor.size=4}{r1}
            \fmfv{decor.shape=circle,decor.size=4}{r2}
            \fmfv{decor.shape=circle,decor.size=4}{l1}
            \fmfv{decor.shape=circle,decor.size=4}{l2}
            \fmf{fermion}{l1,r1}
            \fmf{fermion}{l1,r2}
            \fmf{fermion}{r2,l2}
        \end{fmfgraph*}
        }
    \end{fmffile}
    \;
    \right) + \w \left( \;
    \begin{fmffile}{z8}
        \parbox{25pt}{
        \begin{fmfgraph*}(25,20)
            \fmfright{r1,r2}
            \fmfleft{l1,l2}
            \fmfv{decor.shape=circle,decor.size=4}{r1}
            \fmfv{decor.shape=circle,decor.size=4}{r2}
            \fmfv{decor.shape=circle,decor.size=4}{l1}
            \fmfv{decor.shape=circle,decor.size=4}{l2}
            \fmf{fermion}{r2,l2}
            \fmf{fermion}{r2,l1,r1}
        \end{fmfgraph*}
        }
    \end{fmffile}
    \;
    \right) \right] \nn
    \\[5pt] &= - \left[ \left( - \frac{1}{12} \right) + \left( - \frac{1}{12} \right) + \left( - \frac{1}{12} \right) + \left( - \frac{1}{12} \right) \right] = \frac{1}{3} \,.
\end{align}
\end{subequations}
\endgroup
This nontrivial weighting is important since the combination cancels the IR divergence of the eikonal. The leading $\e$ behaviors of $I_{(4)}^{+\pm}$ are ($D = 4 - 2 \e$)
\begin{gather}
\begin{gathered}
    I_{(4)}^{++} \propto \frac{1}{2 \e^2} + \CO (\e^{-1}) \,,\quad I_{(4)}^{+-} \propto - \frac{1}{\e^2} + \CO (\e^{-1}) \,.
    \\ \therefore \; \frac{2}{3} I_{(4)}^{++} + \frac{1}{3} I_{(4)}^{+-} = \CO (\e^{-1}) \,.
\end{gathered}
\end{gather}
In other words, if the diagrams are computed only using Feynman propagators, the 3PM eikonal will suffer from an IR divergence of the form $\frac{G^3}{\e |b|^2}$ as the weights will become $w_{++} = w_{+-} = \frac{1}{2}$. A similar behavior is also observed in the 0SF sector, where the Magnus expansion provides necessary weighting factors that render the eikonal finite. 

The diagrams \ref{fig:3PMc} and \ref{fig:3PMd} are both finite in $D = 4 - 2\e$ and therefore do not contribute to the eikonal. This is consistent with HEFT computations that the four-graviton cut does not contribute to classical physics; see section 6.3 of ref.~\cite{Brandhuber:2021eyq}. The non-vanishing contributions from the diagram \ref{fig:3PMb} is also consistent, since the diagram contributes to the ``zig-zag'' cuts of HEFT computations; see section 6.2 of ref.~\cite{Brandhuber:2021eyq}.

In principle, the ``mushroom'' family of integrals are needed for the diagram \ref{fig:3PMf} as the two worldline propagators can have differing causality prescriptions. However, the causality prescription for the worldline propagators are irrelevant for this diagram as the integral receives vanishing contributions from the locus of worldline on-shell conditions~\cite{Jakobsen:2023hig}. We can also argue that the integral family \eqref{eq:MIdef} is sufficient since a mixed product of retarded and advanced worldline propagators can be rewritten as linear combinations of products of retarded and products of advanced worldline propagators when appropriate regularization scheme is employed; see the discussion in appendix D of ref.~\cite{Kim:2024grz}.

The computed eikonal,
\begin{align} \label{eq:3PM1SFeik_res}
    \chi_{(3)}^{\text{1SF}} &= \frac{G^3 m_1^2 m_2^2}{|b^2|} \left[ \frac{2 \g (-55 + 6 \g^2 (22 - 19 \g^2 + 6 \g^4) )}{3(\g^2 - 1)^{5/2}} + \frac{4(3 + 12\g^2 - 4\g^4) \operatorname{arccosh}(\gamma)}{(\g^2 - 1)} \right. \nn
    \\ &\phantom{=asdf} \left. - \frac{2(1-2\g^2)^2 (- 8 + 5\g^2)}{3(\g^2 - 1)^2} + \frac{2 \g (1 - 2 \g^2)^2 (-3 + 2\g^2) \operatorname{arccosh}(\gamma)}{(\g^2 - 1)^{5/2}} \right] \,,
\end{align}
exactly matches the corresponding parts computed from amplitudes~\cite{Damgaard:2021ipf,Bjerrum-Bohr:2021din,Brandhuber:2021eyq}; see \eqref{chi3-brandhuber} below. The second line is known as the radiation-reaction contribution, and only receives contributions from diagrams \ref{fig:3PMa} and \ref{fig:3PMf}. One notable aspect of WQFT computations is that the eikonal is real-valued, contrary to amplitudes-based computations where the eikonal is complex-valued.

\subsection{3PM impulse from the eikonal}
We first list the eikonal reported in the literature~\cite{Mogull:2020sak,Damgaard:2021ipf,Bjerrum-Bohr:2021din,Brandhuber:2021eyq},
\begin{subequations} \label{eq:eik_full}
\begingroup
\allowdisplaybreaks
\begin{align}
     \chi_{(1)} &= - G m_1 m_2 \frac{2\gamma^2-1}{\sqrt{\gamma^2-1}} \log|b^2| \,, 
    \\
    \chi_{(2)} &= + \frac{3\pi}{4} G^2 m_1m_2 (m_1+m_2)  \frac{5\gamma^2-1}{\sqrt{\gamma^2-1}} \frac{1}{|b^2|^{1/2}} \,,
    \\
    \chi_{(3)} &= \frac{G^3 m_1 m_2}{3\sqrt{\gamma^2-1}|b^2|} \left[ (m_1^2+m_2^2) X_{(3,0)} + m_1 m_2 (X_{(3,1c)} + X_{(3,1r)}) \right] \,, \label{chi3-brandhuber}
     \\
     X_{(3,0)} &= \frac{\left(64 \gamma^6-120 \gamma^4+60 \gamma^2-5\right)}{\left(\gamma^2-1\right)^{2}} \,, \nn
     \\
     X_{(3,1c)} &= \frac{2\gamma(36\gamma^6-114\gamma^4+132\gamma^2-55)}{(\gamma^2-1)^2} - \frac{12(4 \gamma^4-12 \gamma^2-3)}{\sqrt{\gamma^2-1}} \operatorname{arccosh}(\gamma)\,, \nn
     \\
     X_{(3,1r)} &= (2\gamma^2-1)^2 \left[ -2 \frac{(5\gamma^2-8)}{(\gamma^2-1)^{3/2}} + \frac{6\gamma(2\gamma^2-3)}{(\gamma^2-1)^2} \operatorname{arccosh}(\gamma) \right] \,. \nn
\end{align}
\endgroup
\end{subequations}
For the 3PM eikonal, $0$ and $1$ in the subscripts distinguishes 0SF from 1SF, and $c$ refers to ``conservative" while $r$ refers to ``radiation reaction".

The 3PM impulse reported in the literature can be divided into conservative ($c$), radiation reaction ($rr$), and radiation loss ($rl$) contributions~\cite{Herrmann:2021tct,Kalin:2020fhe,Jakobsen:2022psy,Kalin:2022hph},
\begin{align}
    \D_{(3)} p_1^\m &= \D_{(3,c)} p_1^\m + \D_{(3,rr)} p_1^\m + \D_{(3,rl)} p_1^\m \,. \label{eq:3PM_impulse_full}
\end{align}
The separation of conservative and radiative contributions is based on the integration domain, where conservative/radiative corresponds to the potential/radiative region. The radiation loss $\D_{(3,rl)} p_1^\m$ refers to the momentum lost by bremsstrahlung~\cite{Jakobsen:2022psy},
\begin{align}
    \D_{(3,rl)} p_1^\m &= (v_2 \cdot P_{\text{rad}}) w_2^\m \,,\quad P_{\text{rad}}^\mu := - \D_{(3)} p_1^\m - \D_{(3)} p_2^\m \,, \label{eq:3PM_impulse_rl}
\end{align}
where $w_i^\m$ is the dual velocity vector defined by the conditions $(w_i \cdot v_j) = \delta_{ij}$.
\begin{align}
    w_1^\m := \frac{\g v_2^\m - v_1^\m}{\g^2 - 1} \,,\quad w_2^\m := \frac{\g v_1^\m - v_2^\m}{\g^2 - 1} \,.
\end{align}
The radiation loss \eqref{eq:3PM_impulse_rl} is along the longitudinal direction and does not contribute to the scattering angle. 
We refer to ref.~\cite{Jakobsen:2022psy} for the full impulse.

The impulse computed by inserting \eqref{eq:eik_full} into the scattering generator equation \eqref{eq:eik_scgen} reproduces \eqref{eq:3PM_impulse_full} \emph{except} the radiation loss \eqref{eq:3PM_impulse_rl}, when the Poisson brackets are computed in the phase space spanned by worldline DOFs, i.e. $b^\m \sim x_1^\m - x_2^\m$ and $p_i^\m$.\footnote{JWK would like to thank Gustav Mogull for bringing this point to attention.} 
It is important to recall that the impact parameter space is ``noncommutative''~\cite{Kim:2023vgb,Kim:2024grz},
\begin{align}
    \{ b^\m , b^\n \} &\stackrel{\cdot}{=} \frac{b^{\m} w_1^{\n} - b^{\n} w_1^{\m}}{m_1} - \frac{b^{\m} w_2^{\n} - b^{\n} w_2^{\m}}{m_2} \,,
\end{align}
where $\stackrel{\cdot}{=}$ denotes equality up to mass-shell conditions. Note that the missing radiation loss \eqref{eq:3PM_impulse_rl} \emph{cannot} be attributed to terms missing in the 3PM eikonal \eqref{chi3-brandhuber}, since it is along the ``longitudinal'' direction $w_2^\m$ which can only arise from nested Poisson brackets.

It would be interesting to show if the radiation loss can be attributed to neglected Poisson bracket contributions from the phase space of the gravitational field, e.g. the missing contribution is computed from iteration of leading-order ``radiation eikonal'' $\chi_{(1.5)}^{\text{rad}}$ corresponding to emission ($2 \to 3$) and/or absorption ($3 \to 2$), which may be viewed as a classical version of the eikonal operator considered in ref.~\cite{DiVecchia:2022piu}, such that
\begin{align}
    \D_{(3,rl)} p_1^\m = \frac{1}{2} \{ \chi_{(1.5)}^{\text{rad}} , \{ \chi_{(1.5)}^{\text{rad}} , p_1^\m \}_\circ \}_\circ \,,
\end{align}
where $\{ \bullet , \bullet \}_\circ$ is the extended Poisson bracket that incorporates the symplectic structure of the phase space of the gravitational field. 

\section{Discussion} \label{sec:discussion}

We developed the eikonal expansion mainly in a non-relativistic scattering problem, 
and then applied it to the gravitational binary system using vertex factors from the WQFT literature. 
It would be interesting to derive systematically the eikonal expansion of a single massive particle in an arbitrary electromagnetic and/or gravitational background, paying attention to the gauge/diffeomorphism invariance of the background field.
In refs.~\cite{Kim:2023aff,Kim:2024grz}, it was argued that electromagnetic and gravitational interactions are more naturally described by a deformation of the symplectic form than a deformation of the Hamiltonian. 
The existence of the classical eikonal remains beyond doubt. However, it is not clear whether/how the techniques developed in this paper could be used in a symplectic perturbation theory. 


In the PM computations, the 0SF terms are most efficiently computed by the radial action, 
completely bypassing the need for the eikonal expansion. At 1SF and higher SF orders, 
it would be interesting to see how much one can reduce the need for the eikonal expansion 
by extending the use of the radial action.

A topic that has not been discussed is the extension to diagrams with loops. While examination of tree graphs is enough for the classical limit in worldline-based methods, amplitude-based methods necessitate consideration of loop diagrams. Since amplitude-based methods are still extensively used in post-Minkowskian computations, it would be advantageous to have a diagrammatic representation of the Magnus series in the loop expansion. For example, subtraction of iteration/superclassical contributions when computing the effective Hamiltonian~\cite{Cheung:2018wkq} could be circumvented when diagrammatic representation of the Magnus series is available for loop expansions.

A related topic in this direction would be to study the analytic structure of the ${\chi}$-matrix, defined as the logarithm of the $S$-matrix $\hat{S} = \exp (i \hat{\chi} / \hbar)$. Although the Magnus series makes unitarity of the $S$-matrix manifest at the price of some analyticity, this does not imply that analyticity properties of the $S$-matrix will be completely lost in the $\chi$-matrix. Understanding the analytic structures of the $\chi$-matrix may provide us with an insight into various (non-)exponentiation observed in scattering amplitudes; e.g. exponentiation of IR divergences observed by Weinberg~\cite{Weinberg:1965nx}, exponentiation of $\CN = 4$ amplitudes known as the BDS ansatz~\cite{Bern:2005iz}, and \emph{non-exponentiation} of ladder-type diagrams when spin effects are present~\cite{Cheng:1971gf,Meng:1972xt,Weinberg:1971cdi,Czyz:1975bf,Chen:2022clh}. 
By studying its analyticity properties, we will be able to determine whether the $\chi$-matrix also possesses convenient properties that could benefit computation of collider observables.

\acknowledgments

The authors are grateful to 
Stefano De Angelis, Carlo Heissenberg, Gustav Jakobsen, Gustav Mogull, Julio Parra-Martinez, Radu Roiban, Trevor Scheopner, Jan Steinhoff and Fei Teng
for illuminating discussions. 
JHK is supported by the Department of Energy (Grant No. DE-SC0011632) and the Walter Burke Institute for Theoretical Physics. JHK is also supported by the Ilju Academy and Culture Foundation.  
JWK thanks Gustav Jakobsen for providing Feynman rules for bulk graviton vertices and clarifying various aspects of the master integrals.
JWK also thanks the Munich Institute for Astro-, Particle and BioPhysics (MIAPbP) for their hospitality during the program ``EFT and Multi-Loop Methods for Advancing Precision in Collider and Gravitational Wave Physics'' where part of this work was completed.
JWK was supported in part by MIAPbP which is funded by the Deutsche Forschungsgemeinschaft (DFG, German Research Foundation) under Germany's Excellence Strategy – EXC-2094 – 390783311.
The work of SK and SL is supported in part by the National Research Foundation of Korea (NRF) grants, 
NRF-2023-K2A9A1A0609593811 and NRF RS-2024-00351197.
SL is grateful for the APCTP for hospitality, where part of this work was done. 

\appendix

\section{Radial action} \label{app:radial}

For a generic potential, $\chi_{(n)}$ gives an $n$-fold integral. 
For the central potential, the equivalence between $\chi$ and the radial action allows us to write $\chi_{(n)}$ as a single integral. We discuss the relation and present a perturbation formula for the radial action.

\paragraph{Classical eikonal for a central potential} 
We consider the central potential problem with Hamiltonian 
\begin{align}
    H = \frac{\vec{p}^2}{2m} + \kappa V(r) \,.
\end{align}
The angular momentum $\vec{L} = \vec{x} \times \vec{p}$ is conserved, 
and the motion is confined on the plane perpendicular to $\vec{L}$. 
The scattering acts on the vectors $\vec{b}$ and $\vec{p}$ as an SO(2) rotation. 
The rotation symmetry also implies that the eikonal $\chi$ depends on $\vec{p}$, $\vec{b}$ only through their magnitudes 
($|b| = |\vec{b}|$, $|p| = |\vec{p}|$) 
\begin{align}
\chi = \chi(|b|,|p|) \,.
\end{align}
Recall that the impact parameter vector is defined as
\begin{align}
    \vec{b} = \vec{x} - \frac{\vec{x}\cdot \vec{p}}{p^2} \vec{p} \,,
\end{align}
which satisfies 
\begin{align}
\{ |p|^2 , \vec{b} \} =  0 \,, 
\quad  
\{ |b|^2 , \vec{p} \} =  2\vec{b} \,,
\quad 
\{ |b|^2 , \vec{b} \} = - \frac{2|b|^2}{|p|^2} \vec{p} \,. 
\label{poisson-bb}
\end{align}
Since $|p|$ commutes with everything, the SO$(2)$ Lie algebra action is given by 
\begin{align}
    \{ \chi, \hat{p} \} = + \hat{b}  \left( \frac{1}{|p|} \frac{\partial\chi}{\partial |b|} \right) \,,
    \quad 
    \{ \chi, \hat{b} \} = - \hat{p}  \left(\frac{1}{|p|}   \frac{\partial\chi}{\partial |b|} \right) \,. 
\end{align}
Clearly, the relation between the eikonal and the deflection angle should be 
\begin{align}
  \frac{1}{|p|}  \frac{\partial }{\partial |b|}  \chi(|b|,|p|) = \theta(|b|,|p|) \,.
   \label{eikonal-vs-deflection}
\end{align}
Another way of understanding \eqref{eikonal-vs-deflection} is to view it as the scattering generator equation \eqref{eq:eik_scgen} applied in cylindrical coordinates, $\D \phi = \{ \chi, \phi \} = -\frac{\partial \chi}{\partial J}$, where $J = |p||b|$. Iterated brackets vanish due to $\{ J, J\} = 0$.

\paragraph{Radial action vs. classical eikonal} 

The relation \eqref{eikonal-vs-deflection} is reminiscent of 
how to recover the deflection angle $\theta$ from the radial action.
Let us try to make a precise connection between the two concepts. 
The radial action $\mathcal{I}$ is defined as  
\begin{align}
    \mathcal{I}(E,L; r_1,r_2) = \int_{r_1}^{r_2} p_r(E,L;r) dr \,, 
\end{align}
where $p_r(E,L,r)$ is the solution of the equation 
\begin{align}
    E = \frac{p_r^2}{2m} + \frac{L^2}{2mr} + \kappa V(r) \,. 
\end{align}
For fixed $E$, $\mathcal{I}$ generates the angular motion on the plane as 
\begin{align}
    \frac{\partial \mathcal{I}}{\partial L} = \phi_2 - \phi_1 \,.
    \label{action-angle} 
\end{align}
Assuming that the polar angle $\phi$ increases along the particle trajectory, in a convention that the scattering angle $\theta$ is positive for a repulsive potential, the relation between $\theta$ and the change of $\phi$ is 
\begin{align}
    \theta = \pi - \Delta \phi \,.
\end{align}
In a scattering problem, the particle trajectory connects the past and future asymptotic regions while going through a minimum value of $r$. It is convenient to compute the radial action with $r_\mathrm{min}$ as a reference point, 
\begin{align}
    \mathbb{I}(|p|,|b|;r) = |p| \int_{r_\mathrm{min}}^r 
    \sqrt{1-\frac{b^2}{r^2} - \frac{2m \kappa V(r)}{p^2} } dr \,, \quad  L = |b| |p| \,.
    \label{radial action standard}
\end{align}
Comparing \eqref{eikonal-vs-deflection}, \eqref{action-angle} and \eqref{radial action standard}, 
we note that the relation between the classical eikonal $\chi$ and the radial action should be
\begin{align}
    \chi(|p|,|b|) = 2 \lim_{R\rightarrow \infty} \mathbb{I}(|p|,|b|;R)_\mathrm{finite} \,.
    \label{chi-radial-master}
\end{align}
It is understood that the term linearly divergent in $R$ (and a $\log R$ term for a $1/r$ potential) are subtracted out. 
From here on, we assume that the potential falls off faster than $1/r$.

\paragraph{Perturbation theory}
For a perturbation theory, there is an alternative to the relation \eqref{chi-radial-master}. 
We rewrite \eqref{radial action standard} slightly as
\begin{align}
    \mathbb{I}(\kappa, \tilde{r}) = p \int_{\tilde{r}_\mathrm{min}}^{\tilde{r}} 
    \sqrt{\tilde{r}^2(1- u(\tilde{r})) - b^2} \; \frac{d\tilde{r}}{\tilde{r}} \,, 
    \quad u(\tilde{r}) = \frac{2m \kappa V(\tilde{r})}{p^2} \,.
    \label{radial action variant}
\end{align}
We renamed the original radial coordinate $\tilde{r}$ for later convenience. 
The relation between $\tilde{r}_\mathrm{min}$ and $b$ is non-trivial. 
One way to proceed is to define a new radial coordinate $r$ by
\begin{align}
    \tilde{r}^2(1- u(\tilde{r})) = r^2 \,.
    \label{r-vs-r-tilde}
\end{align}
It follows that 
\begin{align}
    \tilde{r} = \tilde{r}_\mathrm{min} \;\;\Longleftrightarrow \;\; r = b 
    \,,\qquad 
    2\frac{d\tilde{r}}{\tilde{r}} + d\log(1-u) = 2\frac{dr}{r} \,. 
\end{align}
The divergent part is clearly separated from the finite part. 
Doing an integration parts for the latter, we arrive at the alternative expression, 
\begin{align}
    \chi = \frac{p}{2} \int_{-\infty}^\infty 
    \log(1-u(\tilde{r})) dz \,, 
    \quad r = \sqrt{b^2 + z^2} \,.
    \label{chi-radial-new}
\end{align}
The new integration variable is $z$, which quickly determines $\tilde{r}$, but $u$ is still a function of the old $\tilde{r}$. To proceed further, we should solve \eqref{r-vs-r-tilde} to express $r$ as a function of $\tilde{r}$.  
\begin{align}
    \tilde{r} = f(r) = \tilde{r} + f_1(r) + f_2(r) + \cdots \,. 
    \label{r-to-r-tilde}
\end{align}
and use it to expand the integrand of \eqref{chi-radial-new}.
The leading order answer is as expected:
\begin{align}
    \chi_{(1)} = -\frac{p}{2} \int_{-\infty}^\infty u \left(\sqrt{z^2+b^2}\right) dz 
    = -\kappa \int_{-\infty}^\infty 
    V(\vec{b} + \vec{v} t ) dt \,. 
    \label{chi1-central}
\end{align}
Solving \eqref{r-to-r-tilde} order by order becomes cumbersome at higher orders.

\paragraph{All order formula from a contour integral} 

It is convenient to take $r^2$ as an independent variable; define $x=r^2$, $\tilde{x} = \tilde{r}^2$, etc. 
The coordinate change now reads
\begin{align}
    \tilde{x}(1-u(\tilde{x})) = x 
    \quad \Longrightarrow \quad 
    \frac{d\tilde{x}}{\tilde{x}} + d \log(1-u(\tilde{x})) = \frac{dx}{x} \,.
    \label{r-tilde-square}
\end{align}
Our goal is to show that 
\begin{align}
   F(x) :=  - \log\left (1-u[\tilde{x}(x)] \right) = \sum_{n=1}^\infty \frac{1}{n!} \left( \frac{d}{dx} \right)^{n-1} \left[x^{n-1} u(x)^n \right] \,. 
\end{align}
To begin, we use Cauchy's formula to write 
\begin{align}
\begin{split}
   -\log(1-u[\tilde{x}(x)] )   
     &= - \frac{1}{2\pi i} \oint \log(1-u[\widetilde{w}(w)] ) \frac{dw}{w-x}  
     \\
     &= - \frac{1}{2\pi i} \oint \log\left( \frac{w}{\tilde{w}} \right) \frac{dw}{w-x} \,,
\end{split}
\end{align}
where we used \eqref{r-tilde-square}. Now we add to both sides
\begin{align}
      0 =  \frac{1}{2\pi i} \oint \left[ \frac{dw}{w-x} \log(w)- \frac{d\tilde{w}}{\tilde{w}-x} \log(\tilde{w})\right]   \,.
\end{align}
After the $\log(w)$ terms cancel out, we get 
\begin{align}
    F(x) =  \frac{1}{2\pi i} \oint \left[ \frac{dw}{w-x} - \frac{d\tilde{w}}{\tilde{w}-x} \right] \log(\tilde{w}) 
    =  \frac{1}{2\pi i} \oint d \log\left( \frac{w-x}{\tilde{w}-x}\right) \log(\tilde{w}) \,.
    \label{dlog-log-IBP}
\end{align}
The branch cuts are such that we can apply an integration by parts to get  
\begin{align}
    F(x) 
    =  - \frac{1}{2\pi i} \oint \log\left( \frac{w - x }{\tilde{w}-x} \right) \frac{d\tilde{w}}{\tilde{w}} = - \frac{1}{2\pi i} \oint \log\left( 1 - \frac{\tilde{w} u(\tilde{w})}{\tilde{w}-x} \right) \frac{d\tilde{w}}{\tilde{w}}  \,.
    \label{log-dlog}
\end{align}
In the last step, we applied \eqref{r-tilde-square} once again. 
Relabelling the integration variable ($\tilde{w}\rightarrow w$), expanding the log, and exchanging the sum and the integral, we get 
\begin{align}
    F(x) = - \sum_{n=1}^\infty \frac{1}{n} \frac{1}{2\pi i} \oint \frac{w^n u(w)^n}{(w-x)^n} \frac{dw}{w} = \sum_{n=1}^\infty \frac{1}{n!} \left(\frac{d}{dx}\right)^{n-1} [x^{n-1} u(x)^n] \,. 
\end{align}
In the final step, we used Cauchy's formula for derivatives. 
Switching back to the radial coordinates, we obtain the all order formula for $\chi_{(n)}$ in a central potential:
\begin{align}
    \chi_{(n)} = -\left( \frac{p}{2} \right) \frac{1}{2^{n-1} n!} \int \left( \frac{d}{r dr} \right)^{n-1} \left[ r^{2(n-1)} u(r)^n\right] \,.
    \label{all-order-radial-action}
\end{align}
This formula is not new in that an equivalent result, the all order scattering angle formula, was given in~\cite{Bjerrum-Bohr:2019kec,Damgaard:2022jem} (see also~\cite{Cheung:2024byb}). 
But, to the best of our knowledge, our derivation based on a contour integral is new. 

It is a straightforward (but not entirely trivial even for $n=2, 3$) exercise to specialize the general eikonal formula of the main text to a central potential and reproduce \eqref{all-order-radial-action}. The precise values of $\omega(\tau)$ are needed to guarantee the matching.

\section{Diagrammatic representation of Poisson brackets from causality cuts} \label{app:i0cuts}
We consider a worldline in one-dimensional target space whose action takes the form
\begin{align}
S &= \int p dq - \frac{\k}{2} (p^2 + m^2) d\s + S_{\text{int}} [p,q;A] \,,
\end{align}
where $\k$ is the Lagrange multiplier field enforcing mass-shell conditions and $A$ denotes implicit dependence on the background field which will be suppressed in the discussions. Generalization to higher-dimensional target space and multiple worldlines is straightforward.

\subsection{Vertex amputation rules}
We define \emph{fundamental variables} as the variables whose connected correlation functions are constructed from Wick contraction and vertex rules. We make the following definitions for the fundamental variables.
\begin{itemize}
    \item \emph{Momentum variables} refer to fundamental variables whose background-fluctuation separation takes the form $p (\s) = p_0 + \pi (\s)$ where $\pi (\s)$ is the fluctuation field.
    \item \emph{Position variables} refer to fundamental variables whose background-fluctuation separation takes the form $q (\s) = q_0 + \k p_0 \s + \phi (\s)$ where $\phi(\s)$ is the fluctuation field. We refer to $p_0$ as the conjugate momentum of the background position variable $q_0$.
\end{itemize}
This definition allows us to write vertex rules with the fluctuation field $f(\w) \in \{ \pi (\w) , \phi (\w) \}$ attached to the delta constraints $\deltabar(\w)$ or $\deltabar'(\w)$ as amputated vertex rules where the fluctuation field $f(\w)$ is substituted by derivatives of the background fields.

We first consider a pair of canonically conjugate fields $p (\s)$ and $q (\s)$ satisfying the definition of momentum and position variables given above. In frequency space, their expansion is given as
\begin{align}
\begin{aligned}
    p(\w) &= p_0 \deltabar(\w) + \pi (\w) \,,
    \\ q(\w) &= q_0 \deltabar(\w) - i \k p_0 \deltabar' (\w) + \phi (\w) \,,
\end{aligned}
\end{align}
which follows from the free action\footnote{\emph{Free action} refers to terms of the action that are up to quadratic in the fundamental variables.}
\begin{align}
S_{\text{free}} &= \int p dq - \frac{\k}{2} (p^2 + m^2) d\s \,, \label{eq:free_act}
\end{align}
where we use the negative frequency expansion $f (\s) = \int_\w e^{- i \w \s} f(\w)$. The Feynman rules for vertices where one of the legs are specified to be $\pi (\w)$ or $\phi (\w)$ are obtained from the functional derivative of the interaction action,\footnote{\emph{Interaction action} refers to terms of the action that are cubic and higher in the fundamental variables.}
\begin{align}
\begin{aligned}
    \frac{\delta S_{\text{int}}[p,q]}{\delta \pi (\w)} &= \frac{\partial p(\w')}{\partial \pi (\w)} \frac{\delta S_{\text{int}}[p,q]}{\delta p(\w')} = \frac{\delta S_{\text{int}}[p,q]}{\delta p(\w)} \,,
    \\ \frac{\delta S_{\text{int}}[p,q]}{\delta \phi (\w)} &= \frac{\partial q(\w')}{\partial \phi (\w)} \frac{\delta S_{\text{int}}[p,q]}{\delta q(\w')} = \frac{\delta S_{\text{int}}[p,q]}{\delta q(\w)} \,.
\end{aligned}
\end{align}
On the other hand, the vertex rules with derivatives of the background field acting on them can be obtained from the same derivative acting on the interaction action,
\begin{align}
\begin{aligned}
    \frac{\partial S_{\text{int}}[p,q]}{\partial q_0} &= \frac{\partial q(\w)}{\partial q_0} \frac{\delta S_{\text{int}}[p,q]}{\delta q(\w)} = \deltabar (\w) \frac{\delta S_{\text{int}}[p,q]}{\delta \phi (\w)} \,,
    \\ \frac{\partial S_{\text{int}}[p,q]}{\partial p_0} &= \frac{\partial p(\w)}{\partial p_0} \frac{\delta S_{\text{int}}[p,q]}{\delta p(\w)} + \frac{\partial q(\w)}{\partial p_0} \frac{\delta S_{\text{int}}[p,q]}{\delta q(\w)}
    \\ &= \deltabar(\w) \frac{\delta S_{\text{int}}[p,q]}{\delta \pi (\w)} - i \k \deltabar' (\w) \frac{\delta S_{\text{int}}[p,q]}{\delta \phi (\w)} \,,
\end{aligned} \label{eq:intact_bg_deriv}
\end{align}
where the integration $\int_\w$ is implicit on the RHS. Therefore, we find the following \emph{amputation rules} for zero frequency fluctuation fields,
\begin{align}
    V_\phi [\phi (\w)] \deltabar(\w) \, \to \, \frac{\partial V'}{\partial q_0} \,, \quad V_\pi[\pi (\w)] \deltabar(\w) - i \k V_\phi [\phi (\w)] \deltabar'(\w) \, \to \, \frac{\partial V'}{\partial p_0} \,, \label{eq:vertex_amputation_rule}
\end{align}
where $V'$ is the amputated vertex rule and $V_{\pi/\phi}$ is the vertex rule with extra $\pi/ \phi$ fluctuation field attached to the amputated vertex $V'$.

The exceptional cases not covered by the amputation rules are the worldline fluctuation 3pt vertices, which may appear when the 2pt functions are dependent on the background fields, the twistor worldline model~\cite{Kim:2024grz} being an example. Such 3pt vertices should be combined with the 2pt functions and checked separately that the amputation rule reduces to taking a background field derivative of the 2pt functions.\footnote{When taking a product of propagators with differing $i0^+$ prescriptions, an appropriate regularization procedure is required to remove divergences associated to the ``infinite volume'' of the real line. See the discussion in appendix D of ref.~\cite{Kim:2024grz}.}

Generalization of \eqref{eq:vertex_amputation_rule} to higher order background field derivatives is straightforward and can be expressed elegantly using the \emph{augmentation operator} $\Delta$ which acts on a vertex rule $V$ and converts it to the vertex rule $\Delta_{\phi(\w) / \pi(\w)} V$ that has an extra $\phi(\w) / \pi (\w)$ fluctuation field compared to the original vertex rule $V$. The \emph{augmentation rule} is the rewriting of the amputation rule \eqref{eq:vertex_amputation_rule} given by
\begin{align}
    \frac{\partial}{\partial q_0} = \deltabar(\w) \Delta_{\phi (\w)} \,,\quad \frac{\partial}{\partial p_0} = \deltabar(\w) \Delta_{\pi(\w)} - i \k \deltabar'(\w) \Delta_{\phi (\w)} \,, \label{eq:vertex_augmentation_rule}
\end{align}
which can be applied multiple times. The equality \eqref{eq:vertex_augmentation_rule} should be understood as a statement about operators acting on vertex rules where the integration $\int_\w$ is implicit on the RHS.

\subsection{Causality cuts as Poisson brackets}
The free action \eqref{eq:free_act} leads to the Poisson bracket
\begin{align}
\{ q , p \} = 1 = \{ q_0 , p_0 \} \,,
\end{align}
inherited by the background fields and the 2pt functions,
\begin{align}
\langle \pi (\w') \, \phi (\w) \rangle &= - \frac{1}{\w} \deltabar (\w' + \w) \,,\quad \langle \phi (\w') \, \phi (\w) \rangle = \frac{i \k}{\w^2} \deltabar(\w' + \w) \,,
\end{align}
where the $i0^+$ prescription determines the causality flow. The causality flow $\w' \to \w$ corresponds to the $i0^+$ prescription $\w \to \w + i0^+$, and \emph{causality cut} is defined as the difference between the retarded and advanced $i0^+$ prescriptions.\footnote{Compared to the original definition of ``retarded minus time-symmetric'' introduced in ref.~\cite{Kim:2024grz}, this definition differs by a factor of 2.} The cutting rules for the causality cut $[(\w' \to \w) - (\w \to \w')]$ can be written as
\begin{align}
\begin{aligned}
\langle \pi (\w') \, \phi (\w) \rangle \quad &\to \quad + i \deltabar (\w') \deltabar(\w) = - i \left( \deltabar(\w') \, \{ p_0 , q_0 \} \, \deltabar(\w) \right) \,,
\\ \langle \phi (\w') \, \phi (\w) \rangle \quad &\to \quad - \k \deltabar(\w' + \w) \deltabar'(\w) = - i \left( \deltabar(\w') \, \{ q_0 , p_0 \} \left[ - i \k \deltabar'(\w) \right] \right.
\\ &\phantom{\to \quad asdfasdfasdfasdfasdf} \left. \phantom{asdfasdf} + \left[ - i \k \deltabar'(\w') \right] \{ p_0 , q_0 \} \, \deltabar(\w) \right) \,.
\end{aligned} \label{eq:causality_cut_PB_form}
\end{align}
The combinations inside square brackets are intended to reproduce the combination appearing in \eqref{eq:vertex_amputation_rule}.

To show that causality cuts compute Poisson brackets of amputated subdiagrams, we consider the reverse problem of starting from the left/right subdiagrams $G_{L/R}$ and summing over all possible combinations of cuts constructed by selecting a vertex $V_{L/R}$ from each subdiagram $G_{L/R}$ and joining them by the cutting rules \eqref{eq:causality_cut_PB_form}. 
The augmentation operator $\Delta$ defined above \eqref{eq:vertex_augmentation_rule} is useful for this purpose.
Applying the cutting rules \eqref{eq:causality_cut_PB_form} to the sum over propagators joining the augmented vertices $\Delta V_L$ and $\Delta V_R$ generates the Poisson bracket between the original vertices.
\begin{align}
\begin{aligned}
\sum_{f,g = \pi, \phi} [\Delta_{f(\w')} V_L] \langle f(\w') \, g (\w) \rangle [\Delta_{g(\w)} V_R] \quad &\to \sum_{f_0, g_0 = p_0 , q_0} \frac{\partial V_L}{\partial f_0} \times - i \{ f_0 , g_0 \} \times \frac{\partial V_R}{\partial g_0}
\\ &\quad\quad\quad = - i \{ V_L , V_R \} \,.
\end{aligned} \label{eq:vert_cut2PB}
\end{align}
The integrand $I[G_L]$ of the left subdiagram $G_L$ is given by the product of Feynman rules for the vertices $V_i$ and edges $E_i$ of the diagram divided by the symmetry factor $S[G_L]$ associated to the diagram.
\begin{align}
I[G_L] &= \frac{1}{S[G_L]} \prod_{V_i \in G_L} V_i \prod_{E_i \in G_L} E_i \,.
\end{align}
Augmenting a vertex $V_a \in G_L$ converts $G_L$ to the diagram $G_{L \backslash a}$ where the vertex $V_a$ is ``colored'' and considered distinct from other vertices. We join the vertex $V_a$ with the vertex $V_R$ from the right side of the cut using the cutting rules \eqref{eq:vert_cut2PB} and sum over all possible choices of vertices $V_a$ that lead to distinct diagrams. The corresponding integrand is
\begin{align}
\left[ \sum_{V_a} \left( \frac{1}{S[G_{L \backslash a}]} \prod_{\substack{V_i \in G_L \\ V_i \neq V_a}} V_i \prod_{E_i \in G_L} E_i \right) \times - i \{ V_a , V_R \} \right] \times \cdots  = - i \{ I[G_L] , V_R \} \times \cdots \,,
\end{align}
where ellipsis denotes the factors from the right side of the cut. Note that RHS follows from the observation that the ratio $S[G_{L \backslash a}] / S[G_L]$ is the multiplicity of the vertex $V_a$, or the number of vertices equivalent to $V_a$ in $G_L$. Repeating the same process to the right side of the cut yields
\begin{align}
\begin{aligned}
&\sum_{V_a , V_b} \left( \frac{1}{S[G_{L \backslash a}]} \prod_{\substack{V_i \in G_L \\ V_i \neq V_a}} V_i \prod_{E_i \in G_L} E_i \right) \times - i \{ V_a , V_b \} \times \left( \frac{1}{S[G_{R \backslash b}]} \prod_{\substack{V_i \in G_R \\ V_i \neq V_b}} V_i \prod_{E_i \in G_R} E_i \right) 
\\ & \phantom{asdfasdf} = - i \{ I[G_L] , I[G_R] \} \,, 
\end{aligned}\label{eq:subdiagram_cut_prod}
\end{align}
where $V_b \in G_R$ was selected from the right subdiagram $G_R$. Since $S[G_{L \backslash a}] \times S[G_{R \backslash b}]$ is the symmetry factor of the diagram $G_{L \backslash a} \cup G_{R \backslash b}$ formed by adding an edge joining the vertex $V_a \in G_L$ and $V_b \in G_R$, reverting the cut bracket $- i \{ V_a , V_b \}$ on the first line of \eqref{eq:subdiagram_cut_prod} gives a sum over integrands of tree diagrams $G$ that can be bisected into $G_L$ and $G_R$ by cutting a single edge.

Generalization of causality cuts to include second-quantized field degrees of freedom should be relatively straightforward with Hamiltonian action (1st order action) for the field variables in the background field formalism. One caveat is the treatment of trivalent vertices in the amputation rules. Similar to the worldline 3pt vertices, the amputation rule for trivalent vertices is expected to become background field derivatives of 2pt functions with appropriate regularization scheme for the $i0^+$ prescription, which can become involved due to difficulty of computing propagators on nontrivial backgrounds. Moreover, the proof given in this section should be generalized to include Poisson brackets involving edges of subdiagrams.

\section{Operator power series and the CEM Hopf algebra} \label{app:OPS_Hopf}
The CEM Hopf algebra~\cite{Calaque_2011} can be motivated as a diagrammatic representation of \emph{composition of operators} by generalizing the presentation given in section~\ref{sec:CEM_Hopf} to power series of operators. We use a vertex as an abstract representation of an operator\textemdash the interaction Hamiltonian in our context\textemdash given as a (normal-ordered) power series in (abstract) fields\footnote{Abstract in the sense that $\phi$ stands for any (dynamical) field of the theory.}
\begin{align}
\parbox{10pt}{\fmfreuse{dot}}
    \Leftrightarrow \; \CO = \CO_{(0)} + \CO_{(1)} \phi + \cdots + \CN_{(j)} \CO_{(j)} \np{\phi^j} + \cdots \,,
\end{align}
where the colons $\np{AB}$ denote normal ordering and $\CO_{(j)}$ are the $c$-number coefficients attached to the normal-ordered $j$-th power of $\phi$, normalized by the combinatorial factor $\CN_{(j)}$. The monomial $\np{\phi^j}$ stands for any product of fields whose power adds up to $j$; e.g. in a theory with $n$ scalar fields $\Phi_i$, the monomial $\np{\phi^j}$ can stand for any monomial of the form
\begin{align}
    \np{ \Phi_1^{j_1} \cdots \Phi_n^{j_n} } \;\; \subset \;\; \np{\phi^j} \,,\quad j_1 + \cdots + j_n = j \,.
\end{align}
The corresponding combinatorial factor $\CN_{(j)} = \CN_{(j_1,\cdots,j_n)}$ is
\begin{align}
    \np{ \Phi_1^{j_1} \cdots \Phi_n^{j_n} } \quad \to \quad \CN_{(j_1,\cdots,j_n)} = \prod_{i=1}^n \frac{1}{(j_i)!} \,,
\end{align}
such that when the operator $\CN_{(j)} \CO_{(j)} \np{ \phi^j }$ is converted to Feynman rules, the coefficient $\CO_{(j)}$ becomes the corresponding vertex rule.

It is convenient to define the \emph{retarded Wightman functions},
\begin{align}
    G_{R}^\pm (x,y) := \th (x^0 - y^0) \, G^\pm (x,y) \,.
\end{align}
Definition for the Wightman functions $G^\pm (x,y)$ can be found in section~\ref{sec:Wightman}.
Although retarded Wightman functions are not Lorentz invariant, they can be completed into usual propagators which are Lorentz invariant.
\begin{align}
\begin{aligned}
    G_{R} (x,y) &= G_{R}^+ (x,y) - G_{R}^- (x,y) \,,
    \\ G_{F} (x,y) &= G_{R}^+ (x,y) + G_{R}^+ (y,x) \,.
\end{aligned}
\end{align}

We define \emph{directed contraction} between vertices as removing one field from each vertex and substituting by the retarded Wightman function, normal-ordering the remaining products of fields. For example, the directed contraction between $\CO_2$ and $\CO_1$ is,
\begin{align}
    \begin{fmffile}{dc21}
        \parbox{50pt}{
        \begin{fmfgraph*}(40,20)
            \fmfright{i1}
            \fmfleft{o1}
            \fmfv{decor.shape=circle,decor.size=4,label=$\CO_1$,label.angle=90}{i1}
            \fmfv{decor.shape=circle,decor.size=4,label=$\CO_2$,label.angle=90}{o1}
            \fmf{fermion}{i1,o1}
        \end{fmfgraph*}
        }
    \end{fmffile}
    &\Leftrightarrow \;
    \begin{aligned}
        & \CO_{2,(1)} \CO_{1,(1)} G_{R,21}^+ + \cdots
        \\ &\; + \CN_{(j_2-1;j_1-1)} \CO_{2,(j_2)} \CO_{1,(j_1)} G_{R,21}^+ \np{\phi_2^{j_2 - 1} \phi_1^{j_1 - 1}} + \cdots \,,
    \end{aligned}
\end{align}
where $G_{R,21}^+ = \th(t_2 - t_1) \langle 0 | \phi_2 \phi_1 | 0 \rangle$ is the retarded Wightman function. The combinatorial factor $\CN_{(j_2,j_1)}$ only depends on the non-contracted fields in the normal-ordered product, e.g.
\begin{align}
    \np{ \Phi_{2,1}^{j_{2,1}} \cdots \Phi_{2,n}^{j_{2,n}} \Phi_{1,1}^{j_{1,1}} \cdots \Phi_{1,n}^{j_{1,n}} } \quad \to \quad \CN_{(j_{2,1},\cdots,j_{2,n};j_{1,1},\cdots,j_{1,n})} = \prod_{i=1}^n \frac{1}{(j_{2,i})! (j_{1,i})!} \,.
\end{align}
A more nontrivial example is
\begin{align}
    \begin{fmffile}{dc21_31}
        \parbox{47pt}{
        \begin{fmfgraph*}(25,25)
            \fmfright{i1}
            \fmfleft{o1,o2}
            \fmfv{decor.shape=circle,decor.size=4,label=$\CO_1$}{i1}
            \fmfv{decor.shape=circle,decor.size=4,label=$\CO_2$,label.angle=-90}{o1}
            \fmfv{decor.shape=circle,decor.size=4,label=$\CO_3$,label.angle=90}{o2}
            \fmf{fermion}{i1,o1}
            \fmf{fermion}{i1,o2}
        \end{fmfgraph*}
        }
       \end{fmffile}
    \Leftrightarrow \;
    \begin{aligned}
        &\CO_{3,(1)} \CO_{2,(1)} \CO_{1,(2)} G_{R,31}^+ G_{R,21}^+ + \cdots
        \\ &\; + \CN_{(j_3-1;j_2-1;j_1-2)} \CO_{3,(j_3)} \CO_{2,(j_2)} \CO_{1,(j_1)} G_{R,31}^+ G_{R,21}^+ \np{\phi_3^{j_3 - 1} \phi_2^{j_2 - 1} \phi_1^{j_1 - 2}}
        \\ &\; + \cdots \,.
    \end{aligned}
\end{align}
The subscripts of the vertices are intended to aid tracking; we will consider all vertices to correspond to the same operator at different insertion points. In summary, a diagram with $k$ vertices represents the operator $\CO$ raised to the $k$-th power $(\CO)^k$ with fields removed by directed contractions represented by the directed edges.

Given an operator $\CO$, we can construct the ``power series'' operator $F[\CO]$ as a sum over all possible directed contractions, represented as a sum over directed graphs. We define the \emph{coefficient function} $f(\t)$ to the power series operator $F[\CO]$, which associates each directed graph $\t$ with the coefficient of its corresponding directed contraction $f (\t)$. More explicitly,
\begin{align}
\begin{aligned}
    F[\CO] &= \sum_{k = 0}^\infty \left[ \sum_{\t ,\, |\t|=k} f(\t) \, \t \right]
    \\ &= f(\emptyset) + f( \;\,
        \parbox{5pt}{\fmfreuse{dot}}
        ) \;\; 
        \parbox{10pt}{\fmfreuse{dot}}
        \hskip -4pt + \, f( \;\,
         \parbox{25pt}{\fmfreuse{dot2}}
       ) \;\, 
       \parbox{25pt}{\fmfreuse{dot2}}
       \, + \, f( \;\,
       \parbox{25pt}{\fmfreuse{dot3}}
       ) \;\, 
       \parbox{25pt}{\fmfreuse{dot3}}
       \, + \, \cdots \,,
\end{aligned}
\end{align}
where $|\t|$ denotes the number of vertices in the graph $\t$. The coefficient functions are understood to be multiplicative, i.e. for a graph $\t = \t_1 \cup \t_2$ consisting of disjoint graphs $\t_1$ and $\t_2$,
\begin{align}
    f(\t_1 \cup \t_2) &= f(\t_1) f(\t_2) \,,\quad \text{if } \t_1 \cap \t_2 = \emptyset \,.
\end{align}

Now consider the composition (convolution) of the operators $(G \star F) [\CO] = G [F [\CO]]$.
We draw the graphs corresponding to the ``power series'' expansion of $(G \star F) [\CO]$ through the following steps;
\begin{enumerate}
    \item Draw diagrams corresponding to $G[\CO']$, e.g.
    \begin{align}
        G[\CO'] \; &\Leftrightarrow g( \;\,
        \begin{fmffile}{pentagram}
        \parbox{5pt}{
        \begin{fmfgraph*}(5,5)\fmfkeep{pentagram}
            \fmfleft{i1}
            \fmfv{decor.shape=pentagram,decor.size=5}{i1}
        \end{fmfgraph*}
        }
        \end{fmffile}
        ) \;\, 
        \parbox{5pt}{\fmfreuse{pentagram}}
        \, + \, g( \;\,
       \begin{fmffile}{penta2}
        \parbox{25pt}{
        \begin{fmfgraph*}(20,20)\fmfkeep{penta2}
            \fmfright{i1}
            \fmfleft{o1}
            \fmfv{decor.shape=pentagram,decor.size=5}{i1}
            \fmfv{decor.shape=pentagram,decor.size=5}{o1}
            \fmf{fermion}{i1,o1}
        \end{fmfgraph*}
        }
       \end{fmffile}
       ) \;\, 
       \parbox{25pt}{\fmfreuse{penta2}}
       \, + \, g( \;\,
       \parbox{25pt}{\fmfreuse{penta3}} 
       ) \;\, 
       \parbox{25pt}{\fmfreuse{penta3}}
       \, + \, \cdots \,,
    \end{align}
    where the pentagram vertex corresponds to the operator $\CO'$ and we have explicitly written the coefficients $g(\tau)$ corresponding to the graph $\tau$.
    \item Substitute the pentagram vertices by the expansion of $\CO' = F[\CO]$, i.e.
    \begin{align}
    \begin{aligned}
        \CO' = F[\CO] \; \Leftrightarrow \;\, 
        \parbox{5pt}{\fmfreuse{pentagram}}
        &= f( \;\,
        \parbox{5pt}{\fmfreuse{dot}}
        ) \;\; 
        \parbox{10pt}{\fmfreuse{dot}}
        \hskip -4pt + \, f( \;\,
        \parbox{25pt}{\fmfreuse{dot2}}
       ) \;\, 
       \parbox{25pt}{\fmfreuse{dot2}}
       \, + \, f( \;\,
       \parbox{25pt}{\fmfreuse{dot3}}
       ) \;\, 
       \parbox{25pt}{\fmfreuse{dot3}}
       \, + \, \cdots \,,
    \end{aligned}
    \end{align}
    where the circular vertex corresponds to the operator $\CO$. Similarly, the coefficients $f(\tau)$ were explicitly written down for the corresponding graphs $\tau$.
\end{enumerate}
The CEM Hopf algebra relation is obtained by reorganizing the resulting diagrams,
\begin{align}
    (g \star f) (\tau) &= \sum_{p \in \CP(\t)} g ( p_\t ) f (\t \backslash p) \,,
\end{align}
where $(g \star f)$ are the coefficients of the convoluted operator $(G \star F) [\CO]$. The notation for the graphs is slightly different from that of section~\ref{sec:Murua_root} as we do not require the graph to be rooted tree nor the partition $p$ to be connected; $\CP(\t)$ is the set of all partitions (choice of edges) $p$ of a graph $\t$, the remainder $\t \backslash p$ is the (possibly disconnected) graph obtained from $\t$ by removing the partition $p$, and $p_\t$ (called the \emph{skeleton}) is the graph obtained from $\t$ by shrinking connected subgraphs of $\t \backslash p$ to a vertex.

In the Magnus expansion, the tree function $e(\t)$ computes the coefficient of directed contractions in the Dyson series. Therefore, we may understand the CEM Hopf algebra relation as iteratively solving the ``operator differential equation'' $e^{d\O} = \iden + H dt$ to obtain coefficients of the directed contractions appearing in the Magnus expansion. 
We remark that the same Hopf algebra also seems to solve the coefficients of directed contractions for loop graphs when the function $e(\t)$ is appropriately generalized to looped directed graphs. 

Note that terms corresponding to directed contractions only constitute a subset of terms appearing in the Magnus expansion. However, knowing the coefficients of directed contractions is enough to reconstruct the full Magnus expansion for tree graphs, as all retarded Wightman functions $G_{R,jk}^+$ must appear in the combination of retarded Green's functions, $G_{R,jk} = G_{R,jk}^+ - G_{R,jk}^-$, due to the nested commutator structure of the Magnus expansion. The retarded Wightman functions do not follow this pattern at loop level, which is an obstacle for constructing a ``diagrammar'' for the Magnus expansion corresponding to Feynman rules for the Dyson series.

\newpage
\bibliographystyle{JHEP}
\bibliography{biblio}

\providecommand{\href}[2]{#2}\begingroup\raggedright\begin{thebibliography}{10}

\bibitem{Landau:1975pou}
L.D.~Landau and E.M.~Lifschits, \emph{{The Classical Theory of Fields}},
  vol.~Volume 2 of \emph{Course of Theoretical Physics}, Pergamon Press, Oxford
  (1975).

\bibitem{Schrodinger:1926iou}
E.~Schr\"odinger, \emph{{An Undulatory Theory of the Mechanics of Atoms and
  Molecules}}, \href{https://doi.org/10.1103/PhysRev.28.1049}{\emph{Phys. Rev.}
  {\bfseries 28} (1926) 1049}.

\bibitem{Sakurai:2011zz}
J.J.~Sakurai and J.~Napolitano, \emph{{Modern Quantum Mechanics}}, Quantum
  physics, quantum information and quantum computation, Cambridge University
  Press (10, 2020),
  \href{https://doi.org/10.1017/9781108587280}{10.1017/9781108587280}.

\bibitem{Torgerson:1966zz}
R.~Torgerson, \emph{{Field-Theoretic Formulation of the Optical Model at High
  Energies}}, \href{https://doi.org/10.1103/PhysRev.143.1194}{\emph{Phys. Rev.}
  {\bfseries 143} (1966) 1194}.

\bibitem{Cheng:1969eh}
H.~Cheng and T.T.~Wu, \emph{{High-energy elastic scattering in quantum
  electrodynamics}},
  \href{https://doi.org/10.1103/PhysRevLett.22.666}{\emph{Phys. Rev. Lett.}
  {\bfseries 22} (1969) 666}.

\bibitem{Levy:1969cr}
M.~Levy and J.~Sucher, \emph{{Eikonal approximation in quantum field theory}},
  \href{https://doi.org/10.1103/PhysRev.186.1656}{\emph{Phys. Rev.} {\bfseries
  186} (1969) 1656}.

\bibitem{Abarbanel:1969ek}
H.D.I.~Abarbanel and C.~Itzykson, \emph{{Relativistic eikonal expansion}},
  \href{https://doi.org/10.1103/PhysRevLett.23.53}{\emph{Phys. Rev. Lett.}
  {\bfseries 23} (1969) 53}.

\bibitem{Tiktopoulos:1971hi}
G.~Tiktopoulos and S.B.~Treiman, \emph{{Relativistic eikonal approximation}},
  \href{https://doi.org/10.1103/PhysRevD.3.1037}{\emph{Phys. Rev. D} {\bfseries
  3} (1971) 1037}.

\bibitem{Meng:1972xt}
T.-C.~Meng, \emph{{High-energy scattering of a charged vector meson in a static
  field - simple exponentiation and s-channel helicity conservation}},
  \href{https://doi.org/10.1103/PhysRevD.6.1169}{\emph{Phys. Rev. D} {\bfseries
  6} (1972) 1169}.

\bibitem{Weinberg:1971cdi}
S.~Weinberg, \emph{{Exponentiation and sum rules}},
  \href{https://doi.org/10.1016/0370-2693(71)90354-6}{\emph{Phys. Lett. B}
  {\bfseries 37} (1971) 494}.

\bibitem{Czyz:1975bf}
W.~Czyz and P.K.~Kabir, \emph{{High-Energy Scattering of Spinning Particles by
  External Fields and 'Exponentiation'}},
  \href{https://doi.org/10.1103/PhysRevD.11.2219}{\emph{Phys. Rev. D}
  {\bfseries 11} (1975) 2219}.

\bibitem{Kabat:1992pz}
D.N.~Kabat, \emph{{Validity of the Eikonal approximation}}, {\emph{Comments
  Nucl. Part. Phys.} {\bfseries 20} (1992) 325}
  [\href{https://arxiv.org/abs/hep-th/9204103}{{\ttfamily hep-th/9204103}}].

\bibitem{Feynman:1951gn}
R.P.~Feynman, \emph{{An Operator calculus having applications in quantum
  electrodynamics}}, \href{https://doi.org/10.1103/PhysRev.84.108}{\emph{Phys.
  Rev.} {\bfseries 84} (1951) 108}.

\bibitem{Lehmann:1957zz}
H.~Lehmann, K.~Symanzik and W.~Zimmermann, \emph{{On the formulation of
  quantized field theories. II}},
  \href{https://doi.org/10.1007/BF02832508}{\emph{Nuovo Cim.} {\bfseries 6}
  (1957) 319}.

\bibitem{Damgaard:2021ipf}
P.H.~Damgaard, L.~Plante and P.~Vanhove, \emph{{On an exponential
  representation of the gravitational S-matrix}},
  \href{https://doi.org/10.1007/JHEP11(2021)213}{\emph{JHEP} {\bfseries 11}
  (2021) 213} [\href{https://arxiv.org/abs/2107.12891}{{\ttfamily
  2107.12891}}].

\bibitem{Damgaard:2023ttc}
P.H.~Damgaard, E.R.~Hansen, L.~Plant\'e and P.~Vanhove, \emph{{Classical
  observables from the exponential representation of the gravitational
  S-matrix}}, \href{https://doi.org/10.1007/JHEP09(2023)183}{\emph{JHEP}
  {\bfseries 09} (2023) 183}
  [\href{https://arxiv.org/abs/2307.04746}{{\ttfamily 2307.04746}}].

\bibitem{DiVecchia:2021bdo}
P.~Di~Vecchia, C.~Heissenberg, R.~Russo and G.~Veneziano, \emph{{The eikonal
  approach to gravitational scattering and radiation at $ \mathcal{O}
  $(G$^{3}$)}}, \href{https://doi.org/10.1007/JHEP07(2021)169}{\emph{JHEP}
  {\bfseries 07} (2021) 169}
  [\href{https://arxiv.org/abs/2104.03256}{{\ttfamily 2104.03256}}].

\bibitem{Cristofoli:2021vyo}
A.~Cristofoli, R.~Gonzo, D.A.~Kosower and D.~O'Connell, \emph{{Waveforms from
  amplitudes}}, \href{https://doi.org/10.1103/PhysRevD.106.056007}{\emph{Phys.
  Rev. D} {\bfseries 106} (2022) 056007}
  [\href{https://arxiv.org/abs/2107.10193}{{\ttfamily 2107.10193}}].

\bibitem{DiVecchia:2023frv}
P.~Di~Vecchia, C.~Heissenberg, R.~Russo and G.~Veneziano, \emph{{The
  gravitational eikonal: from particle, string and brane collisions to
  black-hole encounters}},  \href{https://arxiv.org/abs/2306.16488}{{\ttfamily
  2306.16488}}.

\bibitem{Kim:2024grz}
J.-H.~Kim, J.-W.~Kim and S.~Lee, \emph{{Massive twistor worldline in
  electromagnetic fields}},
  \href{https://doi.org/10.1007/JHEP08(2024)080}{\emph{JHEP} {\bfseries 08}
  (2024) 080} [\href{https://arxiv.org/abs/2405.17056}{{\ttfamily
  2405.17056}}].

\bibitem{Kosower:2018adc}
D.A.~Kosower, B.~Maybee and D.~O'Connell, \emph{{Amplitudes, Observables, and
  Classical Scattering}},
  \href{https://doi.org/10.1007/JHEP02(2019)137}{\emph{JHEP} {\bfseries 02}
  (2019) 137} [\href{https://arxiv.org/abs/1811.10950}{{\ttfamily
  1811.10950}}].

\bibitem{Gonzo:2024zxo}
R.~Gonzo and C.~Shi, \emph{{Scattering and bound observables for spinning
  particles in Kerr spacetime with generic spin orientations}},
  \href{https://arxiv.org/abs/2405.09687}{{\ttfamily 2405.09687}}.

\bibitem{Magnus:1954zz}
W.~Magnus, \emph{{On the exponential solution of differential equations for a
  linear operator}},
  \href{https://doi.org/10.1002/cpa.3160070404}{\emph{Commun. Pure Appl. Math.}
  {\bfseries 7} (1954) 649}.

\bibitem{Blanes:2008xlr}
S.~Blanes, F.~Casas, J.A.~Oteo and J.~Ros, \emph{{The Magnus expansion and some
  of its applications}},
  \href{https://doi.org/10.1016/j.physrep.2008.11.001}{\emph{Phys. Rept.}
  {\bfseries 470} (2009) 151}.

\bibitem{ebrahimifard2023magnusexpansion}
K.~Ebrahimi-Fard, I.~Mencattini and A.~Quesney, \emph{What is the magnus
  expansion?},  2023.

\bibitem{Mogull:2020sak}
G.~Mogull, J.~Plefka and J.~Steinhoff, \emph{{Classical black hole scattering
  from a worldline quantum field theory}},
  \href{https://doi.org/10.1007/JHEP02(2021)048}{\emph{JHEP} {\bfseries 02}
  (2021) 048} [\href{https://arxiv.org/abs/2010.02865}{{\ttfamily
  2010.02865}}].

\bibitem{Ajith:2024fna}
S.~Ajith, Y.~Du, R.~Rajagopal and D.~Vaman, \emph{{Worldline Formalism, Eikonal
  Expansion and the Classical Limit of Scattering Amplitudes}},
  \href{https://arxiv.org/abs/2409.17866}{{\ttfamily 2409.17866}}.

\bibitem{Du:2024rkf}
Y.~Du, S.~Ajith, R.~Rajagopal and D.~Vaman, \emph{{Worldline Proof of Eikonal
  Exponentiation}},  \href{https://arxiv.org/abs/2409.12895}{{\ttfamily
  2409.12895}}.

\bibitem{Kim:2023vgb}
J.-H.~Kim, \emph{{Asymptotic Spinspacetime}},
  \href{https://arxiv.org/abs/2309.11886}{{\ttfamily 2309.11886}}.

\bibitem{Argeri:2014qva}
M.~Argeri, S.~Di~Vita, P.~Mastrolia, E.~Mirabella, J.~Schlenk, U.~Schubert
  et~al., \emph{{Magnus and Dyson Series for Master Integrals}},
  \href{https://doi.org/10.1007/JHEP03(2014)082}{\emph{JHEP} {\bfseries 03}
  (2014) 082} [\href{https://arxiv.org/abs/1401.2979}{{\ttfamily 1401.2979}}].

\bibitem{taylor2012scattering}
J.~Taylor, \emph{Scattering Theory: The Quantum Theory of Nonrelativistic
  Collisions}, Dover Books on Engineering, Dover Publications (2012).

\bibitem{Murua_2006}
A.~Murua, \emph{The hopf algebra of rooted trees, free lie algebras, and lie
  series}, \href{https://doi.org/10.1007/s10208-003-0111-0}{\emph{Foundations
  of Computational Mathematics} {\bfseries 6} (2006) 387}.

\bibitem{penrose2007road}
R.~Penrose, \emph{The Road to Reality: A Complete Guide to the Laws of the
  Universe}, Knopf Doubleday Publishing Group (2007).

\bibitem{Chartier_2010}
P.~Chartier, E.~Hairer and G.~Vilmart, \emph{Algebraic structures of b-series},
  \href{https://doi.org/10.1007/s10208-010-9065-1}{\emph{Foundations of
  Computational Mathematics} {\bfseries 10} (2010) 407}.

\bibitem{Calaque_2011}
D.~Calaque, K.~Ebrahimi-Fard and D.~Manchon, \emph{Two interacting hopf
  algebras of trees: A hopf-algebraic approach to composition and substitution
  of b-series},
  \href{https://doi.org/https://doi.org/10.1016/j.aam.2009.08.003}{\emph{Advances
  in Applied Mathematics} {\bfseries 47} (2011) 282}.

\bibitem{Connes:1998qv}
A.~Connes and D.~Kreimer, \emph{{Hopf algebras, renormalization and
  noncommutative geometry}},
  \href{https://doi.org/10.1007/s002200050499}{\emph{Commun. Math. Phys.}
  {\bfseries 199} (1998) 203}
  [\href{https://arxiv.org/abs/hep-th/9808042}{{\ttfamily hep-th/9808042}}].

\bibitem{Birrell:1982ix}
N.D.~Birrell and P.C.W.~Davies, \emph{{Quantum Fields in Curved Space}},
  Cambridge Monographs on Mathematical Physics, Cambridge Univ. Press,
  Cambridge, UK (2, 1984),
  \href{https://doi.org/10.1017/CBO9780511622632}{10.1017/CBO9780511622632}.

\bibitem{Peierls:1952cb}
R.E.~Peierls, \emph{{The Commutation laws of relativistic field theory}},
  \href{https://doi.org/10.1098/rspa.1952.0158}{\emph{Proc. Roy. Soc. Lond. A}
  {\bfseries 214} (1952) 143}.

\bibitem{xAct}
{Mart{\'\i}n-Garc{\'i}a, J. M.}, ``{xAct: Efficient tensor computer algebra for
  the Wolfram Language}.'' \url{http://xact.es/}.

\bibitem{Lee:2012cn}
R.N.~Lee, \emph{{Presenting LiteRed: a tool for the Loop InTEgrals REDuction}},
   \href{https://arxiv.org/abs/1212.2685}{{\ttfamily 1212.2685}}.

\bibitem{Jakobsen:2023oow}
G.U.~Jakobsen, \emph{{Gravitational Scattering of Compact Bodies from Worldline
  Quantum Field Theory}}, Ph.D. thesis, Humboldt U., Berlin, Humboldt U.,
  Berlin (main), 2023.
\newblock \href{https://arxiv.org/abs/2308.04388}{{\ttfamily 2308.04388}}.
\newblock 10.18452/27075.

\bibitem{Jakobsen:2022psy}
G.U.~Jakobsen, G.~Mogull, J.~Plefka and B.~Sauer, \emph{{All things retarded:
  radiation-reaction in worldline quantum field theory}},
  \href{https://doi.org/10.1007/JHEP10(2022)128}{\emph{JHEP} {\bfseries 10}
  (2022) 128} [\href{https://arxiv.org/abs/2207.00569}{{\ttfamily
  2207.00569}}].

\bibitem{Brandhuber:2021eyq}
A.~Brandhuber, G.~Chen, G.~Travaglini and C.~Wen, \emph{{Classical
  gravitational scattering from a gauge-invariant double copy}},
  \href{https://doi.org/10.1007/JHEP10(2021)118}{\emph{JHEP} {\bfseries 10}
  (2021) 118} [\href{https://arxiv.org/abs/2108.04216}{{\ttfamily
  2108.04216}}].

\bibitem{Jakobsen:2023hig}
G.U.~Jakobsen, G.~Mogull, J.~Plefka and B.~Sauer, \emph{{Dissipative Scattering
  of Spinning Black Holes at Fourth Post-Minkowskian Order}},
  \href{https://doi.org/10.1103/PhysRevLett.131.241402}{\emph{Phys. Rev. Lett.}
  {\bfseries 131} (2023) 241402}
  [\href{https://arxiv.org/abs/2308.11514}{{\ttfamily 2308.11514}}].

\bibitem{Bjerrum-Bohr:2021din}
N.E.J.~Bjerrum-Bohr, P.H.~Damgaard, L.~Plant\'e and P.~Vanhove, \emph{{The
  amplitude for classical gravitational scattering at third Post-Minkowskian
  order}}, \href{https://doi.org/10.1007/JHEP08(2021)172}{\emph{JHEP}
  {\bfseries 08} (2021) 172}
  [\href{https://arxiv.org/abs/2105.05218}{{\ttfamily 2105.05218}}].

\bibitem{Herrmann:2021tct}
E.~Herrmann, J.~Parra-Martinez, M.S.~Ruf and M.~Zeng, \emph{{Radiative
  classical gravitational observables at $ \mathcal{O} $(G$^{3}$) from
  scattering amplitudes}},
  \href{https://doi.org/10.1007/JHEP10(2021)148}{\emph{JHEP} {\bfseries 10}
  (2021) 148} [\href{https://arxiv.org/abs/2104.03957}{{\ttfamily
  2104.03957}}].

\bibitem{Kalin:2020fhe}
G.~K\"alin, Z.~Liu and R.A.~Porto, \emph{{Conservative Dynamics of Binary
  Systems to Third Post-Minkowskian Order from the Effective Field Theory
  Approach}}, \href{https://doi.org/10.1103/PhysRevLett.125.261103}{\emph{Phys.
  Rev. Lett.} {\bfseries 125} (2020) 261103}
  [\href{https://arxiv.org/abs/2007.04977}{{\ttfamily 2007.04977}}].

\bibitem{Kalin:2022hph}
G.~K\"alin, J.~Neef and R.A.~Porto, \emph{{Radiation-reaction in the Effective
  Field Theory approach to Post-Minkowskian dynamics}},
  \href{https://doi.org/10.1007/JHEP01(2023)140}{\emph{JHEP} {\bfseries 01}
  (2023) 140} [\href{https://arxiv.org/abs/2207.00580}{{\ttfamily
  2207.00580}}].

\bibitem{DiVecchia:2022piu}
P.~Di~Vecchia, C.~Heissenberg, R.~Russo and G.~Veneziano, \emph{{Classical
  gravitational observables from the Eikonal operator}},
  \href{https://doi.org/10.1016/j.physletb.2023.138049}{\emph{Phys. Lett. B}
  {\bfseries 843} (2023) 138049}
  [\href{https://arxiv.org/abs/2210.12118}{{\ttfamily 2210.12118}}].

\bibitem{Kim:2023aff}
J.-H.~Kim and S.~Lee, \emph{{Symplectic Perturbation Theory in Massive
  Ambitwistor Space: A Zig-Zag Theory of Massive Spinning Particles}},
  \href{https://arxiv.org/abs/2301.06203}{{\ttfamily 2301.06203}}.

\bibitem{Cheung:2018wkq}
C.~Cheung, I.Z.~Rothstein and M.P.~Solon, \emph{{From Scattering Amplitudes to
  Classical Potentials in the Post-Minkowskian Expansion}},
  \href{https://doi.org/10.1103/PhysRevLett.121.251101}{\emph{Phys. Rev. Lett.}
  {\bfseries 121} (2018) 251101}
  [\href{https://arxiv.org/abs/1808.02489}{{\ttfamily 1808.02489}}].

\bibitem{Weinberg:1965nx}
S.~Weinberg, \emph{{Infrared photons and gravitons}},
  \href{https://doi.org/10.1103/PhysRev.140.B516}{\emph{Phys. Rev.} {\bfseries
  140} (1965) B516}.

\bibitem{Bern:2005iz}
Z.~Bern, L.J.~Dixon and V.A.~Smirnov, \emph{{Iteration of planar amplitudes in
  maximally supersymmetric Yang-Mills theory at three loops and beyond}},
  \href{https://doi.org/10.1103/PhysRevD.72.085001}{\emph{Phys. Rev. D}
  {\bfseries 72} (2005) 085001}
  [\href{https://arxiv.org/abs/hep-th/0505205}{{\ttfamily hep-th/0505205}}].

\bibitem{Cheng:1971gf}
H.~Cheng and T.T.~Wu, \emph{{High-energy scattering of a fermion with anomalous
  magnetic moment - nonexponentiation}},
  \href{https://doi.org/10.1103/PhysRevD.3.2394}{\emph{Phys. Rev. D} {\bfseries
  3} (1971) 2394}.

\bibitem{Chen:2022clh}
W.-M.~Chen, M.-Z.~Chung, Y.-t.~Huang and J.-W.~Kim, \emph{{Gravitational
  Faraday effect from on-shell amplitudes}},
  \href{https://doi.org/10.1007/JHEP12(2022)058}{\emph{JHEP} {\bfseries 12}
  (2022) 058} [\href{https://arxiv.org/abs/2205.07305}{{\ttfamily
  2205.07305}}].

\bibitem{Bjerrum-Bohr:2019kec}
N.E.J.~Bjerrum-Bohr, A.~Cristofoli and P.H.~Damgaard, \emph{{Post-Minkowskian
  Scattering Angle in Einstein Gravity}},
  \href{https://doi.org/10.1007/JHEP08(2020)038}{\emph{JHEP} {\bfseries 08}
  (2020) 038} [\href{https://arxiv.org/abs/1910.09366}{{\ttfamily
  1910.09366}}].

\bibitem{Damgaard:2022jem}
P.H.~Damgaard, J.~Hoogeveen, A.~Luna and J.~Vines, \emph{{Scattering angles in
  Kerr metrics}},
  \href{https://doi.org/10.1103/PhysRevD.106.124030}{\emph{Phys. Rev. D}
  {\bfseries 106} (2022) 124030}
  [\href{https://arxiv.org/abs/2208.11028}{{\ttfamily 2208.11028}}].

\bibitem{Cheung:2024byb}
C.~Cheung, J.~Parra-Martinez, I.Z.~Rothstein, N.~Shah and J.~Wilson-Gerow,
  \emph{{Gravitational scattering and beyond from extreme mass ratio effective
  field theory}}, \href{https://doi.org/10.1007/JHEP10(2024)005}{\emph{JHEP}
  {\bfseries 10} (2024) 005}
  [\href{https://arxiv.org/abs/2406.14770}{{\ttfamily 2406.14770}}].

\end{thebibliography}\endgroup

\end{document}